\documentclass[prc,twocolumn,showpacs,amsmath,amssymb,superscriptaddress,floatfix,nofootinbib,10pt]{revtex4-1}

\usepackage{slashed}
\usepackage{mathrsfs,bm}
\usepackage{soul}
\usepackage{txfonts}
\usepackage{amssymb}
\usepackage{indentfirst}
\usepackage{graphicx,booktabs}
\usepackage{multirow}
\usepackage{overpic}
\usepackage{color}
\usepackage{amssymb}
\usepackage{bm}
\usepackage{hyperref,bookmark}
\usepackage[utf8]{inputenc}

\newcommand{\mpv}[1]{{#1}}
\newcommand{\mpvrev}[1]{{#1}}

\begin{document}
\title{Spin-parities of the $P_c(4440)$ and $P_c(4457)$  in the One-Boson-Exchange Model}

\author{Ming-Zhu Liu}
\affiliation{School of Space and Environment, \\
Beihang University, Beijing 100191, China} 
\affiliation{School of Physics, Beihang University, Beijing 100191, China} 

\author{Tian-Wei Wu}
\affiliation{School of Physics, Beihang University, Beijing 100191, China} 

\author{Mario S\'anchez S\'anchez}
\affiliation{Centre d'\'Etudes Nucl\'eaires, CNRS/IN2P3, Universit\'e de Bordeaux, 33175 Gradignan, France}

\author{Manuel Pavon Valderrama}\email{mpavon@buaa.edu.cn}
\affiliation{School of Physics, Beihang University, Beijing 100191, China} 
\affiliation{
International Research Center for Nuclei and Particles in the Cosmos and \\
Beijing Key Laboratory of Advanced Nuclear Materials and Physics, \\
Beihang University, Beijing 100191, China} 

\author{Li-Sheng Geng}\email{lisheng.geng@buaa.edu.cn}
\affiliation{School of Physics, Beihang University, Beijing 100191, China} 
\affiliation{
International Research Center for Nuclei and Particles in the Cosmos and \\
Beijing Key Laboratory of Advanced Nuclear Materials and Physics, \\
Beihang University, Beijing 100191, China} 
  \affiliation{School of Physics and Microelectronics,
    Zhengzhou University, Zhengzhou, Henan 450001, China}

\author{Ju-Jun Xie}
\affiliation{Institute of Modern Physics, Chinese Academy of
  Sciences, Lanzhou 730000, China}
  \affiliation{School of Physics and Microelectronics,
Zhengzhou University, Zhengzhou, Henan 450001, China}
  \affiliation{School of Nuclear Science and Technology, University of Chinese Academy of Sciences, Beijing 101408, China}

\date{\today}
\begin{abstract}
  The LHCb collaboration has recently observed three pentaquark peaks,
  the $P_c(4312)$, $P_c(4440)$ and $P_c(4457)$.
  They are very close to a pair of heavy baryon-meson thresholds, with
  the $P_c(4312)$ located $8.9\,{\rm MeV}$ below the $\bar{D} \Sigma_c$
  threshold, and the $P_c(4440)$ and $P_c(4457)$ located $21.8$ and
  $4.8\,{\rm MeV}$ below the $\bar{D}^* \Sigma_c$ one.
  The spin-parities of these three states have not been measured yet.
  In this work we assume that the $P_c(4312)$ is a $J^P = \tfrac{1}{2}^{-}$
  $\bar{D} \Sigma_c$ bound state, while the $P_c(4440)$ and $P_c(4457)$
  are $\bar{D}^* \Sigma_c$ bound states of unknown spin-parity,
  where we notice that the consistent description of the three pentaquarks
  in the one-boson-exchange model can indeed determine the spin and
  parities of the later, i.e. of the two $\bar{D}^* \Sigma_c$
  molecular candidates.
  For this determination we revisit first the one-boson-exchange
  model,
  which in its original formulation contains a short-range delta-like
  contribution in the spin-spin component of the potential.
  We argue that it is better to remove these delta-like contributions
  because, in this way, the one-boson-exchange potential will comply
  with the naive expectation that the form factors should
  not have a significant impact in the long-range part of
  the potential (in particular the one-pion-exchange part).
  Once this is done, we find that it is possible to consistently describe
  the three pentaquarks, to the point that the $P_c(4440)$ and $P_c(4457)$
  can be predicted from the $P_c(4312)$ within a couple of MeV
  with respect to their experimental location.
  In addition the so-constructed one-boson-exchange model predicts the preferred
  quantum numbers of the $P_c(4440)$ and $P_c(4457)$ molecular pentaquarks
  to be $\tfrac{3}{2}^-$ and $\tfrac{1}{2}^-$, respectively.
\end{abstract}


\maketitle
\section{Introduction}

The observation of three pentaquark-like resonances by the
LHCb collaboration~\cite{Aaij:2019vzc} ---
the $P_c(4312)$, $P_c(4440)$ and $P_c(4457)$ ---
provides three of the most robust candidates so far for a hadronic molecule,
a type of exotic hadron conjectured four decades
ago~\cite{Voloshin:1976ap,DeRujula:1976qd}.
As a matter of fact molecular pentaquarks, i.e. bound states
of a charmed antimeson and a charmed baryon, were predicted in a series
of theoretical works~\cite{Wu:2010jy,Wu:2010vk,Xiao:2013yca,Karliner:2015ina,Wang:2011rga,Yang:2011wz}.
Subsequent theoretical analyses
after the experimental observation of
the LHCb pentaquarks~\cite{Aaij:2019vzc}, \mpv{
have further explored the molecular hypothesis~\cite{Chen:2019bip,Chen:2019asm,He:2019ify,Liu:2019tjn,Xiao:2019aya,Shimizu:2019ptd,Guo:2019kdc,Fernandez-Ramirez:2019koa,Wu:2019rog,Valderrama:2019chc} or considered non-molecular explanations~\cite{Eides:2019tgv,Wang:2019got,Cheng:2019obk,Ferretti:2020ewe,Stancu:2020paw},}
indicated the importance of their decays to confirm (or falsify)
their nature~\cite{Guo:2019fdo,Xiao:2019mvs,Voloshin:2019aut,Sakai:2019qph} 
and discussed the existence of new unobserved pentaquark
states~\cite{Shen:2019evi,Xiao:2019gjd}.
{However a crucial piece of information required to determine the nature
   of the LHCb pentaquarks are their quantum numbers, which have not
   been experimentally determined yet, for which molecular and
   non-molecular interpretations usually yield
   different predictions \mpv{(with molecular interpretations overwhelmingly
   preferring negative parity)}.
}
It is interesting to notice that the $P_c(4440)$ and $P_c(4457)$ were previously
identified as a single peak, the $P_c(4450)$~\cite{Aaij:2015tga},
where the later collection of data by the LHCb has finally
uncovered the double peak structure.
In this regard, the previous investigations about the old $P_c(4450)$ peak
are still expected to be largely relevant for the new peaks,
from its \mpv{nature (be it either molecular~\cite{Roca:2015dva,He:2015cea,Xiao:2015fia,Chen:2015loa,Chen:2015moa,Meissner:2015mza} or non-molecular~\cite{Kubarovsky:2015aaa,Diakonov:1997mm,Jaffe:2003sg,Yuan:2012wz,Maiani:2015vwa,Lebed:2015tna,Li:2015gta})}
to its possible partner states~\cite{Liu:2018zzu},
the role of the $\bar{D} \Lambda_{c}(2595)$
threshold~\cite{Burns:2015dwa,Geng:2017hxc}, etc.

Hadronic molecules are bound states among two or more hadrons.
Their existence is contingent on the hadron-hadron potential.
In this regard the one-boson-exchange (OBE) model {provides}
a physically compelling and intuitive picture of
hadron interactions~\cite{Machleidt:1987hj,Machleidt:1989tm},
which can help us to predict prospective molecular states.
According to this model, the potential between two hadrons is a consequence of
the exchange of a series of light mesons, of which the most prominent ones
are the $\pi$, $\sigma$, $\rho$ and $\omega$ mesons.
The OBE model was originally developed to describe the nucleon-nucleon potential,
providing the first quantitative successful {description} of nuclear
scattering observables and
the deuteron~\cite{Machleidt:1987hj,Machleidt:1989tm}.
Besides, it also provided the original theoretical motivation
for the existence of hadronic molecules~\cite{Voloshin:1976ap},
with subsequent explorations often relying {on} this model
to make predictions or to explain already
known states~\cite{Liu:2008tn,Sun:2011uh,Chen:2015loa,Wang:2019nwt}.

The OBE model is endowed with a certain degree of ambiguity though.
The most important limitation of the OBE model is that it requires
the introduction of form factors and cutoffs to mimic
the finite size of the hadrons involved.
The cutoff cannot be determined {\it a priori} and is in principle \mpv{dependent on external information (e.g. experimental measurements).}
Even if the cutoff is required to be of natural-size--
we expect it to lie within the $1-2\,{\rm GeV}$ range ---
this still leaves room for wildly different predictions.
Yet, when applied to hadronic molecules, such a limitation is easy to overcome
provided that there is a clear molecular candidate: the cutoff can be effectively
determined from the condition of reproducing the aforementioned molecular
candidate~\cite{Liu:2018bkx,Liu:2019stu}.

But phenomenological models, even the most successful ones such as
the OBE model, usually end up requiring a certain amount of tweaking
(see for instance Ref.~\cite{Cordon:2009pj} for a lucid exposition of
a few of the quirks of the OBE model).
{
For the OBE model as applied to nuclear physics, it was quickly realized that
the correct description of the deuteron properties requires
the cutoff to be $\Lambda_{\pi} > 1.3\,{\rm GeV}$ for the pion contribution.
The theoretical reason is the distortion of the long-range properties
of the one-pion-exchange (OPE) potential by the form factors, which
can be prevented if the cutoff is hard enough.
The present manuscript indicates that this type of long-range distortion
also happens for hadronic molecules, but proposes a different solution
adapted to the particular circumstances of the application of
the OBE model to the molecular pentaquarks.
}

The problem is as follows: if the $P_c(4312)$ is indeed
a $J^P = \tfrac{1}{2}^{-}$ $\bar{D} \Sigma_c$ molecule
with a binding energy of $8.9\,{\rm MeV}$,
it can be described within the OBE model
with a monopolar form factor and
a cutoff $\Lambda = 1119\,{\rm MeV}$.
If we use the simplest OBE model possible, i.e. we use the same form
factor and cutoff for all the exchanged mesons,
we can  predict the $J^P = \tfrac{1}{2}^{-}$ and
$J^P = \tfrac{3}{2}^{-}$ $\bar{D}^* \Sigma_c$ binding energies
from the cutoff that we already determined from the $P_c(4312)$.
In particular we arrive at
\begin{eqnarray}
  B_E(\tfrac{1}{2}^-) \simeq 74\,{\rm MeV}
  \quad \mbox{and} \quad
  B_E(\tfrac{3}{2}^-) \simeq 3\,{\rm MeV} \, ,
  \label{eq:naive-OBE-Pc}
\end{eqnarray}
which are incompatible with the binding energies of the $P_c(4440)$ and
$P_c(4457)$ as $\bar{D}^* \Sigma_c$ bound states, $B_E = 21.8$ and
$4.8\,{\rm MeV}$ respectively.
This happens regardless of which state we identify with the $J^P = \tfrac{1}{2}^-$
and $J^P = \tfrac{3}{2}^-$ quantum numbers.
The failure of the naive OBE model to naturally explain the $P_c(4312)$,
$P_c(4440)$ and $P_c(4457)$ with the same cutoff can be traced back
to a particular artifact generated by the form factors.
The unregularized spin-spin piece of the OBE potential
contains a contact-range and a finite-range piece, which we write
schematically as
\begin{eqnarray}
  V_S \propto
  \left[ - \delta^{(3)}(m\, \vec{r}) + \frac{e^{-m r}}{4 \pi m r} \right] \, ,
\end{eqnarray}
with $m$ the mass of the exchanged meson.
It happens that the inclusion of a form factor regularizes
the contact-range Dirac-delta piece of the OBE potential,
{making it  finite range}.
The expectation is that the finite range of the regularized delta
will be considerably shorter than the range of the Yukawa-like piece.
However this condition is not actually fulfilled for a monopolar form factor
and a cutoff $\Lambda \sim 1\,{\rm GeV}$.
This is obvious in the OPE contribution of the OBE potential,
which is in fact distorted at distances comparable
with the Compton wavelength of the pion.
In particular the excessive binding of the $\tfrac{1}{2}^{-}$
$\bar{D}^* \Sigma_c$ in Eq.~(\ref{eq:naive-OBE-Pc})
can be traced back to the regularized delta contribution stemming
from the OPE potential: while the Yukawa-like
piece of the OPE potential is repulsive in this system,
the delta-like piece provides the system with a strong,
{probably unphysical}, short-range attraction.
If we remove the delta-like contribution to the OPE potential by hand,
we end up with the set of predictions
\begin{eqnarray}
  B_E(\tfrac{1}{2}^-) = 13.2\,{\rm MeV}
  \quad \mbox{and} \quad
  B_E(\tfrac{3}{2}^-) = 11.6\,{\rm MeV} \, ,
\end{eqnarray}
which are much closer to the expected binding energies of
the molecular pentaquarks.
There are similar delta-like contributions in the spin-spin piece of
the vector-meson-exchange potential.
This piece of the OBE potential is of shorter range than the OPE piece.
The removal of their delta-like contributions is not as crucial as in the OPE
piece, yet it should better be done if we want the OBE model
to be internally consistent,
in which case we arrive at the predictions:
\begin{eqnarray}
  B_E(\tfrac{1}{2}^-) = 4.2\,{\rm MeV}
  \quad \mbox{and} \quad
  B_E(\tfrac{3}{2}^-) = 18.3\,{\rm MeV} \, ,
\end{eqnarray}
where the binding energies are in fact very close (within $1-3\,{\rm MeV}$)
to what we would expect from a molecular $P_c(4440)$ and $P_c(4457)$,
namely $21.8$ and $4.8\,{\rm MeV}$.
Owing to the compatibility of this set of predictions with the current
experimental determination of the $P_c(4440)$ and $P_c(4457)$ masses,
the removal of the Dirac-delta contributions could indeed be considered
as the preferred solution to the form-factor problem.
In this case the OBE model as applied to hadronic molecules ends up having
the phenomenological success of its original nuclear physics version,
modulo the larger experimental uncertainties associated
with hadronic molecules.
The seemingly {\it ad-hoc} removal of the Dirac-delta contributions,
which has also been considered for instance in Ref.~\cite{Meng:2017fwb},
finds a natural explanation within the renormalized
OBE model of Ref.~\cite{Cordon:2009pj}.

The manuscript is structured as follows: in Sect.~\ref{sec:hqqs}
we briefly explain how the heavy hadron-hadron interaction is constrained
by heavy-quark spin symmetry (HQSS), where we also advocate a notation
based on the quark model for heavy-hadron
interactions~\cite{Valderrama:2019sid}.
In Sect.~\ref{sec:OBE} we explain the OBE model,
while in Sect.~\ref{sec:Pc-trio} we explain
the regulator artifact within the OBE model.
Then in Sect.~\ref{sec:Pc-multiplet} we show the predictions
we arrive at after removing this artifact.
We discuss the relation with renormalization and effective field theory ideas in Sect.~\ref{sec:modern}.
Finally in Sect.~\ref{sec:summary} we summarize our results.

\section{Heavy-Quark Spin Symmetry}
\label{sec:hqqs}

In this section we review a few basic consequences of HQSS
for heavy antimeson-baryon molecules.
As applied to molecular states, HQSS refers to the fact that interactions
among heavy hadrons do not depend on the spin of the heavy quarks
within the hadrons.
This can automatically be taken into account by writing
the heavy hadron interactions in a suitable notation.
The standard notation for this purpose is to group heavy hadrons
with the same light-quark structure in a single superfield,
which we review in Sect.~\ref{subsec:superfields}.
Here we advocate instead for a simpler notation in terms of
the light-quark degrees of freedom, which has been
recently presented in Ref.~\cite{Valderrama:2019sid} (though we note
that it has been intermittently used in the literature
for a long time~\cite{Manohar:1992nd}),
which we explain in Sect.~\ref{subsec:light-quark}.

\subsection{Heavy-Superfield Notation}
\label{subsec:superfields}

The $P$ and $P^*$ heavy mesons are $| Q \bar{q} \rangle$ states
with total spin $J=0$ and $1$, respectively.
They can be grouped into the single non-relativistic superfield:
\begin{eqnarray}
  {H}_Q = \frac{1}{\sqrt{2}}\,
  \left[ P + \vec{P}^* \cdot \vec{\sigma} \right] \, ,
\end{eqnarray}
which has been adapted from its relativistic version~\cite{Falk:1992cx}
and has good transformation properties with respect
to heavy-quark spin rotations.
In the formula above $H_Q$ is a 2{$\times$}2 matrix and
$\vec{\sigma}$ are the Pauli matrices.
The $\Sigma_Q$ and $\Sigma_Q^*$ heavy baryons are $| Q q q \rangle$ states
with total spin $J = \tfrac{1}{2}$ and $\tfrac{3}{2}$.
They can be written together as the following
non-relativistic superfield~\cite{Lu:2017dvm}
\begin{eqnarray}
 \vec{S}_Q = \frac{1}{\sqrt{3}}\,\vec{\sigma}\,\Sigma_Q + \vec{\Sigma}_Q^* \, ,
\end{eqnarray}
which corresponds to the relativistic heavy-baryon superfield
written in Ref.~\cite{Cho:1992cf}.
From these superfields, the simplest contact-range, no-derivative Lagrangian involving the heavy (anti-)meson and heavy baryon fields
is~\cite{Liu:2018zzu}
\begin{eqnarray}
  \mathcal{L} &=& 
  C_a \,
  \vec{S}_Q^{\dagger} \cdot \vec{S}_Q\,
      {\rm Tr}\left[ {\bar H}_Q^{\dagger} {\bar H}_Q\right]
  \nonumber \\
  &+& C_b\,\sum_{i = 1}^3\,
  \vec{S}_Q^{\dagger} \cdot (J_i \,\vec{S}_Q)\,
      {\rm Tr}\left[ {\bar H}_Q^{\dagger} \sigma_i {\bar H}_Q\right] \, ,
      \label{eq:contact-range-lagrangian}
\end{eqnarray}
where $J_i$ with $i = 1,2,3$ refers to the spin-1 angular momentum matrices
and with $C_a$ and $C_b$ coupling constants.
If we particularize for the $\bar{D} \Sigma_c$ family of molecules,
we obtain the following potential:
\begin{eqnarray}
  V(\bar{D} \Sigma_c, J = \tfrac{1}{2}) &=& C_a \, , \\
  \label{eq:contact-DSigma-1}
  \nonumber \\
  V(\bar{D} \Sigma_c^*, J = \tfrac{3}{2}) &=& C_a \, , \\
  \nonumber \\
  V(\bar{D}^* \Sigma_c, J = \tfrac{1}{2}) &=& C_a - \tfrac{4}{3}\,C_b \, , \\
  V(\bar{D}^* \Sigma_c, J = \tfrac{3}{2}) &=& C_a + \tfrac{2}{3}\,C_b \, , \\
  \nonumber \\
  V(\bar{D}^* \Sigma_c^*, J = \tfrac{1}{2}) &=& C_a - \tfrac{5}{3}\,C_b \, , \\
  V(\bar{D}^* \Sigma_c^*, J = \tfrac{3}{2}) &=& C_a - \tfrac{2}{3}\,C_b \, , \\
  V(\bar{D}^* \Sigma_c^*, J = \tfrac{5}{2}) &=& C_a + C_b \, .
  \label{eq:contact-DSigma-2}
\end{eqnarray}
We notice that, for simplicity, we have ignored isospin when writing
the Lagrangian of Eq.~(\ref{eq:contact-range-lagrangian})
and the potentials of
Eqs.~(\ref{eq:contact-DSigma-1}-\ref{eq:contact-DSigma-2}).
Isospin can be trivially taken into account by adding a subindex indicating
the isospin of the two-body state in the couplings: $C_{Ia}$, $C_{Ib}$
with $I = \tfrac{1}{2}$, $\tfrac{3}{2}$.

\subsection{Light-Quark Notation}
\label{subsec:light-quark}

Actually the heavy-quark symmetric interactions can be derived in an easier
and more direct way if we consider that the heavy-quark
acts as a spectator~\cite{Valderrama:2019sid}.
Instead of building superfields for the $P$ and $P^*$ heavy mesons,
we can simply express the interactions in terms of
the light-quark subfield within the heavy mesons, $q_L$.
Equivalently, for the $\Sigma_Q$ and $\Sigma_Q^*$ heavy baryons
we can use the light-diquark subfield within them: $d_L$.
After introducing these fields, the lowest-order contact-range Lagrangian
can be simply written as
\begin{eqnarray}
  \mathcal{L} &=& 
  C_a \, (q_L^{\dagger}\,q_L) \, (d_L^{\dagger}\,d_L)
  \nonumber \\
  &+& C_b\,(q_L^{\dagger}\,{\vec{\sigma}_{L}}\,q_L)\,{\cdot} \,
  (d_L^{\dagger}\,{\vec{S}_{L}}\,d_L) \, ,
\end{eqnarray}
where $\vec{\sigma}_{L}$ and $\vec{S}_L$ refer to the spins of
the $q_L$ and $d_L$ subfields, respectively.
This leads to the contact-range potential
\begin{eqnarray}
  V_C(q_L\,d_L) = C_a + C_b\,\vec{\sigma}_{L1} \cdot \vec{S}_{L2} \, ,
  \label{eq:contact-light}
\end{eqnarray}
where we have labeled the heavy meson and baryon {with the light quark
and light diquark inside them} with the subscript $1$ and $2$.
The contact-range potential is now written for the light-quark fields
within the heavy hadrons. The translation into the heavy-hadron
degrees of freedom can be encapsulated in a series of rules.
In particular for the heavy mesons the light-quark spin operators
are translated into 
\begin{eqnarray}
  \langle P | \vec{\sigma}_L | P \rangle &=& 0 \, , \\
  \langle P^* | \vec{\sigma}_L | P^* \rangle &=& \vec{S}_1 \,  ,
\end{eqnarray}
where $\vec{S}_1$ refers to the spin-1 matrices as applied
to the heavy vector meson.
For the heavy baryons the correspondence is
\begin{eqnarray}
  \langle \Sigma_Q | \vec{J}_L | \Sigma_Q \rangle &=&
  \frac{2}{3}\,\vec{\sigma}_2 \, , \\
  \langle \Sigma_Q^* | \vec{J}_L | \Sigma_Q^* \rangle &=&
  \frac{2}{3}\,\vec{S}_2 \,  ,
\end{eqnarray}
where $\vec{\sigma}_2$ refers to the Pauli matrices (applied to the heavy
spin-$\tfrac{1}{2}$ baryon) and $\vec{S}_2$ are the
spin-$\tfrac{3}{2}$ angular momentum matrices.
With these substitutions it is easy to check that the contact-range potential of
Eq.~(\ref{eq:contact-light}) written in the light-quark field basis is
indeed equivalent to the contact-range potential of
Eqs. \eqref{eq:contact-DSigma-1}--\eqref{eq:contact-DSigma-2}
written in the particle basis.
The notation in terms of the light-quark subfields is however much more compact
and we will adopt it for the rest of this work.

\section{The One-Boson-Exchange Potential}
\label{sec:OBE}

In this section we present the OBE model that we use in this work.
In the OBE model the potential between two hadrons is generated by
the exchange of a series of light mesons, which includes the $\pi$,
the $\sigma$, the $\rho$ and the $\omega$ (plus a few extra light
mesons in its more sophisticated versions, \mpv{e.g. the $\eta$,
  the $a_0(980)$ or even the $a_1$~\cite{Durso:1984um}}).
\mpv{We will propose here a minimalistic OBE model containing only
  the four aforementioned light mesons ($\pi$, $\sigma$, $\rho$, $\omega$),
  which \mpvrev{owing to their combination of range and coupling strength}
  are usually assumed to provide the bulk of the hadron-hadron
  interaction (or at least this is the case when we are dealing
  with nucleons).
  Nonetheless, as a cross-check of the previous choice,
  we will consider the effects of additionally including the $\eta$ meson,
  which has a similar range as the other light mesons we are considering
  \mpvrev{(but a smaller coupling)}.
}

\mpv{The OBE model} results in a description of the forces among hadrons
that is both simple and physically compelling,
{representing a natural generalization of
  the original idea by Yukawa for explaining nuclear forces.}
Yet there are disadvantages in the OBE model, which usually include
a large number of coupling constants and the requirement of form
factors and a cutoff to remove the unphysical short-range
components of the potential.
Here the choice of coupling constants will be done in terms of experimentally
known information or by recourse to phenomenological models.
For the form factor we will choose a standard multipolar form,
while the cutoff will be determined by the condition of
reproducing a known molecular candidate,
the $P_c(4312)$ in this case.
By determining the cutoff in this way we are partially {\it renormalizing}
the OBE model, i.e. removing cutoff ambiguities in terms of observable
information.
This concept is based on the fully renormalized OBE model of
Ref.~\cite{Cordon:2009pj}, which in turn helps to understand
a few of the tweaks required in the original OBE model
(e.g. the excessively large coupling to the $\omega$ vector meson
that is usually required in nuclear physics).
We stress however that we have not implemented a renormalized OBE model
in this work, but merely adapted a few of the ideas
of Ref.~\cite{Cordon:2009pj}.

\subsection{The Lagrangian}

First we write down the Lagrangians that encode the couplings between the heavy hadrons and the light mesons
($\pi$, $\sigma$, $\rho$, $\omega$).
We use the light-quark notation introduced in Sect.~\ref{subsec:light-quark}.
For the effective light-quark field degree of freedom
within the heavy mesons the Lagrangian reads as follows
\begin{eqnarray}
  \mathcal{L}_{q_L q_L \pi} &=& \frac{g_1}{\sqrt{2} f_{\pi}}\,
        {q}_{L}^{\dagger}
        \vec{\sigma}_{L} \cdot \nabla ( \vec{\tau} \cdot \vec{\pi})
        q_{L} \, , \\
        \mathcal{L}_{q_L q_L \sigma} &=& {g}_{\sigma 1}\,q_{L}^{\dagger} \sigma q_{L}
        \, , \\
        \mathcal{L}_{q_L q_L \rho} &=&
                {g}_{\rho 1}\,q_{L}^{\dagger} \vec{\tau} \cdot \vec{\rho}_0 q_{L}
                \nonumber \\
                &-& \frac{f_{\rho 1}}{4 {M_1}} \epsilon_{ijk}
                q_{L}^{\dagger}\, \sigma_{L, k} \vec{\tau} \cdot
                ( \partial_i \vec{\rho}_j - \partial_j \vec{\rho}_i )\, q_{L}
                \, , \\
                \mathcal{L}_{q_L q_L \omega} &=&
                {g}_{\omega 1}\,q_{L}^{\dagger} \, {\omega}_0 \, q_{L}
                \nonumber \\
                &-& \frac{f_{\omega 1}}{4 {M_1}} \epsilon_{ijk}
                q_{L}^{\dagger}\, \sigma_{L, k} \,
                ( \partial_i {\omega}_j - \partial_j {\omega}_i )\, q_{L}
                \, ,
\end{eqnarray}
where {$g_1$ is the pion axial coupling, $f_{\pi} = 132\,{\rm MeV}$
  the pion decay constant, $g_{\sigma 1}$ the coupling to the sigma meson},
while $g_{V1}$ and $f_{V1}$ with $V = \rho, \omega$
are the electric- and magnetic-type couplings to the vector mesons;
$M_1$ is a mass scale that we introduce
for $f_{V1}$ to be dimensionless.
{Finally $\vec{\tau}$ refers to the isospin matrices as applied to
  the charmed meson, which coincide with the Pauli matrices.}
For the effective light-diquark field within the heavy baryons we have
\begin{eqnarray}
  \mathcal{L}_{d_L d_L \, \pi} &=& \frac{g_2}{\sqrt{2} f_{\pi}}\,
        {d}_{L}^{\dagger}
        \vec{S}_{L} \cdot \nabla ( \vec{T} \cdot \vec{\pi})
        d_{L} \, , \\
        \mathcal{L}_{d_L d_L \, \sigma} &=& {g}_{\sigma 2}\,d_{L}^{\dagger}
        \sigma d_{L} \, , \\
        \mathcal{L}_{d_L d_L \, \rho} &=&
                {g}_{\rho 2}\,d_{L}^{\dagger} \,
                \vec{T} \cdot \vec{\rho}_0 \, d_{L}
                \nonumber \\
                &-& \frac{f_{\rho 2}}{4 M_2} \epsilon_{ijk}
                d_{L}^{\dagger} \, S_{L, k} \vec{T} \cdot
                ( \partial_i \vec{\rho}_j - \partial_j \vec{\rho}_i ) \, d_{L}
                \nonumber \\
                &+& \frac{h_{\rho 2}}{2 M_2^2} 
                d_{L}^{\dagger} \, Q_{L, ij} \, \vec{T} \cdot
                \partial_i \partial_j \vec{\rho}_0 \, d_{L} \, , \\
        \mathcal{L}_{d_L d_L \, \omega} &=&
                {g}_{\omega 2}\,d_{L}^{\dagger} \,
                {\omega}_0 \, d_{L}
                \nonumber \\
                &-& \frac{f_{\omega 2}}{4 M_2} \epsilon_{ijk}
                d_{L}^{\dagger} \, S_{L, k} \,
                ( \partial_i {\omega}_j - \partial_j {\omega}_i ) \, d_{L}
                \nonumber \\
                &+& \frac{h_{\rho 2}}{2 M_2^2} 
                d_{L}^{\dagger} \, Q_{L, ij} \,
                \partial_i \partial_j {\omega}_0 \, d_{L} \, .
\end{eqnarray}
The spin of the light-diquark field is $S_L = 1$, which means that
there are three possible type of interactions with a vector field:
electric-, magnetic- and quadrupole-type.
They correspond to the $g_{V2}$, $f_{V2}$ and $h_{V2}$ couplings. The mass $M_2$ is introduced to make
the $f_{V2}$ and $h_{V2}$ couplings dimensionless.
{$\vec{T}$ represents the isospin matrices for the charmed baryons,
  which are identical to the spin-1 matrices but applied to isospin space
  instead.}
For the quadrupole-type term we have introduced the spin-2 tensor
\begin{eqnarray}
  Q_{L,ij} &=& \frac{1}{2}\left[ S_{L,i} S_{L,j} + S_{L,j} S_{L,i} \right] -
  \frac{\vec{S}_{L}^2}{3} \delta_{ij} \, ,
  \label{eq:QL}
\end{eqnarray}
which can be translated into its charmed baryon version with
\begin{eqnarray}
  \langle \Sigma_c | Q_{L,ij} | \Sigma_c \rangle &=& 0 \, , \\
  \langle \Sigma_c^* | Q_{L,ij} | \Sigma_c^* \rangle &=&
  \frac{1}{3}\,Q_{2, ij} \, , 
\end{eqnarray}
with $Q_{2, ij}$ analogous to Eq.~\ref{eq:QL}, but written in terms of
the spin-$\tfrac{3}{2}$ angular momentum matrices.
We expect the quadrupole-type term to be small though.

\mpvrev{The previous Lagrangians have been derived in the heavy-quark limit,
  $m_Q \to \infty$.
  But the mass of the charm quark is finite and HQSS-breaking terms
  can in principle be included.
  Yet the relative size of the $1/m_Q$ terms is expected to be of the order of
  $\Lambda_{QCD} / m_Q \sim 15 \%$ in the charm sector.
  In Sect.~\ref{subsec:error-estimations} we will consider the effect of
  these uncertainties in more detail.
}

\mpv{Finally a few comments are in order at this point:
first, the electric-type and quadrupole
interactions of the vector mesons ($\rho$, $\omega$) with the charmed
mesons and baryons depend only on the zero-th component of
the vector meson fields, which at first sight
is anti-intuitive.
This is actually a result of the heavy quark limit and our choice of
parametrization for the heavy hadron fields, which we explain
in Appendix ~\ref{app:zeroth}.
Second, the relation with the multipole expansion can be better understood
from a comparison with the electromagnetic Lagrangian of the heavy hadrons,
which we explain in Appendix ~\ref{app:moments}.
Third, the Lagrangian for the interaction of the heavy hadrons with the $\eta$,
which we do not include in the basic version of the OBE model, can be found
in Appendix ~\ref{app:eta}.
}

\subsection{The OBE Potential}

The OBE potential can be easily derived from the previous Lagrangians
for the light-quark and light-diquark fields.
We write the potential in the following form
\begin{eqnarray}
  V_{\rm OBE} = \zeta\,V_{\pi} + V_{\sigma} + V_{\rho} + \zeta\,V_{\omega} \, ,
  \label{eq:V-OBE}
\end{eqnarray}
where $\zeta = \pm 1$ is a sign, for which the convention is
\begin{eqnarray}
  \zeta = +1 &\quad&
  \mbox{for $q_{L} d_{L}$ (e.g. $\bar{D} \Sigma_c$} ) \, , \\
  \zeta = -1 &\quad&
  \mbox{for $\bar{q}_{L} d_{L}$ (e.g. ${D} \Sigma_c$} ) \, ,
\end{eqnarray}
that is, we take $\zeta = +1$ for the most representative type of molecule,
the (hidden-charm) $\bar{D} \Sigma_c$ in this case.
In momentum space the different components of the OBE potential read
\begin{eqnarray}
  V_{\pi}(\vec{q}) &=& -\frac{g_1 g_2}{2 f_{\pi}^2}\,\vec{\tau}_1 \cdot \vec{T}_2\,
  \frac{\vec{\sigma}_{L1} \cdot \vec{q} \, \vec{S}_{L2} \cdot \vec{q}}
       {{\vec{q}\,}^2 + m_{\pi}^2}
  \, , \\
  V_{\sigma}(\vec{q}) &=& -\frac{g_{\sigma 1} g_{\sigma 2}}
  {{\vec{q}\,}^2 + m_{\sigma}^2}
  \, , \\
  V_{\rho}(\vec{q}) &=& \vec{\tau}_1 \cdot \vec{T}_2\,\Big[
    \frac{g_{\rho 1}}{{\vec{q}\,}^2 + m_{\rho}^2}\,
    (g_{\rho 2} - \frac{h_{\rho 2}}{2 M_2^2}\,\vec{q} \cdot (Q_{L2} q))
    \nonumber \\
    &-& \frac{f_{\rho 1}}{2 M_1}\,\frac{f_{\rho 2}}{2 M_2}\,
    \frac{(\vec{\sigma}_{L1} \times \vec{q}) \cdot (\vec{S}_{L2} \times \vec{q})}
         {{\vec{q}\,}^2 + m_{\rho}^2} \Big] \, , \\
  V_{\omega}(\vec{q}) &=& \frac{g_{\omega 1}}{{\vec{q}\,}^2 + m_{\omega}^2}\,
  (g_{\omega 2} - \frac{h_{\omega 2}}{2 M_2^2}\,\vec{q} \cdot (Q_{L2} q))
    \nonumber \\
    &-& \frac{f_{\omega 1}}{2 M_1}\,\frac{f_{\omega 2}}{2 M_2}\,
    \frac{(\vec{\sigma}_{L1} \times \vec{q}) \cdot (\vec{S}_{L2} \times \vec{q})}
         {{\vec{q}\,}^2 + m_{\omega}^2} \, .
\end{eqnarray}
If we Fourier-transform the previous expressions to coordinate space
we have
\begin{eqnarray}
  V_{\pi}(\vec{r}) &=&
  +\vec{\tau}_1 \cdot \vec{T}_2\,\frac{g_1 g_2}{6 f_{\pi}^2}\,\Big[
    - \vec{\sigma}_{L1} \cdot \vec{S}_{L2}\,\delta(\vec{r})
    \nonumber \\ && \quad
    + \, \vec{\sigma}_{L1} \cdot \vec{S}_{L2}\,m_{\pi}^3\,W_Y(m_{\pi} r)
    \nonumber \\ && \quad
    + \, S_{L12}(\vec{r})\,m_{\pi}^3\,W_T(m_{\pi} r) \Big] \, , \\
  V_{\sigma}(\vec{r}) &=& -{g_{\sigma 1} g_{\sigma 2}}\,m_{\sigma}\,W_Y(m_{\sigma} r)
  \, , \\
  V_{\rho}(\vec{r}) &=& \vec{\tau}_1 \cdot \vec{T}_2\,\Big[
    {g_{\rho 1} g_{\rho_2}}\,m_{\rho}\,W_Y(m_{\rho} r) \nonumber \\
    && \quad + g_{\rho 1}\,{\frac{h_{\rho 2}}{2 M_2^2}}\,Q_{L2}(\hat{r})\,
    m_{\rho}^3\,W_T(m_{\rho} r) \nonumber \\
    && \quad + \frac{f_{\rho 1}}{2 M_1}\,\frac{f_{\rho 2}}{2 M_2}\,\Big(
    -\frac{2}{3}\,\vec{\sigma}_{L1} \cdot \vec{S}_{L2} \delta(\vec{r})
    \nonumber \\ && \qquad
    +\frac{2}{3}\,\vec{\sigma}_{L1} \cdot \vec{S}_{L2}
    \, m_{\rho}^3 \, W_Y(m_{\rho} r)
    \nonumber \\ && \qquad
    -\frac{1}{3}\,S_{L12}(\hat{r})\, m_{\rho}^3 \, W_T(m_{\rho} r) \,\,
    \Big) \, \Big] \, , \\
    V_{\omega}(\vec{r}) &=& 
    {g_{\omega 1} g_{\omega_2}}\,m_{\omega}\,W_Y(m_{\omega} r) \nonumber \\
    && \quad + g_{\omega 1}\,{\frac{h_{\omega 2}}{2 M_2^2}}\,Q_{L2}(\hat{r})\,
    m_{\omega}^3\,W_T(m_{\omega} r) \nonumber \\
    && \quad + \frac{f_{\omega 1}}{2 M_1} \frac{f_{\omega 2}}{2 M_2}\,\,\Big(
    -\frac{2}{3}\,\vec{\sigma}_{L1} \cdot \vec{S}_{L2} \, \delta(\vec{r})
    \nonumber \\ && \qquad
    +\frac{2}{3}\,\vec{\sigma}_{L1} \cdot \vec{S}_{L2} \,
    m_{\omega}^3 \, W_Y(m_{\omega} r)
    \nonumber \\ && \qquad
    -\frac{1}{3}\,S_{L12}(\hat{r})\, m_{\omega}^3 \, W_T(m_{\omega} r) \,\,
    \Big) \, , 
\end{eqnarray}
where we have introduced the dimensionless functions
\begin{align}
W_Y(x) =& \,\frac{e^{-x}}{4\pi x},\\
W_T(x) =& \,\left(1+\frac{3}{x}+\frac{3}{x^2}\right)\frac{e^{-x}}{4\pi x},
\end{align}
while $S_{L12}$ represents the standard tensor operator
\begin{eqnarray}
  S_{L12}(\hat{r}) &=&
  3\, \vec{\sigma}_{L1} \cdot \hat{r} \,  \vec{S}_{L2} \cdot \hat{r} -
  \vec{\sigma}_{L1} \cdot \vec{S}_{L2}
\end{eqnarray}
and $Q_{L2}(\hat{r})$ is a second type of tensor operator
\begin{eqnarray}
  Q_{L2}(\hat{r}) &=& \hat{r} \cdot (Q_{L2} \hat{r}) =
  Q_{L2, ij} \, \hat{r}_i \hat{r}_j \, ,
\end{eqnarray}
with $Q_{L2,ij}$ defined in Eq.~(\ref{eq:QL}).
This second type of tensor operator is theoretically interesting,
but probably not particularly relevant
as the $h_{\omega 2}$ coupling is expected to be small,
see Sect.~\ref{subsec:couplings} for a more detailed discussion.

{
  Finally it is important to notice that we have computed the OBE potentials
  under the assumption that the exchanged light mesons
  ($\pi$, $\sigma$, $\rho$, $\omega$)
  have zero-width.
  This approximation is evident for the $\pi$ and $\omega$ mesons, which are
  narrow, but it is also known to work well for the $\rho$ meson.
  The situation is more subtle for the $\sigma$ meson, yet even
  in this case a broad $\sigma$ meson can be substituted
  with a zero-width one (sometimes called $\sigma'$)
  by a suitable redefinition of its mass and coupling.
  Yet here we will treat the $\sigma$ meson more as an effective degree of
  freedom of the OBE model (i.e. as a ``$\sigma_{\rm OBE}$'')
  than as a physical particle.
  We refer to Ref.~\cite{Machleidt:1987hj} and references therein
  for a complete discussion of this topic.
}

\subsection{Form Factors}
\label{subsec:formfactors}

We have derived the previous OBE potential under the assumption
that the interactions between heavy hadrons and light mesons
are point-like.
Hadrons have however a finite size, which can be taken into account
by the introduction of a form factor for each vertex.
In momentum space we will simply have
\begin{eqnarray}
  V_M(\vec{q}; \Lambda_1, \Lambda_2) =
  V_M(\vec{q})\,F_{M1}(\vec{q}, \Lambda_1)\,F_{M2}(\vec{q}, \Lambda_2) \, .
\end{eqnarray}
We will assume a monopolar form factor for vertices $1$ and $2$:
\begin{eqnarray}
  F_{Mi}(\vec{q}, \Lambda_i) = \frac{\Lambda_i^2 - m_M^2}{\Lambda_i^2 + \vec{q}^2}
  \, .
\end{eqnarray}
In principle we can use different cutoffs for different vertices
to take into account the different internal structure of
the heavy mesons and heavy baryons.
Yet this is only necessary if we want to describe heavy meson-meson,
heavy meson-baryon and heavy baryon-baryon molecules consistently.
If we are only interested in the heavy meson-baryon system
then we can simply assume a unique cutoff for both vertices $1$ and $2$.

If we now Fourier-transform the momentum space potential
with a monopolar form factor into coordinate space, the outcome is
that we simply have to make the following substitutions:
\begin{eqnarray}
  \delta(\vec{x}) &\to& m^3\,d(x, \lambda) \, , \label{eq:d-FF} \\
  W_Y(x) &\to& W_Y(x, \lambda) \, , \label{eq:WY-FF} \\
  W_T(x) &\to& W_T(x, \lambda) \, , \label{eq:WT-FF}
\end{eqnarray}
where
\begin{eqnarray}
  d(x, \lambda) &=& \frac{(\lambda^2 - 1)^2}{2 \lambda}\,
  \frac{e^{-\lambda x}}{4 \pi} \, , \\
  W_Y(x, \lambda) &=& W_Y(x) - \lambda W_Y(\lambda x) \nonumber \\ && -
  \frac{(\lambda^2 - 1)}{2 \lambda}\,\frac{e^{-\lambda x}}{4 \pi} \, , \\
  W_T(x, \lambda) &=& W_T(x) - \lambda^3 W_T(\lambda x) \nonumber \\ && -
  \frac{(\lambda^2 - 1)}{2 \lambda}\,\lambda^2\,
  \left(1 + \frac{1}{\lambda x} \right)\,\frac{e^{-\lambda x}}{4 \pi} \, .
\end{eqnarray}
The corresponding expressions for form factors of higher polarity
(e.g. dipolar) can be consulted in the Appendix of
Ref.~\cite{Liu:2019stu}.

\subsection{Couplings}
\label{subsec:couplings}

\begin{table}[!h]
\caption{Masses and quantum numbers of the light mesons
  of the OBE model ($\pi$, $\sigma$, $\rho$, $\omega$)
  and the heavy hadrons ($D$, $D^*$, $\Sigma_c$, $\Sigma_c^*$).
  Notice that we work in the isospin-symmetric limit and
  take the isospin-averaged {masses of the
    $\pi$, $D$, $D^*$, $\Sigma_c$ and $\Sigma_c^*$
    as listed in the PDG~\cite{Tanabashi:2018oca}.
  }
  {For the $\rho$ and $\omega$ we approximate their masses to
    the closest tens of MeV, while for the $\sigma$ we settle
    for a conventional value of its mass within the OBE model
    (which does not necessarily coincide
    with its physical mass).}
}
\label{tab:masses}
\begin{tabular}{ccc}
  \hline\hline
  Light Meson  & $I^{G}\,(J^{PC})$  & M (MeV) \\
  \hline
  $\pi$ & $1^{-}$ $({0}^{-+})$ & 138 \\
  $\sigma$ & $0^{+}$ $({0}^{++})$ & 600 \\
  $\rho$ & $1^{+}$ $({1}^{--})$ & 770 \\
  $\omega$ & $0^{-}$ $({1}^{--})$ & 780 \\
  \hline \hline \hline
  Heavy Hadron & $I (J^P)$ & M (MeV) \\
  \hline
  $D$ & $\frac{1}{2}(0^-)$ & 1867 \\
  $D^*$ & $\frac{1}{2}(1^-)$ & 2009 \\
  $\Sigma_c$ & $1(\frac{1}{2}^+)$ & 2454 \\
  $\Sigma_c^*$ & $1(\frac{3}{2}^+)$ & 2518 \\
  \hline \hline
\end{tabular}

\end{table}

\begin{table}[!h]
\caption{Couplings of the light mesons of the OBE model
  ($\pi$, $\sigma$, $\rho$, $\omega$) to the heavy-meson
  and heavy-baryon fields.
  For the magnetic-type coupling of the $\rho$ and $\omega$ vector mesons
  we have used the decomposition
  $f_{V} = \kappa_{V}\,g_{V}$, with $V=\rho,\omega$.
  $M$ refers to the mass scale (in MeV) involved
  in the magnetic-type couplings.
}
\label{tab:couplings}
\begin{tabular}{cc}
  \hline \hline
  Coupling  & Value for $P$/$P^*$ \\
  \hline
  $g_1$ & 0.60 \\
  $g_{\sigma 1}$ & 3.4 \\
  $g_{\rho 1}$ & 2.6 \\
  $g_{\omega 1}$ & 2.6 \\
  $\kappa_{\rho 1}$ & 2.3 \\
  $\kappa_{\omega 1}$ & 2.3 \\
  $M_1$ & 940 \\
  \hline \hline
  Coupling  & Value for $\Sigma_Q$/$\Sigma_Q^*$ \\
  \hline
  $g_2$ & 0.84 \\
  $g_{\sigma 2}$ & 6.8 \\
  $g_{\rho 2}$ & 5.8 \\
  $g_{\omega 2}$ & 5.8 \\
  $\kappa_{\rho 2}$ & 1.7 \\
  $\kappa_{\omega 2}$ & 1.7 \\
  $\eta_{\rho 2}$ & 0 \\
  $\eta_{\omega 2}$ & 0 \\
  $M_1$ & 940 \\
  \hline \hline
\end{tabular}
\end{table}

For the axial coupling between the $D$ and $D^*$ heavy mesons and the pion we take
\begin{eqnarray}
  g_1 = 0.60 \, ,
\end{eqnarray}
which is compatible with $g_1 = 0.59 \pm 0.01 \pm 0.07$ as
extracted from the $D^* \to D \pi$ decay~\cite{Ahmed:2001xc,Anastassov:2001cw}.
For the $\Sigma_c$ and $\Sigma_c^*$ heavy baryons the axial coupling is not
experimentally available, but there is a lattice QCD
calculation~\cite{Detmold:2012ge}
\begin{eqnarray}
  g_2 = 0.84 \pm 0.2 \, .
\end{eqnarray}
which is the value we adopt here.
We notice in passing that there are several conventions
for the axial coupling to the heavy baryons in the literature and
here we are effectively using the one
in Ref.~\cite{Detmold:2012ge}.
Other two popular conventions are the ones by Cho~\cite{Cho:1992cf}
and Yan~\cite{Yan:1992gz}, which are related to our convention by
the relations $g_2 = - g_{2,\rm Cho}$ and
$g_2 = \frac{3}{2} g_{1,\rm Yan}$ (in Ref.~\cite{Yan:1992gz}
the axial coupling to the heavy baryons is denoted as $g_1$).

For the couplings to the $\sigma$ meson, in the case of the nucleon-nucleon
interaction it can be determined from the linear sigma
model~\cite{GellMann:1960np} yielding
\begin{eqnarray}
  g_{\sigma NN} = \sqrt{2}\,\frac{M_N}{f_{\pi}} \simeq 10.2 \, .
\end{eqnarray}
For the case of the $D$, $D^*$ mesons and $\Sigma_c$, $\Sigma_c^*$ baryons
we can estimate the coupling to the $\sigma$ from the quark model.
By assuming that the $\sigma$ only couples to the $u$ and $d$ quarks,
we expect
\begin{eqnarray}
  g_{\sigma 1} = \frac{g_{\sigma 2}}{2} = \frac{g_{\sigma NN}}{3} \simeq 3.4 \, .
\end{eqnarray}

The choice of couplings for the $\rho$ and $\omega$ mesons is more laborious.
First, from SU(3)-flavor symmetry and the OZI rule we expect that
\begin{eqnarray}
  g_{\rho 1} = g_{\omega 1} \quad &,& \quad g_{\rho 2} = g_{\omega 2} \, , \\
  f_{\rho 1} = f_{\omega 1} \quad &,& \quad f_{\rho 2} = f_{\omega 2} \, , \\
  h_{\rho 2} &=& h_{\omega 2} \, .
\end{eqnarray}
For the determination of the electric, magnetic and quadrupole couplings
we will use {the vector-meson dominance assumption}.
The original formulation of this idea states that hadrons do not couple
directly to the electromagnetic field, but by means of
the neutral vector meson fields, $\rho_3^{\mu}$ and $\omega^{\mu}$,
where $\mu$ refers to the Lorentz indices of these fields,
and the subindex $_3$ indicates that we are dealing with
the neutral rho meson.
A practical way to apply this idea is to derive the electromagnetic
Lagrangian from the substitutions
\begin{eqnarray}
  \rho_3^{\mu} &\to& e\,\lambda_{\rho} A^{\mu} \, , \\
  \omega^{\mu} &\to& e\,\lambda_{\omega} A^{\mu} \, .
\end{eqnarray}
We can fix $\lambda_{\rho}$ and $\lambda_{\omega}$ from the nucleon case,
in which case we obtain
\begin{eqnarray}
  \lambda_{\rho} &=& \frac{1}{2 g_{\rho NN}} = \frac{1}{2 g_{\rho}} \, , \\
  \lambda_{\omega} &=& \frac{1}{2 g_{\omega NN}} = \frac{1}{6 g_{\rho}} \, ,
\end{eqnarray}
where in the right-hand side we have written $g_{\rho NN}$ and $g_{\omega NN}$
in terms of the universal $\rho$ coupling (Sakurai's
universality~\cite{Sakurai:1960ju})
\begin{eqnarray}
  g_{\rho} = \frac{m_{\rho}}{2 f_{\pi}} \sim 2.9 \, ,
\end{eqnarray}
where we have also made use of the relation $g_{\omega NN} = 3 g_{\rho NN}$,
which is derived from SU(3)-flavor symmetry and the OZI rule.
In can be trivially checked that this choice correctly reproduces that
the proton and neutron charges are $e_p = +e$ and $e_n = 0$, respectively.

The application to the heavy hadrons requires a few modifications.
For instance, vector-meson dominance is expected to reproduce the total
charge of the light quarks only. It does not apply to the heavy quark,
which we consider to couple directly to the electromagnetic field.
Thus the application to the $\bar{D}^{0}$ ($\bar c u$) charmed meson yields
\begin{eqnarray}
  g_{\rho 1}\,\left(\frac{1}{2 g_{\rho}} + \frac{1}{6 g_{\rho}}\right) &=& \frac{2}{3}\,e \, ,
  \label{eq:vmd-q-1}
\end{eqnarray}
from which we deduce
\begin{eqnarray}
  g_{\rho 1} = g_{\rho} \simeq 2.9 \, . \label{eq:grho-D-QM}
\end{eqnarray}
For the magnetic moments we define the following quantity
for the sake of convenience
\begin{eqnarray}
  f_{\rho 1} = \kappa_{\rho 1} g_{\rho 1} \, ,
\end{eqnarray}
which is related to the $\bar{D}^{*0}$ magnetic moment, $\mu({\bar{D}^{*0}})$,
by the relation
\begin{eqnarray}
  \frac{2}{3}\,\kappa_{\rho} &=& \frac{2 M_1}{e}\,\mu({\bar{D}^{*0}}) \, .
\end{eqnarray}
If we set the scaling mass to be the nucleon mass $M_1 = M_N$,
$\kappa_{\rho 1}$ simply coincides with $\mu({\bar{D}^{*0}})$
in units of the nuclear magneton.
If we use the quark model $\mu({\bar{D}^0}) = \mu_u$,
with $\mu_u = 1.85\, \mu_N$, we find
\begin{eqnarray}
  \kappa_{\rho 1}(M_1=M_N) \simeq 2.8\, . \label{eq:kappa-D-QM}
\end{eqnarray}
Notice that the definition of $\kappa_{\rho 1}$ is dependent
on the mass scale $M_1$ in the Lagrangian.
For $M_1 = m_D$ it happens that $\kappa_{\rho 1} \simeq 5.5$.
It should be noticed that the vector-meson dominance model we have presented
here can be further refined to obtain improved determinations
of $g_{\rho 1}$ and $\kappa_{\rho 1}$.
For instance, Ref.~\cite{Casalbuoni:1992dx} applies a more sophisticated
vector-meson dominance model to the weak decays of the charmed mesons,
which translates into the couplings~\cite{Liu:2018bkx}
\begin{eqnarray}
  g_{\rho 1} \simeq 2.6 \quad \mbox{and} \quad \kappa_{\rho 1}(M_1 = M_N)
  \simeq 2.3 \pm 0.4 \, . \label{eq:vmd-charmed}
\end{eqnarray}
As can be appreciated this determination is compatible with the one
in Eqs.~(\ref{eq:grho-D-QM}) and (\ref{eq:kappa-D-QM}) within errors.
We will use the set derived from Ref.~\cite{Casalbuoni:1992dx}, i.e.
the values in Eq.~(\ref{eq:vmd-charmed}), to follow
the same convention as in our previous works.

Now we apply the previous ideas to the $\Sigma_c$ and $\Sigma_c^*$ baryons.
First we define the reduced couplings
\begin{eqnarray}
  f_{\rho 2} = \kappa_{\rho 2} g_{\rho 2}  \quad , \quad
  h_{\rho 2} = \eta_{\rho 2} g_{\rho 2} \, .
\end{eqnarray}
We now apply vector-meson dominance to arrive at the relations
\begin{eqnarray}
  g_{\rho 2} &=& 2 g_{\rho} \, , \\
  \kappa_{\rho 2} &=& \frac{3}{4}\,(\frac{2 M_2}{e})\,
  \mu(\Sigma_c^{*++}) \, , \\
  \eta_{\rho 2} &=& \frac{9}{2}\,(\frac{M_2^2}{e})\,Q(\Sigma_c^{*++})  \, ,
\end{eqnarray}
where $\mu(\Sigma_c^{*++})$ and $Q(\Sigma_c^{*++})$ are the magnetic and
quadrupole moment of the $\Sigma_c^{++}$ baryon.
From the quark model (and the assumption that the charm quark provides a minor
contribution to the magnetic and quadrupole moments)
we expect $\mu(\Sigma_c^{*++}) = 2 \mu_u$ and $Q(\Sigma_c^{*++}) = 0$.
We note that a non-vanishing quadrupole moment will require
the light-diquark wavefunction to have a D-wave component,
which is not the case in the naive quark model.
Thus for $M_2 = M_N$ we arrive at
\begin{eqnarray}
  \kappa_{\rho 2} \simeq 2.8 \quad , \quad \eta_{\rho 2} \simeq 0 \, .
\end{eqnarray}
The fact that the quadrupole vector-meson coupling vanishes
in the naive quark model probably indicates a relatively
small contribution from this piece of the potential.
This is actually good news in the sense that it simplifies the OBE potential.
However the estimations from the quark model have been superseded by
recent lattice QCD calculations, at least for the magnetic moment of
the $\Sigma_c^{++}$ baryon~\cite{Can:2013tna}.
If we use the magnetic moment of the $\Sigma_c^{++}$ to determine
$\kappa_{\rho 2}$, we first note that the vector-meson dominance
relation reads
\begin{eqnarray}
  \kappa_{\rho 2} &=& \frac{9}{8}\,(\frac{2 M_2}{e})\,
  \mu(\Sigma_c^{++}) \, .
\end{eqnarray}
Ref.~\cite{Can:2013tna} obtains $\mu(\Sigma_c^{++}) = 1.499(202)$,
which leads to
\begin{eqnarray}
  \kappa_{\rho 2} \simeq 1.7 \pm 0.2 \, .
\end{eqnarray}
This is the value we will adopt here.
The charmed-antimeson and charmed-baryon masses that we use in this work,
together with the couplings, can be consulted in Tables \ref{tab:masses}
and \ref{tab:couplings}.

\subsection{Wave Functions and Partial Wave Projection}

The wave function for a heavy meson-baryon system is
\begin{eqnarray}
  | \Psi \rangle = \Psi_{J M}(\vec{r}) | I M_I \rangle\, ,
\end{eqnarray}
where $| I M_I \rangle$ is the isospin wave function
and $\Psi_{J M}$ the spin and spatial wave function,
which can be written as a sum over partial waves
\begin{eqnarray}
  \Psi_{JM}(\,\vec{r}\,) = \sum_{LS} \psi_{L SJ}(r) | {}^{2S+1}L_J \rangle \, .
\end{eqnarray}
We use the spectroscopic notation ${}^{2S+1}L_J$, which denotes a partial
wave with total spin $S$, orbital angular momentum $L$ and
total angular momentum $J$:
\begin{eqnarray}
  |{}^{2S+1}L_{J}\rangle &=& \sum_{M_{S},M_L}
  \langle L M_L S M_S | J M \rangle \, | S M_S \rangle \, Y_{L M_{L}}(\hat{r})
  \, , 
\end{eqnarray}
where $\langle L M_L S M_S | J M \rangle$ are the Clebsch-Gordan coefficients,
$| S M_S \rangle$ the spin wavefunction and $Y_{L M_L}(\hat{r})$
the spherical harmonics.
For the $P \Sigma_Q$ and $P\Sigma_Q^*$ systems the spin wave functions
are trivial
\begin{eqnarray}
  | S M_S (P \Sigma_Q) \rangle &=& | \frac{1}{2} M_S \rangle \, , \\
  | S M_S (P \Sigma_Q^*) \rangle &=& | \frac{3}{2} M_S \rangle \, ,
\end{eqnarray}
as they correspond to the spin wave functions of the heavy baryon
(the heavy meson $P$ is a pseudoscalar).
For the $P^* \Sigma_Q$ and $P^* \Sigma_Q^*$ systems,
\begin{eqnarray}
  | S M_S (P^* \Sigma_Q) \rangle &=& \sum_{M_{S1},M_{S2}} 
  \langle 1 M_{S1} \frac{1}{2} M_{S2} | S M_S \rangle \nonumber \\ &\times& 
  | 1 M_{S1} \rangle \, | \frac{1}{2} M_{S2} \rangle \, , \\
  | S M_S (P^* \Sigma_Q^*) \rangle &=& \sum_{M_{S1},M_{S2}} 
  \langle 1 M_{S1} \frac{3}{2} M_{S2} | S M_S \rangle \nonumber \\ &\times& 
  | 1 M_{S1} \rangle \, | \frac{3}{2} M_{S2} \rangle \, ,
\end{eqnarray}
with $| 1 M_{S1} \rangle$, $| J_2 M_{S2} \rangle$ the spin wavefunction of particles $1$ and $2$.

The partial-wave projection of the potential depends on the
matrix elements of the {spin-spin, tensor and quadrupole tensor}
operators, which are independent of $J$ and $M$,
\begin{eqnarray}
  \langle S' L' J' M' | {\bf O}_{12} | S L J M \rangle &=&
  \delta_{J J'} \delta_{M M'} \, {\bf O}^J_{S L, S' L'} \, ,
\end{eqnarray}
with ${\bf O}_{12} = C_{12}$, $S_{12}$, $Q_{2}$, which are in turn defined as
\begin{eqnarray}
  C_{12} &=& \vec{a}_1 \cdot \vec{a}_2 \, , \\
  S_{12} &=& 3 \vec{a}_1 \cdot \hat{r}\,\vec{a}_2 \cdot \hat{r} -
  \vec{a}_1 \cdot \vec{a}_2 \, , \\
  Q_{2,ij} &=& \frac{1}{2}\,\left[ a_{2i} a_{2j} + a_{2j} a_{2i}  \right] -
  \frac{\vec{a}_2^2}{3}\,\delta_{ij} \, ,
\end{eqnarray}
with $\vec{a}_1$ ($\vec{a}_2$) the corresponding spin operator
for the $\bar{D}$, $\bar{D}^*$ mesons ($\Sigma_c$, $\Sigma_c^*$ baryons).
In this work we are using the light-quark notation, which means that we have
written the potentials in terms of the light-quark spin.
The correspondence between the light-quark spin operators and $C_{12}$, $S_{12}$
is given by
\begin{eqnarray}
  \vec{\sigma}_{L1} \cdot \vec{S}_{L2} &=& f_{12}\,C_{12} \, , \\
  S_{L12} &=& f_{12}\,S_{12} \, , \\
  Q_{L2} &=& f_{2}\,Q_{2} \, ,
\end{eqnarray}
where $f_{12}$ and $f_2$ are factors related to the conversion
from the light-quark to the hadron spin degrees of freedom
(for all non-vanishing cases $f_{12} = \tfrac{2}{3}$ and
$f_2 = \tfrac{1}{3}$).
The specific matrix elements of the spin-spin, tensor and quadrupole-tensor
operators can be consulted in Tables \ref{tab:spinspin}, \ref{tab:tensor}
and \ref{tab:quadrupole-tensor} for all the molecular
configurations that contain an S-wave (i.e. the ones that
are more likely to bind).

\begin{table*}[t]
\centering \caption{Matrix elements of the spin-spin operator
  for the partial waves we are considering
  in this work.} \label{tab:spinspin}
\begin{tabular}{c|c|c|c}
\hline\hline
Molecule & Partial Waves & $J^P$ &
$\vec{\sigma}_{L1} \cdot \vec{S}_{L2} = f_{12} \times \, \vec{a}_1 \cdot \vec{a}_2 $ \\ \hline \hline
$\bar{D} \Sigma_c$ & $^2S_{{1}/{2}}$ & $\frac{1}{2}^-$ & $0 \times 0$  \\ \hline
$\bar{D} \Sigma_c^*$ & $^4S_{{3}/{2}}$-$^4D_{{3}/{2}}$ &
$\frac{3}{2}^-$
& $0 \times \left(\begin{matrix}
0 & 0 \\
0 & 0 \\
\end{matrix}\right)$
\\ \hline
$\bar{D}^* \Sigma_c$ & $^2S_{{1}/{2}}$-$^4D_{{1}/{2}}$  & $\frac{1}{2}^-$ & 
$\frac{2}{3} \times
\left(\begin{matrix}
-2 & 0 \\
0 & 1 \\
\end{matrix}\right)$ 
\\ \hline
$\bar{D}^* \Sigma_c$ & $^2D_{{3}/{2}}$-$^4S_{{1}/{2}}$-$^4D_{{1}/{2}}$
& $\frac{3}{2}^-$ & 
$\frac{2}{3} \times
\left(\begin{matrix}
-2 & 0 & 0 \\
0 & 1 & 0\\
0 & 0 & 1
\end{matrix}\right)$ 
\\ \hline
$\bar{D}^* \Sigma_c^*$ & $^2S_{{1}/{2}}$-$^4D_{{1}/{2}}$-$^6D_{{1}/{2}}$
& $\frac{1}{2}^-$ & 
$\frac{2}{3} \times
\left(\begin{matrix}
-\frac{5}{2} & 0 & 0 \\
 0 & -1 & 0 \\
 0 & 0 & \frac{3}{2}
\end{matrix}\right)$ 
\\ \hline
$\bar{D}^* \Sigma_c^*$ &
$^2D_{{3}/{2}}$-$^4S_{{3}/{2}}$-$^4D_{{3}/{2}}$-$^6D_{{3}/{2}}$-$^6G_{{3}/{2}}$ &
$\frac{3}{2}^-$ & 
$\frac{2}{3} \times
\left(\begin{matrix}
-\frac{5}{2} & 0 & 0 & 0 & 0 \\
0 & -1 & 0 & 0 & 0 \\
0 & 0 & -1 & 0 & 0 \\
0 & 0 & 0 & \frac{3}{2} & 0\\
0 & 0 & 0 & 0 & \frac{3}{2} \\
\end{matrix}\right)$
\\ \hline
$\bar{D}^* \Sigma_c^*$ &
$^2D_{{5}/{2}}$-$^4D_{{5}/{2}}$-$^4G_{{5}/{2}}$-$^6S_{{5}/{2}}$-$^6D_{{5}/{2}}$-$^6G_{{5}/{2}}$
& $\frac{5}{2}^-$ &
$\frac{2}{3} \times
\left(\begin{matrix}
-\frac{5}{2} & 0 & 0 & 0 & 0 & 0 \\
0 & -1 & 0 & 0 & 0 & 0\\
0 & 0 & -1 & 0 & 0 & 0\\
0 & 0 & 0 & \frac{3}{2} & 0 & 0\\
0 & 0 & 0 & 0 & \frac{3}{2} & 0\\
0 & 0 & 0 & 0 & 0 & \frac{3}{2} \\
\end{matrix}\right)$
\\ \hline
\hline\hline
\end{tabular}
\end{table*}

\begin{table*}[t]
\centering \caption{Matrix elements of the tensor operator
  for the partial waves we are considering
  in this work.} \label{tab:tensor}
\begin{tabular}{c|c|c|c}
\hline\hline
Molecule & Partial Waves & $J^P$ &
${S}_{L12}(\hat{r}) = f_{12} \times \, S_{12}(\hat{r}) $ \\ \hline \hline
$\bar{D} \Sigma_c$ & $^2S_{{1}/{2}}$ & $\frac{1}{2}^-$ & $0 \times 0$  \\ \hline
$\bar{D} \Sigma_c^*$ & $^4S_{{3}/{2}}$-$^4D_{{3}/{2}}$ &
$\frac{3}{2}^-$
& $0 \times \left(\begin{matrix}
0 & 0 \\
0 & 0 \\
\end{matrix}\right)$
\\ \hline
$\bar{D}^* \Sigma_c$ & $^2S_{{1}/{2}}$-$^4D_{{1}/{2}}$  & $\frac{1}{2}^-$ & 
$\frac{2}{3} \times
\left(\begin{matrix}
0 & \sqrt{2} \\
\sqrt{2} & -2 \\
\end{matrix}\right)
$ 
\\ \hline
$\bar{D}^* \Sigma_c$ & $^2D_{{3}/{2}}$-$^4S_{{1}/{2}}$-$^4D_{{1}/{2}}$
& $\frac{3}{2}^-$ & 
$\frac{2}{3} \times
\left(\begin{matrix}
0 & -1 & 1 \\
-1 & 0 & 2 \\
1 & 2 & 0 
\end{matrix}\right)$ 
\\ \hline
$\bar{D}^* \Sigma_c^*$ & $^2S_{{1}/{2}}$-$^4D_{{1}/{2}}$-$^6D_{{1}/{2}}$
& $\frac{1}{2}^-$ & 
$\frac{2}{3} \times
\left(\begin{matrix}
 0 & -\frac{7}{2 \sqrt{5}} & -\frac{3}{\sqrt{5}} \\
 -\frac{7}{2 \sqrt{5}} & -\frac{8}{5} & -\frac{3}{10} \\
 -\frac{3}{\sqrt{5}} & -\frac{3}{10} & -\frac{12}{5}
\end{matrix}\right)$ 
\\ \hline
$\bar{D}^* \Sigma_c^*$ &
$^2D_{{3}/{2}}$-$^4S_{{3}/{2}}$-$^4D_{{3}/{2}}$-$^6D_{{3}/{2}}$-$^6G_{{3}/{2}}$ &
$\frac{3}{2}^-$ & 
$\frac{2}{3} \times
\left(\begin{matrix}
  0 & \frac{7}{2\sqrt{10}} & -\frac{7}{2\sqrt{10}} & \frac{3}{\sqrt{35}} &
  -3\,\sqrt{\frac{6}{35}} \\
  \frac{7}{2\sqrt{10}} & 0 & \frac{8}{5} & -\frac{3}{10}\sqrt{\frac{7}{2}} &
  0 \\
  -\frac{7}{2\sqrt{10}} & \frac{8}{5} & 0 & -\frac{3}{2 \sqrt{14}} &
  -\frac{3}{5}\,\sqrt{\frac{3}{7}} \\
  \frac{3}{\sqrt{35}} & -\frac{3}{10}\,\sqrt{\frac{7}{2}}
  & -\frac{3}{2\sqrt{14}} & -\frac{6}{7} & \frac{9 \sqrt{6}}{35} \\
  -3\,\sqrt{\frac{6}{35}} & 0 & -\frac{3}{5}\,\sqrt{\frac{3}{7}} &
  \frac{9 \sqrt{6}}{35} & -\frac{15}{7} \\
\end{matrix}\right)$
\\ \hline
$\bar{D}^* \Sigma_c^*$ &
$^2D_{{5}/{2}}$-$^4D_{{5}/{2}}$-$^4G_{{5}/{2}}$-$^6S_{{5}/{2}}$-$^6D_{{5}/{2}}$-$^6G_{{5}/{2}}$
& $\frac{5}{2}^-$ &
$\frac{2}{3} \times
\left(\begin{matrix}
  0 & \frac{1}{2}\sqrt{\frac{7}{5}} & -\sqrt{\frac{21}{10}} &
  -\sqrt{\frac{3}{5}} & 2\sqrt{\frac{6}{35}} & -3\,\sqrt{\frac{2}{35}} \\
  \frac{1}{2}\sqrt{\frac{7}{5}} & \frac{8}{7} & \frac{16\,\sqrt{6}}{35}
  & \frac{\sqrt{21}}{10} & -\frac{1}{7}\sqrt{\frac{3}{2}} &
  - \frac{12 \sqrt{2}}{35} \\
  -\sqrt{\frac{21}{10}} & \frac{16\sqrt{6}}{35} & -\frac{8}{7} &
  0 & \frac{9}{70} & -\frac{3 \sqrt{3}}{14} \\
  -\sqrt{\frac{3}{5}} & \frac{\sqrt{21}}{10} & 0 & 0 &
  \frac{2 \sqrt{14}}{5} & 0\\
  2\sqrt{\frac{6}{35}} & -\frac{1}{7}\sqrt{\frac{3}{2}} & \frac{9}{70} &
  \frac{3 \sqrt{14}}{5} & \frac{6}{7} & \frac{27 \sqrt{3}}{35} \\
  -3\sqrt{\frac{2}{35}} & -\frac{12\sqrt{2}}{35} & -\frac{3\sqrt{3}}{14} &
  0 & \frac{27\sqrt{3}}{35} & -\frac{6}{7} \\
\end{matrix}\right)$
\\ \hline
\hline\hline
\end{tabular}
\end{table*}

\begin{table*}[t]
\centering \caption{Matrix elements of the quadrupole-like tensor operator
  for the partial waves we are considering
  in this work.} \label{tab:quadrupole-tensor}
\begin{tabular}{c|c|c|c}
\hline\hline
Molecule & Partial Waves & $J^P$ &
${Q}_{L2}(\hat{r}) = f_{2} \times \, Q_{2}(\hat{r}) $ \\ \hline \hline
$\bar{D} \Sigma_c$ & $^2S_{{1}/{2}}$ & $\frac{1}{2}^-$ & $0 \times 0$  \\ \hline
$\bar{D} \Sigma_c^*$ & $^4S_{{3}/{2}}$-$^4D_{{3}/{2}}$ &
$\frac{3}{2}^-$
& $\frac{1}{3} \times \left(\begin{matrix}
0 & 1 \\
1 & 0 \\
\end{matrix}\right)$
\\ \hline
$\bar{D}^* \Sigma_c$ & $^2S_{{1}/{2}}$-$^4D_{{1}/{2}}$  & $\frac{1}{2}^-$ & 
$\frac{1}{3} \times
\left(\begin{matrix}
0 & 0 \\
0 & 0 \\
\end{matrix}\right)
$ 
\\ \hline
$\bar{D}^* \Sigma_c$ & $^2D_{{3}/{2}}$-$^4S_{{1}/{2}}$-$^4D_{{1}/{2}}$
& $\frac{3}{2}^-$ & 
$\frac{1}{3} \times
\left(\begin{matrix}
0 & 0 & 0 \\
0 & 0 & 0 \\
0 & 0 & 0 
\end{matrix}\right)$ 
\\ \hline
$\bar{D}^* \Sigma_c^*$ & $^2S_{{1}/{2}}$-$^4D_{{1}/{2}}$-$^6D_{{1}/{2}}$
& $\frac{1}{2}^-$ & 
$\frac{1}{3} \times
\left(\begin{matrix}
 0 & \frac{2}{\sqrt{5}} & \frac{1}{\sqrt{5}} \\
 \frac{2}{\sqrt{5}} & -\frac{1}{5} & \frac{2}{5} \\
 \frac{1}{\sqrt{5}} & \frac{2}{5} & -\frac{4}{5}
\end{matrix}\right)$ 
\\ \hline
$\bar{D}^* \Sigma_c^*$ &
$^2D_{{3}/{2}}$-$^4S_{{3}/{2}}$-$^4D_{{3}/{2}}$-$^6D_{{3}/{2}}$-$^6G_{{3}/{2}}$ &
$\frac{3}{2}^-$ & 
$\frac{1}{3} \times
\left(\begin{matrix}
0 & -\sqrt{\frac{2}{{5}}} & \sqrt{\frac{2}{{5}}} & -\frac{1}{\sqrt{35}} &
\sqrt{\frac{6}{35}} \\
-\sqrt{\frac{2}{{5}}} & 0 & \frac{1}{5} & \frac{\sqrt{14}}{5} & 0 \\
\sqrt{\frac{2}{{5}}} & \frac{1}{5} & 0 & \sqrt{\frac{2}{7}} &
\frac{4}{5}\sqrt{\frac{3}{7}} \\
-\frac{1}{\sqrt{35}} & \frac{\sqrt{14}}{5} & \sqrt{\frac{2}{7}}
& -\frac{2}{7} & \frac{3 \sqrt{6}}{35} \\
\sqrt{\frac{6}{35}} & 0 & \frac{4}{5}\sqrt{\frac{3}{7}} &
\frac{3 \sqrt{6}}{35} & -\frac{5}{7} \\
\end{matrix}\right)
$
\\ \hline
$\bar{D}^* \Sigma_c^*$ &
$^2D_{{5}/{2}}$-$^4D_{{5}/{2}}$-$^4G_{{5}/{2}}$-$^6S_{{5}/{2}}$-$^6D_{{5}/{2}}$-$^6G_{{5}/{2}}$
& $\frac{5}{2}^-$ &
$\frac{1}{3} \times
\left(\begin{matrix}
  0 & -\frac{2}{\sqrt{35}} & 2\sqrt{\frac{6}{35}} & \frac{1}{\sqrt{15}} &
  -2\sqrt{\frac{2}{105}} & \sqrt{\frac{2}{35}} \\
  -\frac{2}{\sqrt{35}} & \frac{1}{7} & \frac{2\sqrt{6}}{35} &
  -\frac{2}{5}\sqrt{\frac{7}{3}} & \frac{2}{7}\sqrt{\frac{2}{3}} &
  \frac{16 \sqrt{2}}{35} \\
  2\sqrt{\frac{6}{35}} & \frac{2\sqrt{6}}{35} & -\frac{1}{7} & 0 &
  -\frac{6}{35} & \frac{2\sqrt{3}}{7} \\
  \frac{1}{\sqrt{15}} & -\frac{2}{5}\sqrt{\frac{7}{3}} & 0 &
  0 & \frac{\sqrt{14}}{5} & 0 \\
  -2 \sqrt{\frac{2}{105}} & \frac{2}{7}\sqrt{\frac{2}{3}} & \frac{6}{35} &
  \frac{\sqrt{14}}{5} & \frac{2}{7} & \frac{9\sqrt{3}}{35} \\
  \sqrt{\frac{2}{35}} & \frac{16\sqrt{2}}{35} & \frac{2\sqrt{3}}{7} & 0 &
  \frac{9 \sqrt{3}}{35} & -\frac{2}{7} \\
\end{matrix}\right)$
\\ 
\hline\hline
\end{tabular}
\end{table*}

\section{The Consistent Description of the Pentaquark Trio}
\label{sec:Pc-trio}

In this section we investigate whether the OBE model can describe
the LHCb pentaquark trio consistently.
We find that the removal of the short-range Dirac-delta contributions
to the OBE potential is a necessary step for achieving this goal.
We discuss the possible interpretations and justifications of
this modification to the OBE model.

\subsection{Predictions of the $P_c(4440)$ and $P_c(4457)$}

In this manuscript, following the ideas of Refs.~\cite{Liu:2018bkx,Liu:2019stu},
we propose the determination of the cutoff from the condition of
reproducing the mass of a known molecular candidate.
As there are three hidden-charm pentaquarks, we are left
with three possibilities:
the $P_c(4312)$ (as a $\bar{D} \Sigma_c$ bound state),
the $P_c(4440)$ and the $P_c(4457)$ (as $\bar{D}^* \Sigma_c$ bound states).
Owing to the aforementioned regulator artifact in the spin-spin piece
of the OBE potential, the most suitable choice is the $P_c(4312)$,
which for the parameters of Table \ref{tab:couplings} is reproduced for
\begin{eqnarray}
  \Lambda_1 = 1119\,{\rm MeV} \, .
\end{eqnarray}
In the naive OBE model, this cutoff leads to the predictions
\begin{eqnarray}
  M(\tfrac{1}{2}) = 4388 \quad \mbox{and} \quad
  M(\tfrac{3}{2}) = 4459 \,{\rm MeV} \, ,
\end{eqnarray}
which are not compatible with the experimental masses of
the $P_c(4440)$ and the $P_c(4457)$, i.e.
\begin{eqnarray}
  M_{P_{c2}} = 4440.3 \pm 1.3 {}^{+4.1}_{-4.6}
  \quad \mbox{and}
  \quad
  M_{P_{c3}} = 4457.3 \pm 0.6 {}^{+4.1}_{-1.7} \, {\rm MeV} \,  . \nonumber \\
\end{eqnarray}
As already explained, the reason for this mismatch is the distortion of
the OBE potential at relatively long distances owing to the delta-like
contribution to the spin-spin interaction, which we will explain
in what follows.

\subsection{The One-Pion-Exchange Potential with a Monopolar Form Factor}

Now if we inspect the OPE contribution to the OBE potential,
it contains a spin-spin and a tensor piece
\begin{eqnarray}
  V_{\pi} = \vec{\sigma}_{L1} \cdot \vec{S}_{L2}\,
  V_{\pi(S)} + S_{L12}(\hat{r})\,V_{\pi(T)} \, ,
\end{eqnarray}
The spin-spin piece reads
\begin{eqnarray}\label{eq:ope}
  V_{\pi(S)} &=&
  \frac{g_1 g_2}{6 f_{\pi}^2}\,\vec{\tau}_1\cdot \vec{T}_2\,m_{\pi}^3\,
  \nonumber \\ &\times&
  \left[ - d(m_{\pi} r, \frac{\Lambda}{m_{\pi}}) +
    W_Y(m_{\pi} r, \frac{\Lambda}{m_{\pi}}) \right] \, ,
\end{eqnarray}
where $d$ and $W_Y$ are the regularized delta-like and Yukawa-like
contributions defined in Eqs.~(\ref{eq:d-FF}) and (\ref{eq:WY-FF}).
\begin{figure}[ttt!]
\begin{center}
\includegraphics[width=8.5cm]{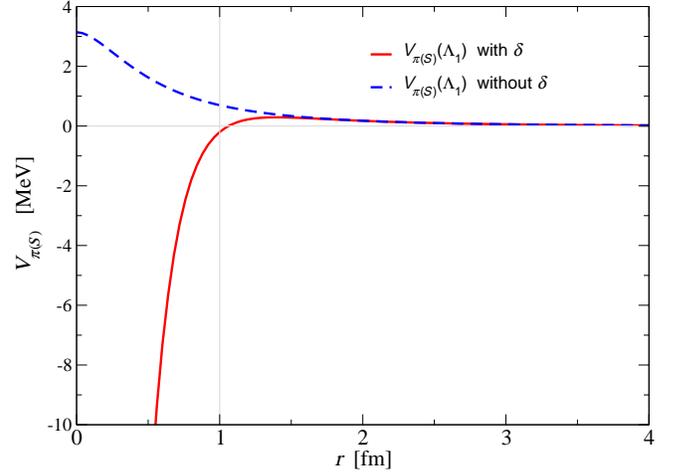}
\end{center}
\caption{The spin-spin piece of the OPE potential as a function of
  the radius $r$ with and without the delta-like contributions.
  For simplicity we show this piece for $\vec{\tau}_1 \cdot \vec{T}_2 = +1$, which corresponds to $I=\tfrac{3}{2}$.
}
\label{fig:ope}
\end{figure}

As can be seen from Eq.~(\ref{eq:ope}) and Fig.~\ref{fig:ope},
these two contributions have opposite sign: the delta-like
contribution will generate a strong short-range attraction/repulsion
that is unphysical.
If the range of this unphysical contribution is short enough,
it will have no observable effect in the predictions of the OBE model.
However the problem is that this is not the case.
If we compute the OPE potential contribution with a monopolar cutoff
$\Lambda_1 = 1119\,{\rm MeV}$, the OPE potential changes sign at
$r = 1.1\,{\rm fm}$, which is comparable with the range of
the OPE potential $R_{\pi} = 1/m_{\pi} = 1.4\,{\rm fm}$.
This is unsettling to say the least: the modification of the form factors
to the OBE potential is expected to be short-ranged, but certainly not of
the order of the pion range.
This indicates that it is better to remove this contribution.
If we remove the delta-like contributions of the pion and
the vector mesons, we end up with the predictions
\begin{eqnarray}
  M(\tfrac{1}{2}) = 4458.0
  \,{\rm MeV}
  \quad \mbox{and} \quad
  M(\tfrac{3}{2}) = 4443.9
  \,{\rm MeV}
  \, ,
\end{eqnarray}
which are basically compatible with the experimental determination
of the masses of the $P_c(4440)$ and $P_c(4457)$.

\section{The Pentaquark Multiplet}
\label{sec:Pc-multiplet}

In this section we compute the predictions of the OBE model
for the hidden-charm molecular pentaquarks.
We determine the cutoff in the calculation from the condition of reproducing
the $P_c(4312)$, as explained in Sect.~\ref{sec:Pc-trio}.
From this condition and the OBE potential we can simply determine
the full hidden-charm molecular spectrum.
We also explain how we estimate the uncertainties of the OBE model.

\subsection{Error Estimations}
\label{subsec:error-estimations}

The OBE model has a series of uncertainties, mostly stemming
from the choice of the coupling constants.
This error source can be dealt with by assigning
a relative uncertainty to the OBE potential:
\begin{eqnarray}
  V({P_c}) = V_{\rm OBE} \, (1 \pm \delta_{\rm OBE}) \, , \label{eq:err-global}
\end{eqnarray}
where $V(P_c)$ is the molecular pentaquark potential in a given channel and
$V_{\rm OBE}$ is the central value of the OBE potential with the central value
of the couplings, see Table \ref{tab:couplings} for details.
We will assume the relative uncertainty to be $\delta_{\rm OBE} = 30\%$,
which is equivalent to assume that the average relative uncertainty
of the coupling constants in Table \ref{tab:couplings} is
$\delta_{\rm coupling} = \delta_{\rm OBE} / {2} \sim 15\%$
(assuming a Gaussian uncertainty distribution
for the couplings)~\footnote{The previous $15\%$ figure is an educated guess,
  as it is markedly difficult to assess the specific errors in the couplings.
  The biggest uncertainty lies probably in the $g_{\sigma}$ coupling,
  which comes from the non-linear sigma model and the quark model:
  though difficult to determine, it could very well be about $20-30\%$.
  For $g_{\rho}$ and $g_{\omega}$ the uncertainty is expected to be smaller,
  maybe at the $10\%$ level, as can be deduced from comparing
  (at least in the heavy meson case) the value deduced from Sakurai's
  universality ($g_{\rho} \simeq 2.9$) with the one calculated
  in the lattice ($g_{\rho} = 2.6 \pm 0.1 \pm 0.4$)~\cite{Detmold:2007wk},
  or the value deduced from semileptonic decays and
  vector meson dominance ($g_{\rho} = 2.6$)~\cite{Casalbuoni:1992dx}.
}.
With the average uncertainty $\delta_{\rm OBE}$ we can recalculate
the cutoff $\Lambda_1$ by determining the location of
the $P_c(4312)$, leading to
\begin{eqnarray}
  \Lambda_1 = 1.119^{+0.190}_{-0.094} \, {\rm GeV} \, .
\end{eqnarray}
The error in the binding energies is simply obtained by propagating
the $(1 \pm \delta_{\rm OBE})$ uncertainty in the OBE potential,
with the condition of recalculating the cutoff
as to reproduce the $P_c(4312)$.
This condition implies the partial renormalization of the OBE model,
which manifests in the fact that the errors derived
from the overall uncertainty in the potential
are rather small.
For the particular case of the $\bar{D}^* \Sigma_c$ bound states
we arrive at
\begin{eqnarray}
  B_E(\tfrac{1}{2}^-) = 4.2^{+0.6}_{-0.7}\,{\rm MeV} \quad \mbox{and} \quad
  B_E(\tfrac{3}{2}^-) = 18.3^{+0.6}_{-0.0}\,{\rm MeV} \, , \nonumber \\
  \label{eq:be-err-OBE}
\end{eqnarray}
where the errors, besides being small, are also asymmetric.
There is a second error source: HQSS is not exact for finite heavy quark masses.
The relative size of HQSS violations are expected to be of the order of
$\delta_{\rm HQSS} \sim \Lambda_{\rm QCD} / m_Q$,
with $\Lambda_{\rm QCD} \sim (200-300)\,{\rm MeV}$ and
$m_Q$ the mass of the heavy quark.
This error manifests in random variations of the OBE potential
around its expected HQSS limit
\begin{eqnarray}
  V(P_c) = V_{\rm OBE}^{(m_Q = \infty)}\,(1 \pm \delta_{\rm HQSS}) \, ,
  \label{eq:err-hqss}
\end{eqnarray}
where in the charm sector we expect $\delta_{\rm HQSS} \sim 15\%$.
It is worth stressing the difference between the OBE error of
Eq.~(\ref{eq:err-global}) and the HQSS error of Eq.~(\ref{eq:err-hqss}):
the OBE error takes into account the error in the coupling constants
but assumes that these couplings are identical for all the possible molecules,
while the HQSS error considers that these couplings might be different
for each of the molecular states.
For the $\bar{D}^* \Sigma_c$ bound states the HQSS uncertainty is
\begin{eqnarray}
  B_E(\tfrac{1}{2}^-) = 4.2^{+5.3}_{-3.3}\,{\rm MeV} \quad \mbox{and} \quad
  B_E(\tfrac{3}{2}^-) = 18.3^{+11.6}_{-9.2}\,{\rm MeV} \, , \nonumber \\
  \label{eq:be-err-hqss}
\end{eqnarray}
which is considerably larger than the OBE uncertainty.
The reason why this happens is that the OBE uncertainty is
{\it renormalized away}:
changes in the couplings of the light mesons to the heavy hadrons
are compensated by a change in the cutoff.
On the contrary HQSS violations imply that the couplings are different
for the ground and excited spin states of a heavy hadron, i.e.
the couplings for the $D$ and $D^*$ (or $\Sigma_c$ and $\Sigma_c^*$)
are a bit different. This uncertainty is not absorbed by the cutoff
variation and results in a larger error.
Finally for the full error we will sum in quadrature the OBE and HQSS errors.

In addition to the binding energies of the molecular pentaquarks,
we also compute the S-wave scattering lengths of
the charmed antimeson-baryon systems.
The reason is to identify molecular configurations
in which the attraction is strong, but not strong enough to bind.
The basis of this idea is a well-known relation between the two-body
scattering length and binding energy, $a_2$ and $B_2$, that works
in the limit in which the bound state is weakly bound
\begin{eqnarray}
  a_2 = \frac{1}{\sqrt{2 \mu B_2}} +
  \mathcal{O}(\frac{\sqrt{2 \mu B_2}}{m_{\pi}}) \, ,
\end{eqnarray}
with $\mu$ the reduced mass of the system and $m_{\pi}$ the pion mass.
For a shallow bound state, i.e. $m_{\pi} > \sqrt{2 \mu B_2} > 0$,
the scattering length is positive and large {($m_{\pi} a_2 \gg 1$)}.
For $B_2 \to 0$ the scattering length diverges and for a system
that almost binds, the scattering length
is negative and large.
We notice that we compute the scattering lengths under the assumption
that the charmed antimeson and the charmed baryon are stable hadrons
with respect to the strong interaction, which is not true in general.
This is not important as we are actually using the scattering length
as a tool to identify configurations that are close to binding.

\subsection{Predictions}

With the OBE model regularized without the delta-like contributions,
we can predict the seven possible S-wave $\bar{D}^{(*)} \Sigma_c^{(*)}$
molecules.
The results are summarized in Table \ref{tab:binding-hidden-Pc}.
For the isodoublet ($I = \tfrac{1}{2}$) molecular pentaquarks,
the states predicted in the OBE model are indeed very similar to
the ones obtained in {\it scenario B} of the contact-range
EFT of Ref.~\cite{Liu:2019tjn}
{and the pionful one of Ref.~\cite{Valderrama:2019chc}}.
Here we note that within {the pionless and pionful} descriptions of
Refs.~\cite{Liu:2019tjn,Valderrama:2019chc}
there are two coupling constants
whose values have to be determined
from experimental information.
Thus two scenarios were considered: scenario A, in which the $P_c(4440)$
and $P_c(4457)$ are the $J=\tfrac{1}{2}$ and $\tfrac{3}{2}$
$\bar{D}^* \Sigma$ molecules, and scenario B
for the opposite identification.
Our OBE model naturally selects scenario B.

For the isoquartet ($I = \tfrac{3}{2}$) molecular pentaquarks,
we find that the $J = \tfrac{1}{2}$ $\bar{D}^* \Sigma_c$ and
the $J = \tfrac{1}{2}$ $\bar{D}^* \Sigma_c^*$ bind.
Yet this conclusion is not particularly strong:
these two molecular pentaquarks are weakly bound and once we consider
the error in the binding energies the outcome is that
there is a fair likelihood that they will not bind.
The isoquartet $J = \tfrac{3}{2}$ $\bar{D}^* \Sigma_c^*$ molecule is close
to binding, as can be inferred from the large negative scattering length.
Conversely, the uncertainties in the OBE model mean that this molecular
pentaquark might be able to bind.
The other isoquartet molecules display mild attraction, a conclusion
which can be deduced from the negative (but natural) values of
the scattering length shown in Table \ref{tab:binding-hidden-Pc}.

\begin{table}[!ttt]
\caption{
  Scattering lengths ($a_2$ in ${\rm fm}$)
  and binding energies ($B_2$ in ${\rm MeV}$) of prospective isodoublet
  and isoquartet hidden-charm antimeson-baryon molecules.
  The column ``Molecule'' refers to the two-body system under consideration,
  while $I$ and $J^P$ denote the isospin and total angular momentum
  and parity of the system.
  The error comes from an estimated relative uncertainty
  for the OBE potential of the order of $30\%$ and
  from HQSS violations of the order of $15\%$,
  where the second error source dominates.
  $M$ refers to the predicted mass (the central value) of a
  particular heavy antimeson-baryon molecule (if it binds).
  The calculation of the scattering length assumes
  that the hadrons are stable.
  }
\label{tab:binding-hidden-Pc}
\begin{tabular}{cccccc}
\hline\hline
Molecule  & $I$ & $J^{P}$ & $a_2$ (fm) & $B_2$ (MeV) & $M$ (MeV) \\
\hline
$\bar{D}\Sigma_c$ & $\tfrac{1}{2}$ & $\tfrac{1}{2}^-$ &
$1.9^{+1.0}_{-0.4}$ & Input & Input \\
\hline
$\bar{D}\Sigma_c^*$ & $\tfrac{1}{2}$ & $\tfrac{3}{2}^-$ &
$1.9^{+0.9}_{-0.4}$
& $9.3^{+7.7}_{-5.7}$ & $4376.0$ \\
\hline
$\bar{D}^*\Sigma_c$ & $\tfrac{1}{2}$ & $\tfrac{1}{2}^-$
& $2.5^{+2.3}_{-0.6}$ & $4.2^{+5.3}_{-3.4}$ & $4458.0$  \\
$\bar{D}^*\Sigma_c$ & $\tfrac{1}{2}$ & $\tfrac{3}{2}^-$
& $1.4^{+0.5}_{-0.3}$ & $18.3^{+11.6}_{-9.2}$ & $4443.9$ \\
\hline
$\bar{D}^*\Sigma_c^*$ & $\tfrac{1}{2}$ & $\tfrac{1}{2}^-$
& $2.6^{+2.5}_{-0.7}$ & $2.9^{+4.5}_{-2.6}$ & $4523.8$ \\
$\bar{D}^*\Sigma_c^*$ & $\tfrac{1}{2}$ & $\tfrac{3}{2}^-$
& $1.9^{+1.0}_{-0.4}$ & $9.2^{+7.9}_{-5.8}$ & $4517.5$  \\
$\bar{D}^*\Sigma_c^*$ & $\tfrac{1}{2}$ & $\tfrac{5}{2}^-$
& $1.3^{+0.4}_{-0.3}$ & $22.4^{+13.1}_{-10.6}$ & $4504.3$ \\
  \hline\hline
Molecule  & $I$ & $J^{P}$ & $a_2$ (fm) & $B_2$ (MeV) & $M$ (MeV) \\
\hline
$\bar{D}\Sigma_c$ & $\tfrac{3}{2}$ & $\tfrac{1}{2}^-$ & $-1.8^{+1.2}_{-2.9}$
& $-$ & $-$ \\
\hline
$\bar{D}\Sigma_c^*$ & $\tfrac{3}{2}$ & $\tfrac{3}{2}^-$ & $-1.8^{+1.2}_{-3.2}$
& $-$ & $-$ \\
\hline
$\bar{D}^*\Sigma_c$ & $\tfrac{3}{2}$ & $\tfrac{1}{2}^-$
& $7.1^{+\infty (-19.5)}_{-4.8}$ & $0.4^{+2.2}_{\dagger}$ & $4461.8$ \\
$\bar{D}^*\Sigma_c$ & $\tfrac{3}{2}$ & $\tfrac{3}{2}^-$ & $-0.8^{+0.5}_{-1.4}$ & $-$ & $-$ \\
\hline
$\bar{D}^*\Sigma_c^*$ & $\tfrac{3}{2}$ & $\tfrac{1}{2}^-$
& $3.9^{+9.8}_{-1.7}$ & $1.4^{+3.2}_{-1.8}$ & $4325.3$ \\
$\bar{D}^*\Sigma_c^*$ & $\tfrac{3}{2}$ & $\tfrac{3}{2}^-$ & $-10.6^{+11.2}_{-\infty (10.0)}$
& $-$ & $-$  \\
$\bar{D}^*\Sigma_c^*$ & $\tfrac{3}{2}$ & $\tfrac{5}{2}^-$ & $-0.6^{+0.4}_{-0.8}$
& $-$ & $-$ \\
  \hline\hline 
\end{tabular}
\end{table}

\subsection{Inclusion of the $\eta$ meson as a theoretical cross-check}

Finally we will consider the effect of including
the $\eta$ meson in the OBE model.
With this we want to check whether our theoretical error estimations
are reliable or not.
As a matter of fact we have a partial check in the prediction of
the $P_c(4440)$ and $P_c(4457)$ pentaquark masses
in Table \ref{tab:binding-hidden-Pc},
which are compatible with the experimental ones.
We find it worth mentioning that the recent theoretical analysis of
Ref.~\cite{Du:2019pij} indicates possible evidence of a narrow $P_c(4380)$
$\bar{D} \Sigma^*$ molecule within the current experimental data,
a result which is also compatible with Table \ref{tab:binding-hidden-Pc}.
Yet a robust experimental confirmation would still require the detection of
the full pentaquark multiplet and their quantum numbers.
Be it as it may, besides experiment, another method to cross-check our results
is a theory-with-theory comparison.
By including the $\eta$ we are indeed comparing the OBE model with itself.

The OBE model is a phenomenological model and its formulation is
dependent on a set of arbitrary choices that might affect its predictions.
The choice of the form-factor cutoff, which we addressed
in Section \ref{sec:Pc-trio}, is the most obvious example.
Yet the selection of which light mesons to include in the description is
equally important.
Here we advocate for a minimalistic OBE in which only the $\pi$, $\sigma$,
$\rho$ and $\omega$ are taken into account.
Interestingly whether this choice is good enough can be explicitly tested
by including additional light mesons.
The obvious candidate is the $\eta$ meson, the mass of which
($m_{\eta} = 548\,{\rm MeV}$) is comparable to the one of the $\sigma$ meson
and thus they will both have a comparable range.
The derivation of the $\eta$-exchange potential is presented
in Appendix \ref{app:eta}, while here we simply notice that
it contains a spin-spin and tensor pieces,
but not a central one.
As a consequence the inclusion of the $\eta$ meson has no effect
for the $\bar{D} \Sigma_c$ and $\bar{D} \Sigma_c^*$ molecules,
in which the OBE potential is purely central.
In particular the $P_c(4312)$ pentaquark, which we use for determining
the form-factor cutoff, is unaffected.
Thus we end up with the same cutoff as before,
i.e. $\Lambda_1 = 1119\,{\rm MeV}$.
However $\eta$-exchange will affect the location of the other pentaquarks,
where in Table \ref{tab:binding-hidden-Pc-eta} we show a detailed comparison
of the $\eta$-less and $\eta$-full OBE models.

In Table \ref{tab:binding-hidden-Pc-eta} we observe that in general
the effects of including the $\eta$ are relatively modest, of
the order of $0.5-1.0\,{\rm MeV}$ in the binding energies in most cases.
That is, the effects of $\eta$-exchange are within the error bands we had
already computed, providing further support for our error estimations.
This also indicates that the inclusion of $\eta$-exchange is
not necessary for the pentaquarks as molecular states.
Yet we warn that this conclusion might be specific to pentaquarks,
as there might be other molecular states in which $\eta$-exchange
could play an important role~\cite{Karliner:2016ith}.
However for the particular case of the pentaquarks it has been argued
that pion exchanges might be perturbative~\cite{Valderrama:2019chc}.
If this were to be the case, then it is not surprising that $\eta$-exchange
plays a minor role in the OBE model, as it is considerably weaker than
pion-exchange owing to SU(3)-flavor geometric factors and also
to $f_{\eta} > f_{\pi}$, which further weakens $\eta$-exchange.

\begin{table*}[!ttt]
\caption{
  Comparison of the scattering lengths and binding energies in our original
  OBE model without $\eta$-exchange ($a_2^{\slashed \eta}$ and $B_2^{\slashed \eta}$,
  in ${\rm fm}$ and ${\rm MeV}$ units, respectively) and after including
  $\eta$-exchange ($a_2^{\eta}$ and $B_2^{\eta}$).
  The columns ``Molecule'', $I$ and $J^P$ read as
  in Table \ref{tab:binding-hidden-Pc},
  while $M^{\slashed \eta}$ and $M^{\eta}$ refer to the masses of the pentaquarks
  in the $\eta$-less and $\eta$-full OBE model.
  The error estimations are estimated in the same way as
  in Table \ref{tab:binding-hidden-Pc} (i.e. assume a $30\%$ uncertainty
  in the strength of the OBE potential and HQSS violations
  at the $15\%$ level).
  The addition of the $\eta$ meson in the OBE model leaves the prediction
  for the $\bar{D} \Sigma$ and $\bar{D} \Sigma^*$ molecules unchanged,
  as these molecules depend only on the central components of
  the potential while $\eta$-exchange only generates
  spin-spin and tensor components.
  In general the predictions of the $\eta$-less and $\eta$-full OBE model
  overlap within the estimated error bands and are thus indistinguishable
  at the theoretical level.
  }
\label{tab:binding-hidden-Pc-eta}
\begin{tabular}{ccccccccc}
\hline\hline
Molecule  & $I$ & $J^{P}$
& $a_2^{\slashed \eta}$ (fm) & $a_2^{\eta}$ (fm) &
$B_2^{\slashed \eta}$ (MeV) & $B_2^{\eta}$ (MeV) &
$M^{\slashed \eta}$ (MeV) & $M^{\eta}$\\
\hline
$\bar{D}\Sigma_c$ & $\tfrac{1}{2}$ & $\tfrac{1}{2}^-$ &
$1.9^{+1.0}_{-0.4}$ & $1.9^{+1.0}_{-0.4}$
& Input & Input
& Input & Input \\
\hline
$\bar{D}\Sigma_c^*$ & $\tfrac{1}{2}$ & $\tfrac{3}{2}^-$ &
$1.9^{+0.9}_{-0.4}$ & $1.9^{+0.9}_{-0.4}$ &
$9.3^{+7.7}_{-5.7}$ & $9.3^{+7.7}_{-5.7}$ &
$4376.0$ & $4376.0$
\\
\hline
$\bar{D}^*\Sigma_c$ & $\tfrac{1}{2}$ & $\tfrac{1}{2}^-$ &
$2.5^{+2.3}_{-0.6}$ & \mpv{$2.4^{+2.0}_{-0.6}$} &
$4.2^{+5.3}_{-3.4}$ & \mpv{$4.8^{+5.7}_{-3.6}$} &
$4458.0$ & \mpv{$4457.4$} \\
$\bar{D}^*\Sigma_c$ & $\tfrac{1}{2}$ & $\tfrac{3}{2}^-$
& $1.4^{+0.5}_{-0.3}$ & \mpv{$1.3^{+0.5}_{-0.2}$} &
$18.3^{+11.6}_{-9.2}$ & \mpv{$17.3^{+11.1}_{-8.8}$} &
$4443.9$ & \mpv{$4442.9$} \\
\hline
$\bar{D}^*\Sigma_c^*$ & $\tfrac{1}{2}$ & $\tfrac{1}{2}^-$
& $2.6^{+2.5}_{-0.7}$ & \mpv{$2.4^{+2.1}_{-0.6}$}& 
$2.9^{+4.5}_{-2.6}$ & \mpv{$3.8^{+5.4}_{-3.2}$} &
$4523.8$ & \mpv{$4522.9$} \\
$\bar{D}^*\Sigma_c^*$ & $\tfrac{1}{2}$ & $\tfrac{3}{2}^-$
& $1.9^{+1.0}_{-0.4}$ & \mpv{$1.9^{+1.0}_{-0.4}$}& 
$9.2^{+7.9}_{-5.8}$ & \mpv{$9.4^{+8.0}_{-5.8}$} &
$4517.5$ & \mpv{$4517.3$} \\
$\bar{D}^*\Sigma_c^*$ & $\tfrac{1}{2}$ & $\tfrac{5}{2}^-$
& $1.3^{+0.4}_{-0.3}$ & \mpv{$1.3^{+0.6}_{-0.3}$} & 
$22.4^{+13.1}_{-10.6}$ & \mpv{$21.1^{+12.3}_{-10.1}$} &
$4504.3$ & \mpv{$4503.0$} \\
  \hline\hline
  Molecule  & $I$ & $J^{P}$
  & $a_2^{\slashed \eta}$ (fm) & $a_2^{\eta}$ (fm) &
  $B_2^{\slashed \eta}$ (MeV) & $B_2^{\eta}$ (MeV) &
  $M^{\slashed \eta}$ (MeV) & $M^{\eta}$\\
\hline
$\bar{D}\Sigma_c$ & $\tfrac{3}{2}$ & $\tfrac{1}{2}^-$ &
$-1.8^{+1.2}_{-2.9}$ & $-1.8^{+1.2}_{-2.9}$
& $-$ & $-$
& $-$ & $-$ \\
\hline
$\bar{D}\Sigma_c^*$ & $\tfrac{3}{2}$ & $\tfrac{3}{2}^-$ &
$-1.8^{+1.2}_{-3.2}$ & $-1.8^{+1.2}_{-3.2}$ 
& $-$ & $-$  
& $-$ & $-$ \\
\hline
$\bar{D}^*\Sigma_c$ & $\tfrac{3}{2}$ & $\tfrac{1}{2}^-$
& $7.1^{+\infty (-19.5)}_{-4.8}$ & \mpv{$5.7^{+\infty (-97.3)}_{-3.4}$}
& $0.4^{+2.2}_{\dagger}$ & \mpv{$0.7^{+2.0}_{\dagger}$}
& $4461.8$ & \mpv{$4461.5$} \\
$\bar{D}^*\Sigma_c$ & $\tfrac{3}{2}$ & $\tfrac{3}{2}^-$ &
$-0.8^{+0.5}_{-1.4}$ & \mpv{$-0.8^{+0.4}_{-1.1}$} &
$-$ & $-$  &
$-$ & $-$\\
\hline
$\bar{D}^*\Sigma_c^*$ & $\tfrac{3}{2}$ & $\tfrac{1}{2}^-$
& $3.9^{+9.8}_{-1.7}$ & \mpv{$3.6^{+5.2}_{-1.6}$} &
$1.4^{+3.2}_{-1.8}$ & \mpv{$2.0^{+3.2}_{-1.7}$} &
$4525.3$ & \mpv{$4524.7$} \\
$\bar{D}^*\Sigma_c^*$ & $\tfrac{3}{2}$ & $\tfrac{3}{2}^-$ &
$-10.6^{+11.2}_{-\infty (10.0)}$ & \mpv{$-15.6^{+17.8}_{-\infty (7.9)}$}
& $-$ & $-$
& $-$ & $-$ \\
$\bar{D}^*\Sigma_c^*$ & $\tfrac{3}{2}$ & $\tfrac{5}{2}^-$ &
$-0.6^{+0.4}_{-0.8}$ & \mpv{$-0.5^{+0.5}_{-0.8}$}
& $-$ & $-$
& $-$ & $-$ \\
  \hline\hline 
\end{tabular}
\end{table*}

\section{The OBE model from a modern perspective}
\label{sec:modern}

In this Section we will consider the OBE model
--- in particular the removal of the Dirac-delta contributions ---
from a modern understanding grounded on renormalization and EFT ideas.
From a phenomenological perspective this removal is motivated because
the range of the regularized Dirac-delta contributions is unexpectedly long,
resulting in the distortion of the OPE potential at distances
comparable to the pion Compton wavelength.
From a modern perspective this distortion will be considered a
{\it regulator artifact}, which should be taken care of by means of
the renormalization procedure.
In the following lines we will explain these points of view
in order to put the OBE model in context.

From a traditional perspective the existence of short-range ambiguities
in the OBE model is apparent from the fact that
the unregularized OBE potential is singular,
with the tensor contributions diverging as $1/r^3$
for distances $m r \ll 1$, with $m$ the mass of
the exchanged boson.
This type of potentials require regularization if we want
to have a unique solution of the Schr\"odinger equation~\cite{Case:1950an}
(for more modern treatments of singular interactions
see Refs.~\cite{Beane:2000wh,PavonValderrama:2005gu,PavonValderrama:2005wv,PavonValderrama:2005uj}).
However we do not expect the behavior of the OBE potential at distances shorter
than the size of the hadrons to be physical.
Thus, we remove these unphysical short-range contributions by regularizing
the OBE potential.
This is the reason that justified the inclusion of form factors
in the original OBE model and this is also the reason
why we removed the Dirac-deltas in Sect.~\ref{sec:Pc-trio}.

Nowadays we know that the removal of short-range ambiguities requires
not only regularization, but also renormalization.
By this we mean the following: we {expect} to trade-off the short-range
ambiguities by observable information.
In the original OBE model we simply regularize the potential by choosing a
sensible form factor and cutoff.
The renormalization process is more systematic: we explicitly
include a contact-range potential to model
the unknown short-range physics.
By fitting the couplings in this contact-range potential to experimental
information we are effectively absorbing the dependence on the form factor
and the cutoff in these couplings.
The price to pay is a reduction in the predictive power,
as we have to include new parameters in the theory
which have to be determined from experimental data.

Yet renormalization helps to understand in hindsight
the success of phenomenological models.
In the particular case of the OBE potential, choosing the form factor and
the cutoff as to reproduce experimental information basically amounts
to an implicit (but usually incomplete) renormalization process
(for an explicit and complete renormalization of the OBE potential
we recommend Ref.~\cite{Cordon:2009pj}).
In particular we will consider in detail the following two perspectives
\begin{enumerate}
\item[(i)] the removal of the Dirac-delta contributions as a renormalization choice, and
\item[(ii)] the relation of the OBE model with EFT descriptions of the pentaquarks.
\end{enumerate}

\subsection{Renormalization and the OBE model}

As we have already mentioned, the most important difference of the OBE model
as presented in this work with its standard implementation is the removal of
the contact-range pieces that appear in the spin-spin component of
the potential.
That is, we have identified that the OBE potential can be divided
into two pieces
\begin{eqnarray}
  V_{\rm OBE} = V_{\rm OBE(F)} + V_{\rm OBE(C)} \, ,
\end{eqnarray}
where $F$ and $C$ refer to the finite- and contact-range piece contributions.
The contact-range piece conforms to the general structure of
Eq.~(\ref{eq:contact-light}), i.e.
\begin{eqnarray}
  V_{\rm OBE(C)} = C_a^{\rm OBE} + C_b^{\rm OBE} \, \vec{\sigma}_{L1} \cdot \vec{S}_{L2}
  \,
\end{eqnarray}
where the couplings are given by
\begin{eqnarray}
  C_a^{\rm OBE} &=& 0 \, , \\
  C_b^{\rm OBE} &=& +
  \left[ C_{b}^{\delta \pi} + C_{b}^{\delta \rho} + C_{b}^{\delta \omega} \right] \, , 
\end{eqnarray}
where $C_b^{\delta \pi}$, $C_b^{\delta \rho}$ and $C_b^{\delta \omega}$ are determined by
the Dirac-delta contributions to the spin-spin component of
the OBE potential (hence the notation $\delta \pi$,
$\delta \rho$, $\delta \omega$).
These components come from $\pi$ exchange and the magnetic-like piece of $\rho$
and $\omega$ exchange~\footnote{Notice
  that there is no Dirac-delta contribution stemming
  from the central pieces of the potential, i.e.
  from the exchange of the scalar meson $\sigma$ or
  from the electric-type couplings of the $\rho$ and $\omega$.
}:
\begin{eqnarray}
  C_{b}^{\delta \pi} &=& -\frac{g_1 g_2}{6 f_{\pi}^2}\,\vec{\tau}_1 \cdot \vec{T}_2\,
  \, , \label{eq:pion-delta-contrib} \\
  C_{b}^{\delta \rho} &=& -\frac{2}{3}\,
  \frac{f_{\rho 1}}{2 M_1}\,\frac{f_{\rho 2}}{2 M_2}\,\vec{\tau}_1 \cdot \vec{T}_2
  \, ,
  \\
  C_{b}^{\delta \omega} &=& -\frac{2}{3}\,
  \frac{f_{\omega 1}}{2 M_1}\,\frac{f_{\omega 2}}{2 M_2} \, .
\end{eqnarray}
The removal of these contact-range pieces is equivalent
to the following substitution:
\begin{eqnarray}
  V_{\rm OBE} = V_{\rm OBE(F)} + V_{\rm OBE(C)}
  \quad \rightarrow \quad
  V_{\rm OBE} = V_{\rm OBE(F)} \, , \nonumber
\end{eqnarray}
that is, instead of using the complete OBE potential we restrict ourselves to
the finite-range piece of the OBE potential, where the justification
we have provided so far is the unexpected long-range distortion of
the finite-range pieces when the $V_{\rm OBE(C)}$ piece is kept.

Although there is indeed a physical justification for this removal,
it would be interesting whether this can be justified
from a renormalization perspective.
A straightforward justification is the explicit inclusion of a contact-range
component in the potential
\begin{eqnarray}
  V = \delta V_C + V_{\rm OBE(F)} + V_{\rm OBE(C)} \, ,
\end{eqnarray}
where $\delta V_C$ merely represents the unknown short-range physics
not explicitly included in the OBE model, where the structure of 
$\delta V_C$ still follows Eq.~(\ref{eq:contact-light}):
\begin{eqnarray}
  \delta V_C = \delta C_a + \delta C_b \,
  \vec{\sigma}_{L1} \cdot \vec{S}_{L2}
  \, ,
\end{eqnarray}
Obviously the choice
\begin{eqnarray}
  \delta V_c = - V_{\rm OBE(C)} \, ,
\end{eqnarray}
will do the trick, where this choice corresponds to
\begin{eqnarray}
  \delta C_a = 0 \quad \mbox{and} \quad \delta C_b = - C_b^{\rm OBE} \, .
  \label{eq:delta-Vc-OBE}
\end{eqnarray}
From this point of view the removal of the Dirac-delta contributions
merely amounts to playing a shell game between unknown short-range
interactions and the Dirac-delta contributions already present
in the OBE potential.
Complementarily the observation that with this choice we end up with a correct
prediction of the masses of the $P_c(4440)$ and $P_c(4457)$ pentaquarks
motivates and provides physical content to Eq.~(\ref{eq:delta-Vc-OBE}) 

Here it is worth noticing that the rationale of renormalization
is very different from that in the original OBE model, where
this strong distortion of the pion contribution to
the potential at long distances was avoided by the use of a
large enough cutoff, usually $\Lambda_{\pi} > 1.3\,{\rm GeV}$.
Besides, the finite-range piece of the spin-spin piece of
the OPE potential is attractive in the S-wave singlet
and triplet partial waves, which in turn leads to
a repulsive Dirac-delta contribution.
But for the heavy antimeson-baryon system it is difficult to have a large
enough cutoff that still reproduces the  three pentaquark
poles~\footnote{This will require making the $\omega$-meson contribution
  considerably more repulsive by breaking the SU(3) relation
  $g_{\rho} = g_{\omega}$, which is what happens in the OBE model
  as applied to the nucleon-nucleon system.}.
In any case it would be interesting to check whether the present
identification of the quantum numbers of the $P_c(4440)$ and $P_c(4457)$
will still be {correct} in a fully renormalized OBE model
as the one presented in Ref.~\cite{Cordon:2009pj},
i.e. with a cutoff that is not fixed but floats
within a given range.

\subsection{The EFT description and the OBE model}

The previous interpretation is rather formal, merely showing
that it is possible to renormalize away the Dirac-deltas
appearing in the OBE potential.
It is more interesting though to compare the OBE model with the EFT approach,
as this would provide us with some interesting insights.

Within EFT we divide physics into long- and short-range contributions,
where the long-range contributions are assumed to be known
while the short-range ones are not.
If we are dealing with a non-relativistic problem, what we can do is to build
an effective potential which can be decomposed into two pieces
\begin{eqnarray}
  V_{\rm EFT} = V_C + V_F \, .
\end{eqnarray}
Here $V_C$ and $V_F$ refers to the contact- and a finite-range piece,
which represent the long- and short-range physics respectively.
Actual EFTs are arranged in terms of a power counting, which orders
the different contributions to $V_C$ and $V_F$
from more to less relevant.
However the issue of power counting is not particularly relevant to
the current discussion: thus we will implicitly assume that
we are working at leading order (${\rm LO}$),
i.e. we only take into account the most important contributions
to the effective potential.
Notice that the decomposition into long- and short-range physics is not unique
and neither is the choice of power counting.
Thus the meaning and interpretation of the  contact-range potential $V_C$
will depend on these choices.

Here we will compare the OBE model
with the EFTs of Refs.~\cite{Liu:2019tjn,Valderrama:2019chc},
which differ on their choices for the finite-range part of
the effective potential.
In Ref.~\cite{Liu:2019tjn} pion exchanges are considered to be perturbative and
thus subleading, which means that the LO potential reads
\begin{eqnarray}
  V_F^{\slashed{\pi}} = 0 \, .
\end{eqnarray}
In Ref.~\cite{Valderrama:2019chc} pion exchanges are considered to be
non-perturbative and included at LO
\begin{eqnarray}
  V_F^{{\pi}} = V_{\pi} \, ,
\end{eqnarray}
where $V^{\pi}$ coincides with the OPE potential calculated here.
For these two EFTs~\cite{Liu:2019tjn,Valderrama:2019chc}
the LO contact-range potential takes the form
\begin{eqnarray}
  V_C = C_a + C_b\,{\vec{\sigma}_{L1}} \cdot \vec{S}_{L2} \, ,
\end{eqnarray}
where $C_a$ and $C_b$ are couplings that have to be determined
from experimental information, in particular
from the masses of the pentaquarks.
It happens that there are three pentaquarks to which to fit $C_a$ and $C_b$.
Besides, the coupling $C_b$ represents a spin-dependent interaction and its
determination depends on which is the spin of
the $P_c(4440)$ and $P_c(4457)$ pentaquarks.
Thus these EFTs consider two possible scenarios ($A$ and $B$)
for which the spin of these two pentaquarks is
\begin{enumerate}
\item[(A)] the $P_c(4440)$ ($P_c(4457)$) is $J=1/2$ ($J=3/2$),
\item[(B)] the $P_c(4457)$ ($P_c(4440)$) is $J=1/2$ ($J=3/2$).
\end{enumerate}
Scenario $A$ corresponds to the standard quark model expectation that hadrons
with higher spin should have higher mass, while scenario $B$ describes
the opposite situation.
These EFTs cannot discriminate {\it a priori} between these two possibilities,
but only {\it a posteriori} if the predictions of each scenario
are different enough.

The OBE model operates differently: the OBE potential is assumed to be
a complete description of the hadron-hadron interaction.
Thus we do not include a contact-range potential.
Indeed we could interpret the OBE model as a EFT for which at LO
\begin{eqnarray}
  V_C = 0 \quad \mbox{and} \quad V_F = V_{\rm OBE(F)} \, .
\end{eqnarray}
Yet this interpretation is admittedly liberal and we have included it
for illustrative purposes only: it would be difficult to construct
an EFT when the mass of the vector mesons is so close to
the expected breakdown scale.
Besides, the fact that predictions depend on the cutoff shows
that it is indeed not an EFT.
Be it as it may, from an operational standpoint we have a free parameter
in the OBE model, the form-factor cutoff $\Lambda$, 
which we have determined from the $P_c(4312)$.
From this cutoff we can predict the masses and spins of the other
two pentaquarks, while within EFT we could not predict
their spins.
We present a summary of the EFT and OBE model predictions
in Table \ref{tab:binding-hidden-Pc-OBE-vs-EFT},
where it turns out that the predictions of the OBE model coincide with those of
scenario B in Refs.~\cite{Liu:2019tjn,Valderrama:2019chc}.

\mpv{
Further support towards scenario B from the OBE model can be found
from analyzing the values of the $C_a$ and $C_b$ couplings.
Though these are running couplings, they still might carry information
about the short-range physics not explicitly included
within the EFT description, i.e. about $\sigma$,
$\rho$ and $\omega$ exchange.
The mechanism for this is the saturation hypothesis, i.e. that the value of
the EFT couplings are saturated by light meson exchange.
This idea works in pion-nucleon~\cite{Ecker:1988te} and
nucleon-nucleon~\cite{Epelbaum:2001fm} scattering,
though its application in a purely non-perturbative context
such as nucleon-nucleon scattering (or hadronic molecules)
is considerably less clean and depends on the EFT cutoff
being of the same order of magnitude as the exchanged mesons
from which saturation comes.

Following the formalism of Ref.~\cite{Peng:2020xrf}, we expect the EFT couplings
in Refs.~\cite{Liu:2019tjn,Valderrama:2019chc} to be saturated
primarily by scalar and meson vector exchange
\begin{eqnarray}
  C_a^{\rm sat}(\Lambda \sim m_{\sigma}, m_V) &\propto& C_a^{S} + C_a^V \, , \\
  C_b^{\rm sat}(\Lambda \sim m_{\sigma}, m_V) &\propto& C_b^V \, , 
\end{eqnarray}
where the superscript $^{\rm sat}$ indicates that we are considering the part of
the couplings given by saturation and with $\Lambda \sim m_{\sigma}, m_V$
indicating that saturation is expected to work only at EFT cutoffs
close to the mass of the exchanged mesons, in this case $\sigma$,
$\rho$ and $\omega$, suggesting $\Lambda \sim 0.6-0.8\,{\rm GeV}$.
The value of the saturated couplings is expected to be proportional to the
potential contribution $V_M(\vec{q})$ of the light mesons $M$
evaluated at $|\vec{q}| = 0$ once we have removed
the spurious Dirac-delta contributions~\cite{Peng:2020xrf}.
This gives us
\begin{eqnarray}
  C_a^{{\rm sat} (\sigma)}(\Lambda \sim m_{\sigma})
  &\propto& - \frac{g_{\sigma 1} g_{\sigma 2}}{m_{\sigma}^2} \, , \\
  C_a^{{\rm sat } (V)}(\Lambda \sim m_{\rho})
  &\propto& + \frac{g_{V 1} g_{V 2}}{m_{V}^2}\,(1 + \vec{\tau}_1 \cdot \vec{T}_2)
  \, , \\
  C_b^{{\rm sat} (V)}(\Lambda \sim m_{\rho})
  &\propto& + \frac{f_{V 1} f_{V 2}}{6 M^2}
  (1 + \vec{\tau}_1 \cdot \vec{T}_2)
  \, , 
\end{eqnarray}
where $V = \rho, \omega$ and we have made the simplification
that $m_\rho = m_\omega = m_V$, and we do not include the pion
as it is not expected to contribute
for saturation at $\Lambda \gg m_{\pi}$.
The proportionality constant is unknown and will depend on the details of
the renormalization process.
However, assuming this proportionality constant is the same for $C_a^{\rm sat}$
and $C_b^{\rm sat}$, we can compute their ratio
\begin{eqnarray}
  \frac{C_b^{{\rm sat} (V)}}{C_a^{{\rm sat} (V+\sigma)}} \simeq 0.123 \, .
\end{eqnarray}
We will compare this reference value with the ratio we obtain in EFT at a
momentum space cutoff $\Lambda = 750\,{\rm MeV}$, which is close to the
vector meson masses.
For the EFT of Ref.~\cite{Liu:2019tjn} in which pions are subleading,
we obtain the ratios
\begin{eqnarray}
  \frac{C_b^{\rm EFT(\slashed \pi)}}{C_a^{\rm EFT(\slashed \pi)}}
       {\Big|}^{A}_{\Lambda = 750\,{\rm MeV}} &=& -0.176 \, ,
  \\
  \frac{C_b^{\rm EFT(\slashed \pi)}}{C_a^{\rm EFT(\slashed \pi)}}
       {\Big|}^{B}_{\Lambda = 750\,{\rm MeV}} &=& +0.158 \, ,
\end{eqnarray}
which for scenario B is qualitatively compatible with the OBE estimate.
For the EFT of Ref.~\cite{Valderrama:2019chc}, in which pions are included
at leading order, the comparison is not direct.
The reason is that the momentum space version of the one pion exchange
potential used in Ref.~\cite{Valderrama:2019chc} contains
the spurious Dirac-delta contributions that we have
removed in the OBE model here.
Of course within the EFT framework these spurious contributions are not a
problem because they can be reshuffled into the contact-range couplings.
However, if we want to make a comparison with the saturated couplings
then we have to explicitly remove this spurious pion contribution
given by Eq.~(\ref{eq:pion-delta-contrib}) and which evaluation
yields $C_b^{\delta \pi} = 0.375\,{\rm fm}^2$.
Now, if we remove $C_b^{\delta \pi}$, we will obtain the ratios
\begin{eqnarray}
  \frac{C_b^{\rm EFT(\pi)} - C_b^{\delta \pi}}{C_a^{\rm EFT(\pi)}}
       {\Big|}^{A}_{\Lambda = 750\,{\rm MeV}} &=& -0.229\, ,  \label{eq:rat-A}
       \\ 
       \frac{C_b^{\rm EFT(\pi)} - C_b^{\delta \pi}}{C_a^{\rm EFT(\pi)}}
            {\Big|}^{B}_{\Lambda = 750\,{\rm MeV}} &=& +0.136 \, , \label{eq:rat-B}
\end{eqnarray}
where scenario B is again closer to the ratio expected within the OBE model
(in fact it basically reproduces the expected ratio).

It is interesting to notice that the EFT couplings can also be used to check
the necessity of including certain light mesons within the OBE model.
For instance, if we remove the $\sigma$ meson, the $C_a^{\rm sat}$ coupling will
not be saturated by this meson, and we will end up with the ratio
\begin{eqnarray}
  \frac{C_b^{{\rm sat} (V)}}{C_a^{{\rm sat} (V)}} \simeq 0.426 \, ,
\end{eqnarray}
which is not compatible with scenario A or B
in the two EFTs of Refs.~\cite{Liu:2019tjn,Valderrama:2019chc}
by a factor of three, give or take.
This favors the inclusion of the $\sigma$ in our OBE model.
\mpvrev{
Coversely, if we remove the $\omega$ or the $\rho$
the ratios will change to
\begin{eqnarray}
  \frac{C_b^{{\rm sat} (\rho)}}{C_a^{{\rm sat} (\rho+\sigma)}} \simeq 0.193
  \quad , \quad
  \frac{C_b^{{\rm sat} (\omega)}}{C_a^{{\rm sat} (\omega+\sigma)}} \simeq -0.282
\end{eqnarray}
which, surprisingly, are somewhat similar to the EFT ratios for scenarios A
and B, respectively, see Eqs.~(\ref{eq:rat-A}) and (\ref{eq:rat-B}).
Removing the $\omega$ ($\rho$) leads to a version of
scenario B (A) with larger hyperfine splittings than
in Refs.~\cite{Liu:2019tjn,Valderrama:2019chc}.
However from flavor symmetry we expect the $\rho$ and $\omega$ to have similar
couplings and thus we do not consider removing one without the other.
Removing the vector mesons completely would lead to $C_b / C_a \simeq 0$,
which results in a spectrum were the isospin splitting comes from OPE only:
it would be similar to scenario B but with smaller hyperfine splittings.
}
If we now turn our attention to the $\eta$ meson, whose exchange potential
is computed in Appendix ~\ref{app:eta}, this meson will be able
to saturate the $C_b$ coupling:
\begin{eqnarray}
    C_a^{{\rm sat} (\eta)}(\Lambda \sim m_{\eta})
    &\propto& 0 \, \\
    C_b^{{\rm sat} (\eta)}(\Lambda \sim m_{\eta})
    &\propto& \frac{g_1 g_2}{18 f_{\eta}^2} \, 
\end{eqnarray}
with $g_1$ and $g_2$ the axial couplings for the charmed antimeson and baryon
and $f_{\eta} \simeq 150\,{\rm MeV}$ the weak decay constant of
the $\eta$ meson.
In this case we obtain the ratio
\begin{eqnarray}
  \frac{C_b^{{\rm sat} (V+\eta)}}{C_a^{{\rm sat} (V+\sigma)}} \simeq 0.109 \, ,
\end{eqnarray}
which is compatible with the $\eta$-less ratio within the $10\%$ level and
with the EFT ratios in scenario B within the $(25-45)\%$ level.
Owing to the qualitative nature of saturation, it is probably difficult to
argue strongly for or against the inclusion of the $\eta$ in the OBE as
applied to the molecular pentaquarks,
as it effects seems to be small in this case.
This is in contrast with the $\sigma$ meson,
for which the evidence is more conclusive.
\mpvrev{As already mentioned, the previous conclusions about $\eta$ exchange
  are {\it molecule-specific}: while in the pentaquarks it probably
  is a minor effect, in other two-hadron systems (particularly
  if they involve strange hadrons~\cite{Karliner:2016ith})
  it might represent an important contribution to binding.
}
}

\begin{table*}[!ttt]
\caption{
  Comparison of the predicted masses of the pentaquarks in the OBE model
  and the EFT frameworks of Refs.~\cite{Liu:2019tjn,Valderrama:2019chc}.
  Following the notation of Table \ref{tab:binding-hidden-Pc}, the columns
  ``Molecule'', $I$ and $J^P$ refer to the type of two-body system,
  its isospin and total spin and parity.
  For the masses, $M_{\rm OBE}$ is the predicted mass in the OBE model,
  $M_{\rm EFT(\slashed \pi)}$ to the mass within the EFT of Ref.~\cite{Liu:2019tjn}
  (in which pions are subleading) while $M_{\rm EFT (\pi)}$ to the EFT
  of Ref.~\cite{Valderrama:2019chc} (in which pions are leading).
  Both EFTs are unable to determine the quantum numbers of
  the $P_c(4440)$ and $P_c(4457)$ from first principles and
  thus a choice has to be made: in scenario A the $P_c(4440)$
  is the $J^P = \tfrac{1}{2}^-$ $\bar{D}^* \Sigma_c$ molecule,
  while in scenario B it is the $P_c(4457)$ which is
  the $J^P = \tfrac{1}{2}^-$ $\bar{D}^* \Sigma_c$
  molecule.
  The scenarios are indicated by a superscript in the mass
  ($M_{\rm EFT}^A$ and $M_{\rm EFT}^B$).
  The mass ranges are derived from varying the cutoff $\Lambda$
  within the EFT descriptions, where for the EFT of Ref.~\cite{Liu:2019tjn}
  (Ref.~\cite{Valderrama:2019chc}) the cutoff window is
  $\Lambda = 0.5-1.0\,{\rm GeV}$ ($\Lambda = 0.75-1.5\,{\rm GeV}$).
  In general the predictions of the OBE model agree well with scenario B,
  but not with scenario A.
  }
\label{tab:binding-hidden-Pc-OBE-vs-EFT}
\begin{tabular}{cccccccc}
\hline\hline
Molecule  & $I$ & $J^{P}$ & $M_{\rm OBE}$ (MeV) &
$M_{\rm EFT(\slashed \pi)}^A$ (MeV)~\cite{Liu:2019tjn} & $M_{\rm EFT(\slashed \pi)}^B$ (MeV)~\cite{Liu:2019tjn} &
$M_{\rm EFT(\pi)}^A$ (MeV)~\cite{Valderrama:2019chc} & $M_{\rm EFT(\pi)}^B$ (MeV)~\cite{Valderrama:2019chc} \\
\\
\hline
$\bar{D}\Sigma_c$ & $\tfrac{1}{2}$ & $\tfrac{1}{2}^-$ & Input & $4312-4313$ & $4306-4308$  & $4314-4320$ & $4313-4320$ \\
\hline
$\bar{D}\Sigma_c^*$ & $\tfrac{1}{2}$ & $\tfrac{3}{2}^-$ & $4376.0$ & $4371-4372$   & $4376-4377$ & $4378-4383$ & $4373-4385$  \\
\hline
$\bar{D}^*\Sigma_c$ & $\tfrac{1}{2}$ & $\tfrac{1}{2}^-$ & $4458.0$ & Input & Input  & Input & Input \\
$\bar{D}^*\Sigma_c$ & $\tfrac{1}{2}$ & $\tfrac{3}{2}^-$ & $4443.9$ & Input & Input & Input & Input \\
\hline
$\bar{D}^*\Sigma_c^*$ & $\tfrac{1}{2}$ & $\tfrac{1}{2}^-$ & $4523.8$ & $4500-4501$ & $4523-4524$ & $4483-4500$ & $4513-4523$ \\
$\bar{D}^*\Sigma_c^*$ & $\tfrac{1}{2}$ & $\tfrac{3}{2}^-$ & $4517.5$ & $4511$ & $4517$ & $4507-4512$ & $4511-4516$ \\
$\bar{D}^*\Sigma_c^*$ & $\tfrac{1}{2}$ & $\tfrac{5}{2}^-$ & $4504.3$  & $4523-4524$ & $4500-4501$ & $4520-4523$ & $4497-4501$ \\
  \hline\hline
\end{tabular}
\end{table*}

\section{Summary}
\label{sec:summary}

In this manuscript we have investigated the spectroscopy of
the hidden-charm pentaquarks from the point of view of the OBE model.
In particular we considered the impact of the short-range delta-like
contributions in the OBE potentials.
The removal of these contributions, in combination
with the condition of reproducing the mass of
the $P_c(4312)$ pentaquark as a $\bar{D} \Sigma_c$ bound state,
leads to the following predictions for the $\bar{D}^* \Sigma_c$ molecules:
\begin{eqnarray}
  M(\tfrac{1}{2}^-) = 4458.0^{+3.4}_{-5.3} \,{\rm MeV} \quad \mbox{and} \quad
  M(\tfrac{3}{2}^-) = 4443.9^{+9.2}_{-11.6} \,{\rm MeV} \, ,
  \nonumber \\
\end{eqnarray}
which are close to the experimental masses of the $P_c(4440)$
and $P_c(4457)$ pentaquarks.
This suggests the identification of the $P_c(4440)$ with the $J=\tfrac{3}{2}$
$\bar{D}^* \Sigma_c$ bound state and the $P_c(4457)$
with the $J=\tfrac{1}{2}$ one.
In fact the expectation from OPE alone is that the $J=\tfrac{3}{2}$ molecule
should be more bound than the $J=\tfrac{1}{2}$
one~\cite{Karliner:2015ina}, as a consequence of (the spin-spin component of)
OPE being attractive (repulsive) in the $J=\tfrac{3}{2}$
($\tfrac{1}{2}$) channel.
The combination of OPE with short-range physics,
as in Ref.~\cite{Uchino:2015uha} (which uses the hidden-gauge approach
to which it adds pion-exchange diagrams), leads to the same conclusion. 
The recent work of Ref.~\cite{Yamaguchi:2019seo} also explains the molecular
pentaquark spectrum on the basis of OPE and proposes the same spin-parity
identification as here, but suggest that the reason why
the $J=\tfrac{3}{2}$ molecule is more bound is
the tensor component of the OPE potential
(instead of the spin-spin component,
as in Ref.~\cite{Karliner:2015ina}).
Be it as it may, we warn that theoretical predictions in the OBE model
have significant uncertainties and that these uncertainties
cannot be systematically estimated, as we are dealing
with a model (instead of an effective field theory).
Other aspect to consider are the decays of the pentaquarks:
within the molecular picture the natural expectation is that the pentaquarks
would predominantly decay into $\bar{D} \Lambda_c$ and $\bar{D}^* \Lambda_c$,
with the decay mediated by pion and vector meson exchange.
In principle it could be possible to extend the OBE model to
the $\Sigma_c \to \Lambda_c$ transition to calculate this decay width,
but this is beyond the scope of this work.
\mpv{It is also worth mentioning that the decays of molecular pentaquarks
  into $J / \Psi N$ and $\eta_c N$ are expected to be suppressed,
  which raises the question of whether this is compatible with
  the original detection of the $P_c$'s in the $J / \Psi p$
  channel by the LHCb~\cite{Aaij:2019vzc}.
  However the non-observation of the $P_c$'s by
  the GlueX collaboration~\cite{Ali:2019lzf}
  sets higher limits on the branching ratios of the pentaquarks
  into $J / \Psi p$, which are in the singlet digit percent level and
  thus might be compatible with the molecular picture.
  Yet this remains to be confirmed by concrete calculations of
  both charmed antimeson-baryon and charmonium decays.
}

Besides proposing a possible identification for the quantum numbers of the
three hidden-charm pentaquarks, we predict the existence of
other four molecular pentaquarks with $I=\tfrac{1}{2}$.
This prediction indeed confirms the conclusion of Ref.~\cite{Liu:2019tjn},
which used a contact-range effective field theory to describe the molecular
pentaquarks, and of Ref.~\cite{Xiao:2019aya},
which used the hidden-gauge formalism
(constrained by HQSS) instead.
Among the predicted states there is the $J=\tfrac{5}{2}$
$\bar{D}^* \Sigma_c^*$ molecule, which was conjectured
in Refs.~\cite{Xiao:2013yca,Liu:2018zzu} 
and reproduced in a few recent theoretical
works~\cite{Liu:2019tjn,Xiao:2019aya,Shimizu:2019ptd,Mutuk:2019snd}.
Finally, in the isoquartet sector ($I=\tfrac{3}{2}$)
there might be two or three molecular pentaquarks
that bind.
We note that this will impact the size of the proposed isospin-breaking
decay $\Gamma(P_c \to J/\Psi \Delta^+) / \Gamma(P_c \to J/\Psi p)$
calculated in Ref.~\cite{Guo:2019fdo} (in a similar way,
for instance, as the presence of a bound or virtual state
in the $D\bar{D}$ system will affect the decay of the $X(3872)$
to $D^0 \bar{D}^0 \pi^0$~\cite{Guo:2014hqa}).
Conversely, the experimental measurement of the isospin-breaking decay ratio
proposed in  Ref.~\cite{Guo:2019fdo} might provide important clues
regarding the existence of isoquartet molecular pentaquarks.

\section*{Acknowledgments}
We would like to thank Atsushi Hosaka for useful discussions,
Eulogio Oset for suggesting a few interesting references and
Ruprecht Machleidt for his clarifications and
explanations regarding the OBE model.
M.P.V. thanks the IJCLab of Orsay, where part of this work was done,
for its long-term hospitality
This work is partly supported by the National Natural Science Foundation of
China under Grant Nos.11735003, 11975041, 11961141004, 11961141012, and
the fundamental Research Funds for the Central Universities,
the Youth Innovation Promotion Association CAS (No. 2016367)
and the Thousand Talents Plan for Young Professionals.

\appendix

\section{Lagrangian for vector meson exchange for heavy hadrons and the choice
  of a velocity parameter}
\label{app:zeroth}

The Lagrangians describing the interaction of the vector meson
with the charmed mesons and baryons contains electric-,
magnetic- and quadrupole-like components, i.e.
it is a multipole expansion.
It is worth noticing that the electric- and quadrupole-like terms depend only
on the zero-th component of the vector meson field,
which is expected to be suppressed for non-relativistic processes.
If we take the electric-type term as an example,
the Lagrangian can be generically written as
\begin{eqnarray}
  \mathcal{L}_{hhV} = g_V \, h^{\dagger} V^0 h \, , \label{eq:hhV}
\end{eqnarray}
with $g_V$ a coupling, $h$ a non-relativistic field representing
the heavy hadron (e.g. $h = H_Q, S_Q$ for superfield notation
ot $h_Q = q_L$, $d_L$ for subfield notation) and $V^0$
the zero-th component of the vector meson field.
This prompts the question of how it is possible to derive a potential
from a Lagrangian that generates vanishing amplitudes 
in the heavy quark limit (i.e. when the heavy hadron
masses go to infinity).
Indeed from the previous Lagrangian we can derive the non-relativistic amplitude
\begin{eqnarray}
  \mathcal{A}(h \to h \, V^{\mu} ) = g_V \, \epsilon^{\mu} \, \delta_{\mu 0} \, ,
\end{eqnarray}
where $\epsilon^{\mu}$ refers to the polarization of the vector meson and
$\delta_{\mu \nu}$ is the Kronecker delta.
If we evaluate this amplitude for a physical vector meson, for which
the polarization vector obeys the relation $q^\mu \epsilon_{\mu} = 0$
with $q^{\mu}$ the four-momentum of the vector meson,
and assume that the initial and final hadrons are on-shell,
then we have $q^{\mu} = p'^{\mu} - p^{\mu}$,
i.e. the difference of the final and initial heavy hadron momenta.
It is apparent that in the heavy quark limit $q^{0}$ scales as $1/m_Q$
with $m_Q$ the mass of the heavy quark inside the heavy hadron.
For $m_Q \to \infty$ the relation
$q^\mu \epsilon_{\mu} = 0$ simplifies to
$\vec{q} \cdot \vec{\epsilon} = 0$, which implies that $\epsilon^0 = 0$
for a physical vector meson in this limit, thus resulting
in the aforementioned vanishing amplitude.

There are however two factors to consider here:
(i) the potential is a non-observable quantity which actually results from
the exchange of a virtual vector meson instead of a physical one (i.e. we
end up with a non-vanishing potential in the $m_Q \to \infty$ limit
because $q^\mu \epsilon_{\mu} = 0$ only applies to physical states),
(ii) the fact that the Lagrangian vanishes for on-shell heavy hadrons
in this limit does not mean that the $g_V$ coupling is detached
from physical processes, with this detachment being
an artifact of the notation instead.
To specifically address this second point, we notice that
the full Lagrangians for a relativistic hadron field
and a vector meson are
\begin{eqnarray}
  \mathcal{L}_{MMV} &=& g_V \, \left( i\, {M}^{\dagger} \overleftrightarrow{\partial_\mu}  M \right) \, V^{\mu} \, , \\
  \mathcal{L}_{BBV} &=& g_V \, \bar{B} \, \gamma_\mu V^{\mu} B \, , 
\end{eqnarray}
where $M$ and $B$ are meson and baryon fields, respectively,
$V^{\mu}$ the vector meson field and $\gamma_\mu$ the Dirac matrices.
As we are dealing with heavy hadrons, we consider them to move
with constant speed $v = (v^0, \vec{v})$, with $v^{\mu} v_{\mu} = 1$,
prompting the (schematic) field redefinitions
\begin{eqnarray}
  M_v(x) &\propto& e^{i m_Q\, v \cdot x} M(x) \, , \\
  B_v(x) &\propto& e^{i m_Q\, v \cdot x} B(x) \, , 
\end{eqnarray}
with $m_Q$ the mass of the heavy quark inside the heavy hadron,
which in the heavy quark limit can be basically identified
with the heavy hadron mass, and $v$ the velocity parameter.
After a few manipulations we end up with the Lagrangians
\begin{eqnarray}
  \mathcal{L}_{MMV} &=& g_V \, {M}_v^\dagger v \cdot V \, M_v \, , \\
  \mathcal{L}_{BBV} &=& g_V \, {B}_v^\dagger v \cdot V \, B_v \, , 
\end{eqnarray}
which are formally identical, leading to the more simple notation:
\begin{eqnarray}
  \mathcal{L}_{hhV} = g_V \, h_v^{\dagger} v \cdot V \, h_v \, , 
\end{eqnarray}
where $h_v$ stands for a heavy hadron field with velocity $v$, for which
we obtain the non-relativistic amplitude
\begin{eqnarray}
  \mathcal{A}(h_v \to h_v \, V^{\mu} ) = g_V \, \epsilon \cdot v \, .
\end{eqnarray}
If we choose $v = (1, \vec{0})$, we end up with Eq.~(\ref{eq:hhV}).

Finally, for deriving the potential for the exchange of a vector meson
we first define 
\begin{eqnarray}
  \mathcal{A} = \epsilon^\mu \mathcal{A}_{\mu} \,
  \rightarrow \mathcal{A}_{\mu} = g_V v_{\mu}
  \, , \label{eq:hhV-v}
\end{eqnarray}
With this new amplitude plus the convenient normalization we have used in
Eqs.~(\ref{eq:hhV}) and (\ref{eq:hhV-v}), we can write the potential as
\begin{eqnarray}
  V(\vec{q}) &=&
  \frac{\mathcal{A}^\mu(\vec{q}) \mathcal{A}_\mu(-\vec{q})}{\vec{q}^2 + m_V^2}
  \nonumber \\
  &=& \frac{g_V^2}{\vec{q}^2 + m_V^2} \, ,
\end{eqnarray}
where $\vec{q}$ is the exchanged momentum, $m_V$ the mass of the vector meson
and with the second line being a consequence of $v^2 = 1$, meaning that
the potential is independent of the velocity parameter.
The addition of the isospin degrees of freedom in the case of
the $\rho$ is trivial.
Further details about the mechanics of the calculation of the potential
for heavy hadrons can be consulted in Ref.~\cite{Valderrama:2012jv}.

\section{Lagrangians for the Magnetic and Quadrupole Moments}
\label{app:moments}

In this appendix we discuss the magnetic and quadrupole couplings of
a heavy hadron to the electromagnetic field.
This is useful for the derivation of the magnetic- and quadrupole-like
couplings to the vector mesons in the vector-meson dominance model.
In particular we write
\begin{eqnarray}
  \mathcal{L}_{\mu} &=& 
  \mu(h)\,h^{\dagger}\left[ \frac{1}{|S_3|}\,\epsilon_{ijk}\,S_i \partial_j A_k
    \right] h \, , \\
  \mathcal{L}_{Q} &=& Q(h)\,h^{\dagger}\left[ \frac{1}{|Q_{33}|}\,
    Q_{ij}\,\partial_i \partial_j A_0
    \right] h \, ,
\end{eqnarray}
for the magnetic-dipole and electric-quadrupole coupling of a heavy hadron
field $h$ to the photon field $A_{\mu} = ( A_0, \vec{A} )$.
In the magnetic term,
$\mu(h)$ is the magnetic-dipole moment of the heavy hadron $h$ and
$\vec{S}$ represents the spin operator of this heavy hadron,
which we assume to be spin-$S$ (with $S \geq \tfrac{1}{2}$
if we want the magnetic moment to be non-vanishing).
In the quadrupole term, $Q(h)$ is the electric-quadrupole moment of the heavy
hadron and $Q_{ij}$ is a spin-2 tensor that can be constructed
from the spin operator $\vec{S}$:
\begin{eqnarray}
  Q_{ij} = \frac{1}{2} \left[ S_i S_j + S_j S_i \right] -
  \frac{1}{3} S(S+1)\,\delta_{ij} \, ,
\end{eqnarray}
which requires $S \geq 1$ to be non-vanishing.
\mpvrev{
Finally $|S_3| = S$ and $|Q_{33}| = \tfrac{1}{3} S (2 S-1)$ refer to the
evaluation of the $S_3$ and $Q_{33}$ operators for a state
with maximum third component of the spin.
}
These definitions ensure that
\begin{eqnarray}
  \langle S S | \,\hat{\mu}_3 | S S \rangle &=& \mu(h) \, , \\
  \langle S S | \,\hat{Q}_{33} | S S \rangle &=& Q(h) \, , 
\end{eqnarray}
where $| S S \rangle$ represents a spin state of the heavy hadron $h$
where the third component is $S_3 = + S$, while $\hat{\mu}_3$ and
$\hat{Q}_{33}$ are the $i=3$ and $ij = 33$ components
of the magnetic and tensor operators,
which can be identified with
\begin{eqnarray}
  \hat{\mu}_i &=& \mu(h)\,\frac{1}{|S_3|}\,S_i \, , \\
  \hat{Q}_{ij} &=& Q(h)\,\frac{1}{|Q_{33}|}\,Q_{ij} \, .
\end{eqnarray}
Conversely, the moments of order $n$ can be defined analogously as
\begin{eqnarray}
  M^{(n)}_{i_1 \dots i_n} = M^{(n)}(h)\,\frac{1}{|T^{(n)}_{3 \dots 3}|}
  T^{(n)}_{i_1 \dots i_n} \, ,
\end{eqnarray}
with $M^{(n)}(h)$ the $n$-polar moment of hadron $h$, $T^{(n)}$ a spin
$n$-th order tensor constructed from the hadron spin operator $\vec{S}$ and
${|T^{(n)}_{3 \dots 3}|}$ the evaluation of its $i_2 = i_2 = \dots = i_3 = 3$
component for the $| S S \rangle$ spin state.

For the heavy-baryon sextet,
assuming that the multipole moments are dominated
by the light-quarks, only the magnetic-dipole and electric-quadrupole
moments will be relevant; as discussed, the quadrupole moment
is expected to be small (it requires either HQSS breaking or
a sizable D-wave component for the light quark pair).

\section{Lagrangian for the $\eta$ meson}
\label{app:eta}

In this appendix we discuss the inclusion of the $\eta$ meson in the OBE model.
Even though it is well known within the standard OBE model as applied
to nucleon-nucleon interactions that the contribution of
the $\eta$ meson is not particularly important,
it is nonetheless interesting to include it explicitly
not only to test this hypothesis but also to assess
the systematic uncertainties of the OBE model.

To include the $\eta$ we first notice that the pion, kaon and $\eta$ mesons
are pseudo Nambu-Goldstone bosons associated with the breakdown of
chiral symmetry, which can be grouped together into the field
\begin{eqnarray}
  M = 
  \begin{pmatrix}
    \frac{\pi^0}{\sqrt{2}} + \frac{\eta}{\sqrt{6}} & \pi^{+} & K^{+} \\
    \pi^{-} & - \frac{\pi^0}{\sqrt{2}} + \frac{\eta}{\sqrt{6}} &  K^{0} \\
    K^{-} & \bar{K}^0 & - \sqrt{\frac{2}{3}}\,\eta
  \end{pmatrix} \, . \label{eq:NG-mesons}
\end{eqnarray}
Analogously the charmed meson and baryon fields can also be arranged
into SU(3)-flavor matrices
\begin{eqnarray}
  H &=&
  \begin{pmatrix}
    \bar{D}^0 \\
    D^- \\
    \bar{D}_s
  \end{pmatrix} \, , \\
  S &=&
  \begin{pmatrix}
    \Sigma_c^{++} & \frac{1}{\sqrt{2}}\,\Sigma_c^{+} & 
    \frac{1}{\sqrt{2}}\,\Xi_c^{+'} \\
    \frac{1}{\sqrt{2}}\,\Sigma_c^{+} & \Sigma_c^{0} & \frac{1}{\sqrt{2}}\,\Xi_c^{0'} \\
    \frac{1}{\sqrt{2}}\,\Xi_c^{+'}  & \frac{1}{\sqrt{2}}\,\Xi_c^{0'} & \Omega_c^0
\end{pmatrix} \, .
\end{eqnarray}
The flavor structure of the interaction between the pseudo Nambu-Goldstone and
charmed meson and baryon fields will take the form
\begin{eqnarray}
  H_a M_{ab} H_b \quad \mbox{and} \quad S_{ab} M_{bc} S_{ca} \, ,
\end{eqnarray}
from which we can deduce that the Lagrangians will take the form
\begin{eqnarray}
  \mathcal{L}_{q_L q_L \eta} &=& \frac{1}{\sqrt{3}}\,\frac{g_1}{\sqrt{2} f_{\eta}}\,
        {q}_{L}^{\dagger}
        \vec{\sigma}_{L} \cdot \vec{\nabla} \eta \,
        q_{L} \, , \\
        \mathcal{L}_{d_L d_L \, \eta} &=& \frac{1}{\sqrt{3}}\,
        \frac{g_2}{\sqrt{2} f_{\eta}}\,
        {d}_{L}^{\dagger}
        \vec{S}_{L} \cdot \vec{\nabla} \eta \,
        d_{L} \, , 
\end{eqnarray}
where $g_1$ and $g_2$ are the axial coupling to the charmed mesons and baryons,
respectively, which are expected to be identical to those for the pion,
while for the weak decay constant we use $f_{\eta} \simeq 150\,{\rm MeV}$
instead of $f_{\pi}$.
It is interesting to notice that the previous Lagrangians can also be obtained
from the substitution rules
\begin{eqnarray}
  \tau^a \pi^a \,\, \mbox{and} \,\, T^a \pi^a  \quad \rightarrow
  \quad \frac{\eta}{\sqrt{3}}
  \, ,
\end{eqnarray}
which can be traced back to the form of the pseudo Nambu-Goldstone meson
octet matrix, Eq.~(\ref{eq:NG-mesons}).
From the previous Lagrangians we deduce the momentum space potential
\begin{eqnarray}
  V_{\eta}(\vec{q}) &=& -\frac{g_1 g_2}{2 f_{\eta}^2}\,\frac{1}{3}\,
  \frac{\vec{\sigma}_{L1} \cdot \vec{q} \, \vec{S}_{L2} \cdot \vec{q}}
       {{\vec{q}\,}^2 + m_{\eta}^2}
\end{eqnarray}
or, if we Fourier-transform into coordinate space
\begin{eqnarray}
  V_{\eta}(\vec{r}) &=&
  +\frac{1}{3}\,\frac{g_1 g_2}{6 f_{\eta}^2}\,\Big[
    - \vec{\sigma}_{L1} \cdot \vec{S}_{L2}\,\delta(\vec{r})
    \nonumber \\ && \quad
    + \, \vec{\sigma}_{L1} \cdot \vec{S}_{L2}\,m_{\eta}^3\,W_Y(\mu_{\eta} r)
    \nonumber \\ && \quad
    + \, S_{L12}(\vec{r})\,m_{\eta}^3\,W_T(m_{\eta} r) \Big] \, ,
\end{eqnarray}
which is to be regularized with a suitable form-factor and cutoff.
To include it into the full OBE potential, we notice
\begin{eqnarray}
  V_{\rm OBE} = \zeta\,V_{\pi} + V_{\eta} + V_{\sigma} + V_{\rho} +
  \zeta\,V_{\omega} \, ,
\end{eqnarray}
with $\zeta = \pm 1$  the sign for distinguishing
the $q_L d_L$ and $\bar{q}_L d_L$ cases, and where we notice that
the G-parity of the $\eta$-meson is $G=+1$, and thus this piece of
the potential does not depend on whether we have particles or antiparticles.


\begin{thebibliography}{88}%
\makeatletter
\providecommand \@ifxundefined [1]{%
 \@ifx{#1\undefined}
}%
\providecommand \@ifnum [1]{%
 \ifnum #1\expandafter \@firstoftwo
 \else \expandafter \@secondoftwo
 \fi
}%
\providecommand \@ifx [1]{%
 \ifx #1\expandafter \@firstoftwo
 \else \expandafter \@secondoftwo
 \fi
}%
\providecommand \natexlab [1]{#1}%
\providecommand \enquote  [1]{``#1''}%
\providecommand \bibnamefont  [1]{#1}%
\providecommand \bibfnamefont [1]{#1}%
\providecommand \citenamefont [1]{#1}%
\providecommand \href@noop [0]{\@secondoftwo}%
\providecommand \href [0]{\begingroup \@sanitize@url \@href}%
\providecommand \@href[1]{\@@startlink{#1}\@@href}%
\providecommand \@@href[1]{\endgroup#1\@@endlink}%
\providecommand \@sanitize@url [0]{\catcode `\\12\catcode `\$12\catcode
  `\&12\catcode `\#12\catcode `\^12\catcode `\_12\catcode `\%12\relax}%
\providecommand \@@startlink[1]{}%
\providecommand \@@endlink[0]{}%
\providecommand \url  [0]{\begingroup\@sanitize@url \@url }%
\providecommand \@url [1]{\endgroup\@href {#1}{\urlprefix }}%
\providecommand \urlprefix  [0]{URL }%
\providecommand \Eprint [0]{\href }%
\providecommand \doibase [0]{http://dx.doi.org/}%
\providecommand \selectlanguage [0]{\@gobble}%
\providecommand \bibinfo  [0]{\@secondoftwo}%
\providecommand \bibfield  [0]{\@secondoftwo}%
\providecommand \translation [1]{[#1]}%
\providecommand \BibitemOpen [0]{}%
\providecommand \bibitemStop [0]{}%
\providecommand \bibitemNoStop [0]{.\EOS\space}%
\providecommand \EOS [0]{\spacefactor3000\relax}%
\providecommand \BibitemShut  [1]{\csname bibitem#1\endcsname}%
\let\auto@bib@innerbib\@empty
\bibitem [{\citenamefont {Aaij}\ \emph {et~al.}(2019)\citenamefont {Aaij} \emph
  {et~al.}}]{Aaij:2019vzc}%
  \BibitemOpen
  \bibfield  {author} {\bibinfo {author} {\bibfnamefont {R.}~\bibnamefont
  {Aaij}} \emph {et~al.} (\bibinfo {collaboration} {LHCb}),\ }\href {\doibase
  10.1103/PhysRevLett.122.222001} {\bibfield  {journal} {\bibinfo  {journal}
  {Phys. Rev. Lett.}\ }\textbf {\bibinfo {volume} {122}},\ \bibinfo {pages}
  {222001} (\bibinfo {year} {2019})},\ \Eprint
  {http://arxiv.org/abs/1904.03947} {arXiv:1904.03947 [hep-ex]} \BibitemShut
  {NoStop}%
\bibitem [{\citenamefont {Voloshin}\ and\ \citenamefont
  {Okun}(1976)}]{Voloshin:1976ap}%
  \BibitemOpen
  \bibfield  {author} {\bibinfo {author} {\bibfnamefont {M.}~\bibnamefont
  {Voloshin}}\ and\ \bibinfo {author} {\bibfnamefont {L.}~\bibnamefont
  {Okun}},\ }\href@noop {} {\bibfield  {journal} {\bibinfo  {journal} {JETP
  Lett.}\ }\textbf {\bibinfo {volume} {23}},\ \bibinfo {pages} {333} (\bibinfo
  {year} {1976})}\BibitemShut {NoStop}%
\bibitem [{\citenamefont {De~Rujula}\ \emph {et~al.}(1977)\citenamefont
  {De~Rujula}, \citenamefont {Georgi},\ and\ \citenamefont
  {Glashow}}]{DeRujula:1976qd}%
  \BibitemOpen
  \bibfield  {author} {\bibinfo {author} {\bibfnamefont {A.}~\bibnamefont
  {De~Rujula}}, \bibinfo {author} {\bibfnamefont {H.}~\bibnamefont {Georgi}}, \
  and\ \bibinfo {author} {\bibfnamefont {S.}~\bibnamefont {Glashow}},\ }\href
  {\doibase 10.1103/PhysRevLett.38.317} {\bibfield  {journal} {\bibinfo
  {journal} {Phys.Rev.Lett.}\ }\textbf {\bibinfo {volume} {38}},\ \bibinfo
  {pages} {317} (\bibinfo {year} {1977})}\BibitemShut {NoStop}%
\bibitem [{\citenamefont {Wu}\ \emph {et~al.}(2010)\citenamefont {Wu},
  \citenamefont {Molina}, \citenamefont {Oset},\ and\ \citenamefont
  {Zou}}]{Wu:2010jy}%
  \BibitemOpen
  \bibfield  {author} {\bibinfo {author} {\bibfnamefont {J.-J.}\ \bibnamefont
  {Wu}}, \bibinfo {author} {\bibfnamefont {R.}~\bibnamefont {Molina}}, \bibinfo
  {author} {\bibfnamefont {E.}~\bibnamefont {Oset}}, \ and\ \bibinfo {author}
  {\bibfnamefont {B.~S.}\ \bibnamefont {Zou}},\ }\href {\doibase
  10.1103/PhysRevLett.105.232001} {\bibfield  {journal} {\bibinfo  {journal}
  {Phys. Rev. Lett.}\ }\textbf {\bibinfo {volume} {105}},\ \bibinfo {pages}
  {232001} (\bibinfo {year} {2010})},\ \Eprint {http://arxiv.org/abs/1007.0573}
  {arXiv:1007.0573 [nucl-th]} \BibitemShut {NoStop}%
\bibitem [{\citenamefont {Wu}\ \emph {et~al.}(2011)\citenamefont {Wu},
  \citenamefont {Molina}, \citenamefont {Oset},\ and\ \citenamefont
  {Zou}}]{Wu:2010vk}%
  \BibitemOpen
  \bibfield  {author} {\bibinfo {author} {\bibfnamefont {J.-J.}\ \bibnamefont
  {Wu}}, \bibinfo {author} {\bibfnamefont {R.}~\bibnamefont {Molina}}, \bibinfo
  {author} {\bibfnamefont {E.}~\bibnamefont {Oset}}, \ and\ \bibinfo {author}
  {\bibfnamefont {B.~S.}\ \bibnamefont {Zou}},\ }\href {\doibase
  10.1103/PhysRevC.84.015202} {\bibfield  {journal} {\bibinfo  {journal} {Phys.
  Rev.}\ }\textbf {\bibinfo {volume} {C84}},\ \bibinfo {pages} {015202}
  (\bibinfo {year} {2011})},\ \Eprint {http://arxiv.org/abs/1011.2399}
  {arXiv:1011.2399 [nucl-th]} \BibitemShut {NoStop}%
\bibitem [{\citenamefont {Xiao}\ \emph {et~al.}(2013)\citenamefont {Xiao},
  \citenamefont {Nieves},\ and\ \citenamefont {Oset}}]{Xiao:2013yca}%
  \BibitemOpen
  \bibfield  {author} {\bibinfo {author} {\bibfnamefont {C.~W.}\ \bibnamefont
  {Xiao}}, \bibinfo {author} {\bibfnamefont {J.}~\bibnamefont {Nieves}}, \ and\
  \bibinfo {author} {\bibfnamefont {E.}~\bibnamefont {Oset}},\ }\href {\doibase
  10.1103/PhysRevD.88.056012} {\bibfield  {journal} {\bibinfo  {journal} {Phys.
  Rev.}\ }\textbf {\bibinfo {volume} {D88}},\ \bibinfo {pages} {056012}
  (\bibinfo {year} {2013})},\ \Eprint {http://arxiv.org/abs/1304.5368}
  {arXiv:1304.5368 [hep-ph]} \BibitemShut {NoStop}%
\bibitem [{\citenamefont {Karliner}\ and\ \citenamefont
  {Rosner}(2015)}]{Karliner:2015ina}%
  \BibitemOpen
  \bibfield  {author} {\bibinfo {author} {\bibfnamefont {M.}~\bibnamefont
  {Karliner}}\ and\ \bibinfo {author} {\bibfnamefont {J.~L.}\ \bibnamefont
  {Rosner}},\ }\href {\doibase 10.1103/PhysRevLett.115.122001} {\bibfield
  {journal} {\bibinfo  {journal} {Phys. Rev. Lett.}\ }\textbf {\bibinfo
  {volume} {115}},\ \bibinfo {pages} {122001} (\bibinfo {year} {2015})},\
  \Eprint {http://arxiv.org/abs/1506.06386} {arXiv:1506.06386 [hep-ph]}
  \BibitemShut {NoStop}%
\bibitem [{\citenamefont {Wang}\ \emph {et~al.}(2011)\citenamefont {Wang},
  \citenamefont {Huang}, \citenamefont {Zhang},\ and\ \citenamefont
  {Zou}}]{Wang:2011rga}%
  \BibitemOpen
  \bibfield  {author} {\bibinfo {author} {\bibfnamefont {W.~L.}\ \bibnamefont
  {Wang}}, \bibinfo {author} {\bibfnamefont {F.}~\bibnamefont {Huang}},
  \bibinfo {author} {\bibfnamefont {Z.~Y.}\ \bibnamefont {Zhang}}, \ and\
  \bibinfo {author} {\bibfnamefont {B.~S.}\ \bibnamefont {Zou}},\ }\href
  {\doibase 10.1103/PhysRevC.84.015203} {\bibfield  {journal} {\bibinfo
  {journal} {Phys. Rev.}\ }\textbf {\bibinfo {volume} {C84}},\ \bibinfo {pages}
  {015203} (\bibinfo {year} {2011})},\ \Eprint {http://arxiv.org/abs/1101.0453}
  {arXiv:1101.0453 [nucl-th]} \BibitemShut {NoStop}%
\bibitem [{\citenamefont {Yang}\ \emph {et~al.}(2012)\citenamefont {Yang},
  \citenamefont {Sun}, \citenamefont {He}, \citenamefont {Liu},\ and\
  \citenamefont {Zhu}}]{Yang:2011wz}%
  \BibitemOpen
  \bibfield  {author} {\bibinfo {author} {\bibfnamefont {Z.-C.}\ \bibnamefont
  {Yang}}, \bibinfo {author} {\bibfnamefont {Z.-F.}\ \bibnamefont {Sun}},
  \bibinfo {author} {\bibfnamefont {J.}~\bibnamefont {He}}, \bibinfo {author}
  {\bibfnamefont {X.}~\bibnamefont {Liu}}, \ and\ \bibinfo {author}
  {\bibfnamefont {S.-L.}\ \bibnamefont {Zhu}},\ }\href {\doibase
  10.1088/1674-1137/36/1/002, 10.1088/1674-1137/36/3/006} {\bibfield  {journal}
  {\bibinfo  {journal} {Chin. Phys.}\ }\textbf {\bibinfo {volume} {C36}},\
  \bibinfo {pages} {6} (\bibinfo {year} {2012})},\ \Eprint
  {http://arxiv.org/abs/1105.2901} {arXiv:1105.2901 [hep-ph]} \BibitemShut
  {NoStop}%
\bibitem [{\citenamefont {Chen}\ \emph
  {et~al.}(2019{\natexlab{a}})\citenamefont {Chen}, \citenamefont {Chen},\ and\
  \citenamefont {Zhu}}]{Chen:2019bip}%
  \BibitemOpen
  \bibfield  {author} {\bibinfo {author} {\bibfnamefont {H.-X.}\ \bibnamefont
  {Chen}}, \bibinfo {author} {\bibfnamefont {W.}~\bibnamefont {Chen}}, \ and\
  \bibinfo {author} {\bibfnamefont {S.-L.}\ \bibnamefont {Zhu}},\ }\href
  {\doibase 10.1103/PhysRevD.100.051501} {\bibfield  {journal} {\bibinfo
  {journal} {Phys. Rev. D}\ }\textbf {\bibinfo {volume} {100}},\ \bibinfo
  {pages} {051501} (\bibinfo {year} {2019}{\natexlab{a}})},\ \Eprint
  {http://arxiv.org/abs/1903.11001} {arXiv:1903.11001 [hep-ph]} \BibitemShut
  {NoStop}%
\bibitem [{\citenamefont {Chen}\ \emph
  {et~al.}(2019{\natexlab{b}})\citenamefont {Chen}, \citenamefont {Sun},
  \citenamefont {Liu},\ and\ \citenamefont {Zhu}}]{Chen:2019asm}%
  \BibitemOpen
  \bibfield  {author} {\bibinfo {author} {\bibfnamefont {R.}~\bibnamefont
  {Chen}}, \bibinfo {author} {\bibfnamefont {Z.-F.}\ \bibnamefont {Sun}},
  \bibinfo {author} {\bibfnamefont {X.}~\bibnamefont {Liu}}, \ and\ \bibinfo
  {author} {\bibfnamefont {S.-L.}\ \bibnamefont {Zhu}},\ }\href {\doibase
  10.1103/PhysRevD.100.011502} {\bibfield  {journal} {\bibinfo  {journal}
  {Phys. Rev. D}\ }\textbf {\bibinfo {volume} {100}},\ \bibinfo {pages}
  {011502} (\bibinfo {year} {2019}{\natexlab{b}})},\ \Eprint
  {http://arxiv.org/abs/1903.11013} {arXiv:1903.11013 [hep-ph]} \BibitemShut
  {NoStop}%
\bibitem [{\citenamefont {He}(2019)}]{He:2019ify}%
  \BibitemOpen
  \bibfield  {author} {\bibinfo {author} {\bibfnamefont {J.}~\bibnamefont
  {He}},\ }\href {\doibase 10.1140/epjc/s10052-019-6906-1} {\bibfield
  {journal} {\bibinfo  {journal} {Eur. Phys. J.}\ }\textbf {\bibinfo {volume}
  {C79}},\ \bibinfo {pages} {393} (\bibinfo {year} {2019})},\ \Eprint
  {http://arxiv.org/abs/1903.11872} {arXiv:1903.11872 [hep-ph]} \BibitemShut
  {NoStop}%
\bibitem [{\citenamefont {Liu}\ \emph {et~al.}(2019{\natexlab{a}})\citenamefont
  {Liu}, \citenamefont {Pan}, \citenamefont {Peng}, \citenamefont
  {Sánchez~Sánchez}, \citenamefont {Geng}, \citenamefont {Hosaka},\ and\
  \citenamefont {Pavon~Valderrama}}]{Liu:2019tjn}%
  \BibitemOpen
  \bibfield  {author} {\bibinfo {author} {\bibfnamefont {M.-Z.}\ \bibnamefont
  {Liu}}, \bibinfo {author} {\bibfnamefont {Y.-W.}\ \bibnamefont {Pan}},
  \bibinfo {author} {\bibfnamefont {F.-Z.}\ \bibnamefont {Peng}}, \bibinfo
  {author} {\bibfnamefont {M.}~\bibnamefont {Sánchez~Sánchez}}, \bibinfo
  {author} {\bibfnamefont {L.-S.}\ \bibnamefont {Geng}}, \bibinfo {author}
  {\bibfnamefont {A.}~\bibnamefont {Hosaka}}, \ and\ \bibinfo {author}
  {\bibfnamefont {M.}~\bibnamefont {Pavon~Valderrama}},\ }\href {\doibase
  10.1103/PhysRevLett.122.242001} {\bibfield  {journal} {\bibinfo  {journal}
  {Phys. Rev. Lett.}\ }\textbf {\bibinfo {volume} {122}},\ \bibinfo {pages}
  {242001} (\bibinfo {year} {2019}{\natexlab{a}})},\ \Eprint
  {http://arxiv.org/abs/1903.11560} {arXiv:1903.11560 [hep-ph]} \BibitemShut
  {NoStop}%
\bibitem [{\citenamefont {Xiao}\ \emph
  {et~al.}(2019{\natexlab{a}})\citenamefont {Xiao}, \citenamefont {Nieves},\
  and\ \citenamefont {Oset}}]{Xiao:2019aya}%
  \BibitemOpen
  \bibfield  {author} {\bibinfo {author} {\bibfnamefont {C.}~\bibnamefont
  {Xiao}}, \bibinfo {author} {\bibfnamefont {J.}~\bibnamefont {Nieves}}, \ and\
  \bibinfo {author} {\bibfnamefont {E.}~\bibnamefont {Oset}},\ }\href {\doibase
  10.1103/PhysRevD.100.014021} {\bibfield  {journal} {\bibinfo  {journal}
  {Phys. Rev. D}\ }\textbf {\bibinfo {volume} {100}},\ \bibinfo {pages}
  {014021} (\bibinfo {year} {2019}{\natexlab{a}})},\ \Eprint
  {http://arxiv.org/abs/1904.01296} {arXiv:1904.01296 [hep-ph]} \BibitemShut
  {NoStop}%
\bibitem [{\citenamefont {Shimizu}\ \emph {et~al.}(2019)\citenamefont
  {Shimizu}, \citenamefont {Yamaguchi},\ and\ \citenamefont
  {Harada}}]{Shimizu:2019ptd}%
  \BibitemOpen
  \bibfield  {author} {\bibinfo {author} {\bibfnamefont {Y.}~\bibnamefont
  {Shimizu}}, \bibinfo {author} {\bibfnamefont {Y.}~\bibnamefont {Yamaguchi}},
  \ and\ \bibinfo {author} {\bibfnamefont {M.}~\bibnamefont {Harada}},\
  }\href@noop {} {\  (\bibinfo {year} {2019})},\ \Eprint
  {http://arxiv.org/abs/1904.00587} {arXiv:1904.00587 [hep-ph]} \BibitemShut
  {NoStop}%
\bibitem [{\citenamefont {Guo}\ and\ \citenamefont
  {Oller}(2019)}]{Guo:2019kdc}%
  \BibitemOpen
  \bibfield  {author} {\bibinfo {author} {\bibfnamefont {Z.-H.}\ \bibnamefont
  {Guo}}\ and\ \bibinfo {author} {\bibfnamefont {J.~A.}\ \bibnamefont
  {Oller}},\ }\href {\doibase 10.1016/j.physletb.2019.04.053} {\bibfield
  {journal} {\bibinfo  {journal} {Phys. Lett.}\ }\textbf {\bibinfo {volume}
  {B793}},\ \bibinfo {pages} {144} (\bibinfo {year} {2019})},\ \Eprint
  {http://arxiv.org/abs/1904.00851} {arXiv:1904.00851 [hep-ph]} \BibitemShut
  {NoStop}%
\bibitem [{\citenamefont {Fern\'andez-Ram\'\i{}rez}\ \emph
  {et~al.}(2019)\citenamefont {Fern\'andez-Ram\'\i{}rez}, \citenamefont
  {Pilloni}, \citenamefont {Albaladejo}, \citenamefont {Jackura}, \citenamefont
  {Mathieu}, \citenamefont {Mikhasenko}, \citenamefont {Silva-Castro},\ and\
  \citenamefont {Szczepaniak}}]{Fernandez-Ramirez:2019koa}%
  \BibitemOpen
  \bibfield  {author} {\bibinfo {author} {\bibfnamefont {C.}~\bibnamefont
  {Fern\'andez-Ram\'\i{}rez}}, \bibinfo {author} {\bibfnamefont
  {A.}~\bibnamefont {Pilloni}}, \bibinfo {author} {\bibfnamefont
  {M.}~\bibnamefont {Albaladejo}}, \bibinfo {author} {\bibfnamefont
  {A.}~\bibnamefont {Jackura}}, \bibinfo {author} {\bibfnamefont
  {V.}~\bibnamefont {Mathieu}}, \bibinfo {author} {\bibfnamefont
  {M.}~\bibnamefont {Mikhasenko}}, \bibinfo {author} {\bibfnamefont
  {J.}~\bibnamefont {Silva-Castro}}, \ and\ \bibinfo {author} {\bibfnamefont
  {A.}~\bibnamefont {Szczepaniak}} (\bibinfo {collaboration} {JPAC}),\ }\href
  {\doibase 10.1103/PhysRevLett.123.092001} {\bibfield  {journal} {\bibinfo
  {journal} {Phys. Rev. Lett.}\ }\textbf {\bibinfo {volume} {123}},\ \bibinfo
  {pages} {092001} (\bibinfo {year} {2019})},\ \Eprint
  {http://arxiv.org/abs/1904.10021} {arXiv:1904.10021 [hep-ph]} \BibitemShut
  {NoStop}%
\bibitem [{\citenamefont {Wu}\ and\ \citenamefont {Chen}(2019)}]{Wu:2019rog}%
  \BibitemOpen
  \bibfield  {author} {\bibinfo {author} {\bibfnamefont {Q.}~\bibnamefont
  {Wu}}\ and\ \bibinfo {author} {\bibfnamefont {D.-Y.}\ \bibnamefont {Chen}},\
  }\href {\doibase 10.1103/PhysRevD.100.114002} {\bibfield  {journal} {\bibinfo
   {journal} {Phys. Rev. D}\ }\textbf {\bibinfo {volume} {100}},\ \bibinfo
  {pages} {114002} (\bibinfo {year} {2019})},\ \Eprint
  {http://arxiv.org/abs/1906.02480} {arXiv:1906.02480 [hep-ph]} \BibitemShut
  {NoStop}%
\bibitem [{\citenamefont {Pavon~Valderrama}(2019)}]{Valderrama:2019chc}%
  \BibitemOpen
  \bibfield  {author} {\bibinfo {author} {\bibfnamefont {M.}~\bibnamefont
  {Pavon~Valderrama}},\ }\href {\doibase 10.1103/PhysRevD.100.094028}
  {\bibfield  {journal} {\bibinfo  {journal} {Phys. Rev.}\ }\textbf {\bibinfo
  {volume} {D100}},\ \bibinfo {pages} {094028} (\bibinfo {year} {2019})},\
  \Eprint {http://arxiv.org/abs/1907.05294} {arXiv:1907.05294 [hep-ph]}
  \BibitemShut {NoStop}%
\bibitem [{\citenamefont {Eides}\ \emph {et~al.}(2020)\citenamefont {Eides},
  \citenamefont {Petrov},\ and\ \citenamefont {Polyakov}}]{Eides:2019tgv}%
  \BibitemOpen
  \bibfield  {author} {\bibinfo {author} {\bibfnamefont {M.~I.}\ \bibnamefont
  {Eides}}, \bibinfo {author} {\bibfnamefont {V.~Y.}\ \bibnamefont {Petrov}}, \
  and\ \bibinfo {author} {\bibfnamefont {M.~V.}\ \bibnamefont {Polyakov}},\
  }\href {\doibase 10.1142/S0217732320501515} {\bibfield  {journal} {\bibinfo
  {journal} {Mod. Phys. Lett. A}\ }\textbf {\bibinfo {volume} {35}},\ \bibinfo
  {pages} {2050151} (\bibinfo {year} {2020})},\ \Eprint
  {http://arxiv.org/abs/1904.11616} {arXiv:1904.11616 [hep-ph]} \BibitemShut
  {NoStop}%
\bibitem [{\citenamefont {Wang}(2020)}]{Wang:2019got}%
  \BibitemOpen
  \bibfield  {author} {\bibinfo {author} {\bibfnamefont {Z.-G.}\ \bibnamefont
  {Wang}},\ }\href {\doibase 10.1142/S0217751X20500037} {\bibfield  {journal}
  {\bibinfo  {journal} {Int. J. Mod. Phys. A}\ }\textbf {\bibinfo {volume}
  {35}},\ \bibinfo {pages} {2050003} (\bibinfo {year} {2020})},\ \Eprint
  {http://arxiv.org/abs/1905.02892} {arXiv:1905.02892 [hep-ph]} \BibitemShut
  {NoStop}%
\bibitem [{\citenamefont {Cheng}\ and\ \citenamefont
  {Liu}(2019)}]{Cheng:2019obk}%
  \BibitemOpen
  \bibfield  {author} {\bibinfo {author} {\bibfnamefont {J.-B.}\ \bibnamefont
  {Cheng}}\ and\ \bibinfo {author} {\bibfnamefont {Y.-R.}\ \bibnamefont
  {Liu}},\ }\href {\doibase 10.1103/PhysRevD.100.054002} {\bibfield  {journal}
  {\bibinfo  {journal} {Phys. Rev.}\ }\textbf {\bibinfo {volume} {D100}},\
  \bibinfo {pages} {054002} (\bibinfo {year} {2019})},\ \Eprint
  {http://arxiv.org/abs/1905.08605} {arXiv:1905.08605 [hep-ph]} \BibitemShut
  {NoStop}%
\bibitem [{\citenamefont {Ferretti}\ and\ \citenamefont
  {Santopinto}(2020)}]{Ferretti:2020ewe}%
  \BibitemOpen
  \bibfield  {author} {\bibinfo {author} {\bibfnamefont {J.}~\bibnamefont
  {Ferretti}}\ and\ \bibinfo {author} {\bibfnamefont {E.}~\bibnamefont
  {Santopinto}},\ }\href {\doibase 10.1007/JHEP04(2020)119} {\bibfield
  {journal} {\bibinfo  {journal} {JHEP}\ }\textbf {\bibinfo {volume} {04}},\
  \bibinfo {pages} {119} (\bibinfo {year} {2020})},\ \Eprint
  {http://arxiv.org/abs/2001.01067} {arXiv:2001.01067 [hep-ph]} \BibitemShut
  {NoStop}%
\bibitem [{\citenamefont {Stancu}(2020)}]{Stancu:2020paw}%
  \BibitemOpen
  \bibfield  {author} {\bibinfo {author} {\bibfnamefont {F.}~\bibnamefont
  {Stancu}},\ }\href {\doibase 10.1103/PhysRevD.101.094007} {\bibfield
  {journal} {\bibinfo  {journal} {Phys. Rev. D}\ }\textbf {\bibinfo {volume}
  {101}},\ \bibinfo {pages} {094007} (\bibinfo {year} {2020})},\ \Eprint
  {http://arxiv.org/abs/2004.06009} {arXiv:2004.06009 [hep-ph]} \BibitemShut
  {NoStop}%
\bibitem [{\citenamefont {Guo}\ \emph {et~al.}(2019)\citenamefont {Guo},
  \citenamefont {Jing}, \citenamefont {Meißner},\ and\ \citenamefont
  {Sakai}}]{Guo:2019fdo}%
  \BibitemOpen
  \bibfield  {author} {\bibinfo {author} {\bibfnamefont {F.-K.}\ \bibnamefont
  {Guo}}, \bibinfo {author} {\bibfnamefont {H.-J.}\ \bibnamefont {Jing}},
  \bibinfo {author} {\bibfnamefont {U.-G.}\ \bibnamefont {Meißner}}, \ and\
  \bibinfo {author} {\bibfnamefont {S.}~\bibnamefont {Sakai}},\ }\href
  {\doibase 10.1103/PhysRevD.99.091501} {\bibfield  {journal} {\bibinfo
  {journal} {Phys. Rev.}\ }\textbf {\bibinfo {volume} {D99}},\ \bibinfo {pages}
  {091501} (\bibinfo {year} {2019})},\ \Eprint
  {http://arxiv.org/abs/1903.11503} {arXiv:1903.11503 [hep-ph]} \BibitemShut
  {NoStop}%
\bibitem [{\citenamefont {Xiao}\ \emph
  {et~al.}(2019{\natexlab{b}})\citenamefont {Xiao}, \citenamefont {Huang},
  \citenamefont {Dong}, \citenamefont {Geng},\ and\ \citenamefont
  {Chen}}]{Xiao:2019mvs}%
  \BibitemOpen
  \bibfield  {author} {\bibinfo {author} {\bibfnamefont {C.-J.}\ \bibnamefont
  {Xiao}}, \bibinfo {author} {\bibfnamefont {Y.}~\bibnamefont {Huang}},
  \bibinfo {author} {\bibfnamefont {Y.-B.}\ \bibnamefont {Dong}}, \bibinfo
  {author} {\bibfnamefont {L.-S.}\ \bibnamefont {Geng}}, \ and\ \bibinfo
  {author} {\bibfnamefont {D.-Y.}\ \bibnamefont {Chen}},\ }\href {\doibase
  10.1103/PhysRevD.100.014022} {\bibfield  {journal} {\bibinfo  {journal}
  {Phys. Rev. D}\ }\textbf {\bibinfo {volume} {100}},\ \bibinfo {pages}
  {014022} (\bibinfo {year} {2019}{\natexlab{b}})},\ \Eprint
  {http://arxiv.org/abs/1904.00872} {arXiv:1904.00872 [hep-ph]} \BibitemShut
  {NoStop}%
\bibitem [{\citenamefont {Voloshin}(2019)}]{Voloshin:2019aut}%
  \BibitemOpen
  \bibfield  {author} {\bibinfo {author} {\bibfnamefont {M.}~\bibnamefont
  {Voloshin}},\ }\href {\doibase 10.1103/PhysRevD.100.034020} {\bibfield
  {journal} {\bibinfo  {journal} {Phys. Rev. D}\ }\textbf {\bibinfo {volume}
  {100}},\ \bibinfo {pages} {034020} (\bibinfo {year} {2019})},\ \Eprint
  {http://arxiv.org/abs/1907.01476} {arXiv:1907.01476 [hep-ph]} \BibitemShut
  {NoStop}%
\bibitem [{\citenamefont {Sakai}\ \emph {et~al.}(2019)\citenamefont {Sakai},
  \citenamefont {Jing},\ and\ \citenamefont {Guo}}]{Sakai:2019qph}%
  \BibitemOpen
  \bibfield  {author} {\bibinfo {author} {\bibfnamefont {S.}~\bibnamefont
  {Sakai}}, \bibinfo {author} {\bibfnamefont {H.-J.}\ \bibnamefont {Jing}}, \
  and\ \bibinfo {author} {\bibfnamefont {F.-K.}\ \bibnamefont {Guo}},\ }\href
  {\doibase 10.1103/PhysRevD.100.074007} {\bibfield  {journal} {\bibinfo
  {journal} {Phys. Rev. D}\ }\textbf {\bibinfo {volume} {100}},\ \bibinfo
  {pages} {074007} (\bibinfo {year} {2019})},\ \Eprint
  {http://arxiv.org/abs/1907.03414} {arXiv:1907.03414 [hep-ph]} \BibitemShut
  {NoStop}%
\bibitem [{\citenamefont {Shen}\ \emph {et~al.}(2019)\citenamefont {Shen},
  \citenamefont {Wu},\ and\ \citenamefont {Zou}}]{Shen:2019evi}%
  \BibitemOpen
  \bibfield  {author} {\bibinfo {author} {\bibfnamefont {C.-W.}\ \bibnamefont
  {Shen}}, \bibinfo {author} {\bibfnamefont {J.-J.}\ \bibnamefont {Wu}}, \ and\
  \bibinfo {author} {\bibfnamefont {B.-S.}\ \bibnamefont {Zou}},\ }\href
  {\doibase 10.1103/PhysRevD.100.056006} {\bibfield  {journal} {\bibinfo
  {journal} {Phys. Rev. D}\ }\textbf {\bibinfo {volume} {100}},\ \bibinfo
  {pages} {056006} (\bibinfo {year} {2019})},\ \Eprint
  {http://arxiv.org/abs/1906.03896} {arXiv:1906.03896 [hep-ph]} \BibitemShut
  {NoStop}%
\bibitem [{\citenamefont {Xiao}\ \emph
  {et~al.}(2019{\natexlab{c}})\citenamefont {Xiao}, \citenamefont {Nieves},\
  and\ \citenamefont {Oset}}]{Xiao:2019gjd}%
  \BibitemOpen
  \bibfield  {author} {\bibinfo {author} {\bibfnamefont {C.}~\bibnamefont
  {Xiao}}, \bibinfo {author} {\bibfnamefont {J.}~\bibnamefont {Nieves}}, \ and\
  \bibinfo {author} {\bibfnamefont {E.}~\bibnamefont {Oset}},\ }\href {\doibase
  10.1016/j.physletb.2019.135051} {\bibfield  {journal} {\bibinfo  {journal}
  {Phys. Lett. B}\ }\textbf {\bibinfo {volume} {799}},\ \bibinfo {pages}
  {135051} (\bibinfo {year} {2019}{\natexlab{c}})},\ \Eprint
  {http://arxiv.org/abs/1906.09010} {arXiv:1906.09010 [hep-ph]} \BibitemShut
  {NoStop}%
\bibitem [{\citenamefont {Aaij}\ \emph {et~al.}(2015)\citenamefont {Aaij} \emph
  {et~al.}}]{Aaij:2015tga}%
  \BibitemOpen
  \bibfield  {author} {\bibinfo {author} {\bibfnamefont {R.}~\bibnamefont
  {Aaij}} \emph {et~al.} (\bibinfo {collaboration} {LHCb}),\ }\href {\doibase
  10.1103/PhysRevLett.115.072001} {\bibfield  {journal} {\bibinfo  {journal}
  {Phys. Rev. Lett.}\ }\textbf {\bibinfo {volume} {115}},\ \bibinfo {pages}
  {072001} (\bibinfo {year} {2015})},\ \Eprint
  {http://arxiv.org/abs/1507.03414} {arXiv:1507.03414 [hep-ex]} \BibitemShut
  {NoStop}%
\bibitem [{\citenamefont {Roca}\ \emph {et~al.}(2015)\citenamefont {Roca},
  \citenamefont {Nieves},\ and\ \citenamefont {Oset}}]{Roca:2015dva}%
  \BibitemOpen
  \bibfield  {author} {\bibinfo {author} {\bibfnamefont {L.}~\bibnamefont
  {Roca}}, \bibinfo {author} {\bibfnamefont {J.}~\bibnamefont {Nieves}}, \ and\
  \bibinfo {author} {\bibfnamefont {E.}~\bibnamefont {Oset}},\ }\href {\doibase
  10.1103/PhysRevD.92.094003} {\bibfield  {journal} {\bibinfo  {journal} {Phys.
  Rev.}\ }\textbf {\bibinfo {volume} {D92}},\ \bibinfo {pages} {094003}
  (\bibinfo {year} {2015})},\ \Eprint {http://arxiv.org/abs/1507.04249}
  {arXiv:1507.04249 [hep-ph]} \BibitemShut {NoStop}%
\bibitem [{\citenamefont {He}(2016)}]{He:2015cea}%
  \BibitemOpen
  \bibfield  {author} {\bibinfo {author} {\bibfnamefont {J.}~\bibnamefont
  {He}},\ }\href {\doibase 10.1016/j.physletb.2015.12.071} {\bibfield
  {journal} {\bibinfo  {journal} {Phys. Lett.}\ }\textbf {\bibinfo {volume}
  {B753}},\ \bibinfo {pages} {547} (\bibinfo {year} {2016})},\ \Eprint
  {http://arxiv.org/abs/1507.05200} {arXiv:1507.05200 [hep-ph]} \BibitemShut
  {NoStop}%
\bibitem [{\citenamefont {Xiao}\ and\ \citenamefont
  {Meißner}(2015)}]{Xiao:2015fia}%
  \BibitemOpen
  \bibfield  {author} {\bibinfo {author} {\bibfnamefont {C.~W.}\ \bibnamefont
  {Xiao}}\ and\ \bibinfo {author} {\bibfnamefont {U.~G.}\ \bibnamefont
  {Meißner}},\ }\href {\doibase 10.1103/PhysRevD.92.114002} {\bibfield
  {journal} {\bibinfo  {journal} {Phys. Rev.}\ }\textbf {\bibinfo {volume}
  {D92}},\ \bibinfo {pages} {114002} (\bibinfo {year} {2015})},\ \Eprint
  {http://arxiv.org/abs/1508.00924} {arXiv:1508.00924 [hep-ph]} \BibitemShut
  {NoStop}%
\bibitem [{\citenamefont {Chen}\ \emph
  {et~al.}(2015{\natexlab{a}})\citenamefont {Chen}, \citenamefont {Liu},
  \citenamefont {Li},\ and\ \citenamefont {Zhu}}]{Chen:2015loa}%
  \BibitemOpen
  \bibfield  {author} {\bibinfo {author} {\bibfnamefont {R.}~\bibnamefont
  {Chen}}, \bibinfo {author} {\bibfnamefont {X.}~\bibnamefont {Liu}}, \bibinfo
  {author} {\bibfnamefont {X.-Q.}\ \bibnamefont {Li}}, \ and\ \bibinfo {author}
  {\bibfnamefont {S.-L.}\ \bibnamefont {Zhu}},\ }\href {\doibase
  10.1103/PhysRevLett.115.132002} {\bibfield  {journal} {\bibinfo  {journal}
  {Phys. Rev. Lett.}\ }\textbf {\bibinfo {volume} {115}},\ \bibinfo {pages}
  {132002} (\bibinfo {year} {2015}{\natexlab{a}})},\ \Eprint
  {http://arxiv.org/abs/1507.03704} {arXiv:1507.03704 [hep-ph]} \BibitemShut
  {NoStop}%
\bibitem [{\citenamefont {Chen}\ \emph
  {et~al.}(2015{\natexlab{b}})\citenamefont {Chen}, \citenamefont {Chen},
  \citenamefont {Liu}, \citenamefont {Steele},\ and\ \citenamefont
  {Zhu}}]{Chen:2015moa}%
  \BibitemOpen
  \bibfield  {author} {\bibinfo {author} {\bibfnamefont {H.-X.}\ \bibnamefont
  {Chen}}, \bibinfo {author} {\bibfnamefont {W.}~\bibnamefont {Chen}}, \bibinfo
  {author} {\bibfnamefont {X.}~\bibnamefont {Liu}}, \bibinfo {author}
  {\bibfnamefont {T.~G.}\ \bibnamefont {Steele}}, \ and\ \bibinfo {author}
  {\bibfnamefont {S.-L.}\ \bibnamefont {Zhu}},\ }\href {\doibase
  10.1103/PhysRevLett.115.172001} {\bibfield  {journal} {\bibinfo  {journal}
  {Phys. Rev. Lett.}\ }\textbf {\bibinfo {volume} {115}},\ \bibinfo {pages}
  {172001} (\bibinfo {year} {2015}{\natexlab{b}})},\ \Eprint
  {http://arxiv.org/abs/1507.03717} {arXiv:1507.03717 [hep-ph]} \BibitemShut
  {NoStop}%
\bibitem [{\citenamefont {Meißner}\ and\ \citenamefont
  {Oller}(2015)}]{Meissner:2015mza}%
  \BibitemOpen
  \bibfield  {author} {\bibinfo {author} {\bibfnamefont {U.-G.}\ \bibnamefont
  {Meißner}}\ and\ \bibinfo {author} {\bibfnamefont {J.~A.}\ \bibnamefont
  {Oller}},\ }\href {\doibase 10.1016/j.physletb.2015.10.015} {\bibfield
  {journal} {\bibinfo  {journal} {Phys. Lett.}\ }\textbf {\bibinfo {volume}
  {B751}},\ \bibinfo {pages} {59} (\bibinfo {year} {2015})},\ \Eprint
  {http://arxiv.org/abs/1507.07478} {arXiv:1507.07478 [hep-ph]} \BibitemShut
  {NoStop}%
\bibitem [{\citenamefont {Kubarovsky}\ and\ \citenamefont
  {Voloshin}(2015)}]{Kubarovsky:2015aaa}%
  \BibitemOpen
  \bibfield  {author} {\bibinfo {author} {\bibfnamefont {V.}~\bibnamefont
  {Kubarovsky}}\ and\ \bibinfo {author} {\bibfnamefont {M.~B.}\ \bibnamefont
  {Voloshin}},\ }\href {\doibase 10.1103/PhysRevD.92.031502} {\bibfield
  {journal} {\bibinfo  {journal} {Phys. Rev.}\ }\textbf {\bibinfo {volume}
  {D92}},\ \bibinfo {pages} {031502} (\bibinfo {year} {2015})},\ \Eprint
  {http://arxiv.org/abs/1508.00888} {arXiv:1508.00888 [hep-ph]} \BibitemShut
  {NoStop}%
\bibitem [{\citenamefont {Diakonov}\ \emph {et~al.}(1997)\citenamefont
  {Diakonov}, \citenamefont {Petrov},\ and\ \citenamefont
  {Polyakov}}]{Diakonov:1997mm}%
  \BibitemOpen
  \bibfield  {author} {\bibinfo {author} {\bibfnamefont {D.}~\bibnamefont
  {Diakonov}}, \bibinfo {author} {\bibfnamefont {V.}~\bibnamefont {Petrov}}, \
  and\ \bibinfo {author} {\bibfnamefont {M.~V.}\ \bibnamefont {Polyakov}},\
  }\href {\doibase 10.1007/s002180050406} {\bibfield  {journal} {\bibinfo
  {journal} {Z. Phys.}\ }\textbf {\bibinfo {volume} {A359}},\ \bibinfo {pages}
  {305} (\bibinfo {year} {1997})},\ \Eprint
  {http://arxiv.org/abs/hep-ph/9703373} {arXiv:hep-ph/9703373 [hep-ph]}
  \BibitemShut {NoStop}%
\bibitem [{\citenamefont {Jaffe}\ and\ \citenamefont
  {Wilczek}(2003)}]{Jaffe:2003sg}%
  \BibitemOpen
  \bibfield  {author} {\bibinfo {author} {\bibfnamefont {R.~L.}\ \bibnamefont
  {Jaffe}}\ and\ \bibinfo {author} {\bibfnamefont {F.}~\bibnamefont
  {Wilczek}},\ }\href {\doibase 10.1103/PhysRevLett.91.232003} {\bibfield
  {journal} {\bibinfo  {journal} {Phys. Rev. Lett.}\ }\textbf {\bibinfo
  {volume} {91}},\ \bibinfo {pages} {232003} (\bibinfo {year} {2003})},\
  \Eprint {http://arxiv.org/abs/hep-ph/0307341} {arXiv:hep-ph/0307341 [hep-ph]}
  \BibitemShut {NoStop}%
\bibitem [{\citenamefont {Yuan}\ \emph {et~al.}(2012)\citenamefont {Yuan},
  \citenamefont {Wei}, \citenamefont {He}, \citenamefont {Xu},\ and\
  \citenamefont {Zou}}]{Yuan:2012wz}%
  \BibitemOpen
  \bibfield  {author} {\bibinfo {author} {\bibfnamefont {S.~G.}\ \bibnamefont
  {Yuan}}, \bibinfo {author} {\bibfnamefont {K.~W.}\ \bibnamefont {Wei}},
  \bibinfo {author} {\bibfnamefont {J.}~\bibnamefont {He}}, \bibinfo {author}
  {\bibfnamefont {H.~S.}\ \bibnamefont {Xu}}, \ and\ \bibinfo {author}
  {\bibfnamefont {B.~S.}\ \bibnamefont {Zou}},\ }\href {\doibase
  10.1140/epja/i2012-12061-2} {\bibfield  {journal} {\bibinfo  {journal} {Eur.
  Phys. J.}\ }\textbf {\bibinfo {volume} {A48}},\ \bibinfo {pages} {61}
  (\bibinfo {year} {2012})},\ \Eprint {http://arxiv.org/abs/1201.0807}
  {arXiv:1201.0807 [nucl-th]} \BibitemShut {NoStop}%
\bibitem [{\citenamefont {Maiani}\ \emph {et~al.}(2015)\citenamefont {Maiani},
  \citenamefont {Polosa},\ and\ \citenamefont {Riquer}}]{Maiani:2015vwa}%
  \BibitemOpen
  \bibfield  {author} {\bibinfo {author} {\bibfnamefont {L.}~\bibnamefont
  {Maiani}}, \bibinfo {author} {\bibfnamefont {A.~D.}\ \bibnamefont {Polosa}},
  \ and\ \bibinfo {author} {\bibfnamefont {V.}~\bibnamefont {Riquer}},\ }\href
  {\doibase 10.1016/j.physletb.2015.08.008} {\bibfield  {journal} {\bibinfo
  {journal} {Phys. Lett.}\ }\textbf {\bibinfo {volume} {B749}},\ \bibinfo
  {pages} {289} (\bibinfo {year} {2015})},\ \Eprint
  {http://arxiv.org/abs/1507.04980} {arXiv:1507.04980 [hep-ph]} \BibitemShut
  {NoStop}%
\bibitem [{\citenamefont {Lebed}(2015)}]{Lebed:2015tna}%
  \BibitemOpen
  \bibfield  {author} {\bibinfo {author} {\bibfnamefont {R.~F.}\ \bibnamefont
  {Lebed}},\ }\href {\doibase 10.1016/j.physletb.2015.08.032} {\bibfield
  {journal} {\bibinfo  {journal} {Phys. Lett.}\ }\textbf {\bibinfo {volume}
  {B749}},\ \bibinfo {pages} {454} (\bibinfo {year} {2015})},\ \Eprint
  {http://arxiv.org/abs/1507.05867} {arXiv:1507.05867 [hep-ph]} \BibitemShut
  {NoStop}%
\bibitem [{\citenamefont {Li}\ \emph {et~al.}(2015)\citenamefont {Li},
  \citenamefont {He},\ and\ \citenamefont {He}}]{Li:2015gta}%
  \BibitemOpen
  \bibfield  {author} {\bibinfo {author} {\bibfnamefont {G.-N.}\ \bibnamefont
  {Li}}, \bibinfo {author} {\bibfnamefont {X.-G.}\ \bibnamefont {He}}, \ and\
  \bibinfo {author} {\bibfnamefont {M.}~\bibnamefont {He}},\ }\href {\doibase
  10.1007/JHEP12(2015)128} {\bibfield  {journal} {\bibinfo  {journal} {JHEP}\
  }\textbf {\bibinfo {volume} {12}},\ \bibinfo {pages} {128} (\bibinfo {year}
  {2015})},\ \Eprint {http://arxiv.org/abs/1507.08252} {arXiv:1507.08252
  [hep-ph]} \BibitemShut {NoStop}%
\bibitem [{\citenamefont {Liu}\ \emph {et~al.}(2018{\natexlab{a}})\citenamefont
  {Liu}, \citenamefont {Peng}, \citenamefont {Sánchez~Sánchez},\ and\
  \citenamefont {Valderrama}}]{Liu:2018zzu}%
  \BibitemOpen
  \bibfield  {author} {\bibinfo {author} {\bibfnamefont {M.-Z.}\ \bibnamefont
  {Liu}}, \bibinfo {author} {\bibfnamefont {F.-Z.}\ \bibnamefont {Peng}},
  \bibinfo {author} {\bibfnamefont {M.}~\bibnamefont {Sánchez~Sánchez}}, \
  and\ \bibinfo {author} {\bibfnamefont {M.~P.}\ \bibnamefont {Valderrama}},\
  }\href {\doibase 10.1103/PhysRevD.98.114030} {\bibfield  {journal} {\bibinfo
  {journal} {Phys. Rev.}\ }\textbf {\bibinfo {volume} {D98}},\ \bibinfo {pages}
  {114030} (\bibinfo {year} {2018}{\natexlab{a}})},\ \Eprint
  {http://arxiv.org/abs/1811.03992} {arXiv:1811.03992 [hep-ph]} \BibitemShut
  {NoStop}%
\bibitem [{\citenamefont {Burns}(2015)}]{Burns:2015dwa}%
  \BibitemOpen
  \bibfield  {author} {\bibinfo {author} {\bibfnamefont {T.~J.}\ \bibnamefont
  {Burns}},\ }\href {\doibase 10.1140/epja/i2015-15152-6} {\bibfield  {journal}
  {\bibinfo  {journal} {Eur. Phys. J.}\ }\textbf {\bibinfo {volume} {A51}},\
  \bibinfo {pages} {152} (\bibinfo {year} {2015})},\ \Eprint
  {http://arxiv.org/abs/1509.02460} {arXiv:1509.02460 [hep-ph]} \BibitemShut
  {NoStop}%
\bibitem [{\citenamefont {Geng}\ \emph {et~al.}(2018)\citenamefont {Geng},
  \citenamefont {Lu},\ and\ \citenamefont {Valderrama}}]{Geng:2017hxc}%
  \BibitemOpen
  \bibfield  {author} {\bibinfo {author} {\bibfnamefont {L.}~\bibnamefont
  {Geng}}, \bibinfo {author} {\bibfnamefont {J.}~\bibnamefont {Lu}}, \ and\
  \bibinfo {author} {\bibfnamefont {M.~P.}\ \bibnamefont {Valderrama}},\ }\href
  {\doibase 10.1103/PhysRevD.97.094036} {\bibfield  {journal} {\bibinfo
  {journal} {Phys. Rev.}\ }\textbf {\bibinfo {volume} {D97}},\ \bibinfo {pages}
  {094036} (\bibinfo {year} {2018})},\ \Eprint
  {http://arxiv.org/abs/1704.06123} {arXiv:1704.06123 [hep-ph]} \BibitemShut
  {NoStop}%
\bibitem [{\citenamefont {Machleidt}\ \emph {et~al.}(1987)\citenamefont
  {Machleidt}, \citenamefont {Holinde},\ and\ \citenamefont
  {Elster}}]{Machleidt:1987hj}%
  \BibitemOpen
  \bibfield  {author} {\bibinfo {author} {\bibfnamefont {R.}~\bibnamefont
  {Machleidt}}, \bibinfo {author} {\bibfnamefont {K.}~\bibnamefont {Holinde}},
  \ and\ \bibinfo {author} {\bibfnamefont {C.}~\bibnamefont {Elster}},\ }\href
  {\doibase 10.1016/S0370-1573(87)80002-9} {\bibfield  {journal} {\bibinfo
  {journal} {Phys. Rept.}\ }\textbf {\bibinfo {volume} {149}},\ \bibinfo
  {pages} {1} (\bibinfo {year} {1987})}\BibitemShut {NoStop}%
\bibitem [{\citenamefont {Machleidt}(1989)}]{Machleidt:1989tm}%
  \BibitemOpen
  \bibfield  {author} {\bibinfo {author} {\bibfnamefont {R.}~\bibnamefont
  {Machleidt}},\ }\href@noop {} {\bibfield  {journal} {\bibinfo  {journal}
  {Adv. Nucl. Phys.}\ }\textbf {\bibinfo {volume} {19}},\ \bibinfo {pages}
  {189} (\bibinfo {year} {1989})}\BibitemShut {NoStop}%
\bibitem [{\citenamefont {Liu}\ \emph {et~al.}(2009)\citenamefont {Liu},
  \citenamefont {Luo}, \citenamefont {Liu},\ and\ \citenamefont
  {Zhu}}]{Liu:2008tn}%
  \BibitemOpen
  \bibfield  {author} {\bibinfo {author} {\bibfnamefont {X.}~\bibnamefont
  {Liu}}, \bibinfo {author} {\bibfnamefont {Z.-G.}\ \bibnamefont {Luo}},
  \bibinfo {author} {\bibfnamefont {Y.-R.}\ \bibnamefont {Liu}}, \ and\
  \bibinfo {author} {\bibfnamefont {S.-L.}\ \bibnamefont {Zhu}},\ }\href
  {\doibase 10.1140/epjc/s10052-009-1020-4} {\bibfield  {journal} {\bibinfo
  {journal} {Eur. Phys. J.}\ }\textbf {\bibinfo {volume} {C61}},\ \bibinfo
  {pages} {411} (\bibinfo {year} {2009})},\ \Eprint
  {http://arxiv.org/abs/0808.0073} {arXiv:0808.0073 [hep-ph]} \BibitemShut
  {NoStop}%
\bibitem [{\citenamefont {Sun}\ \emph {et~al.}(2011)\citenamefont {Sun},
  \citenamefont {He}, \citenamefont {Liu}, \citenamefont {Luo},\ and\
  \citenamefont {Zhu}}]{Sun:2011uh}%
  \BibitemOpen
  \bibfield  {author} {\bibinfo {author} {\bibfnamefont {Z.-F.}\ \bibnamefont
  {Sun}}, \bibinfo {author} {\bibfnamefont {J.}~\bibnamefont {He}}, \bibinfo
  {author} {\bibfnamefont {X.}~\bibnamefont {Liu}}, \bibinfo {author}
  {\bibfnamefont {Z.-G.}\ \bibnamefont {Luo}}, \ and\ \bibinfo {author}
  {\bibfnamefont {S.-L.}\ \bibnamefont {Zhu}},\ }\href {\doibase
  10.1103/PhysRevD.84.054002} {\bibfield  {journal} {\bibinfo  {journal} {Phys.
  Rev.}\ }\textbf {\bibinfo {volume} {D84}},\ \bibinfo {pages} {054002}
  (\bibinfo {year} {2011})},\ \Eprint {http://arxiv.org/abs/1106.2968}
  {arXiv:1106.2968 [hep-ph]} \BibitemShut {NoStop}%
\bibitem [{\citenamefont {Wang}\ \emph {et~al.}(2020)\citenamefont {Wang},
  \citenamefont {Chen}, \citenamefont {Liu},\ and\ \citenamefont
  {Liu}}]{Wang:2019nwt}%
  \BibitemOpen
  \bibfield  {author} {\bibinfo {author} {\bibfnamefont {F.-L.}\ \bibnamefont
  {Wang}}, \bibinfo {author} {\bibfnamefont {R.}~\bibnamefont {Chen}}, \bibinfo
  {author} {\bibfnamefont {Z.-W.}\ \bibnamefont {Liu}}, \ and\ \bibinfo
  {author} {\bibfnamefont {X.}~\bibnamefont {Liu}},\ }\href {\doibase
  10.1103/PhysRevC.101.025201} {\bibfield  {journal} {\bibinfo  {journal}
  {Phys. Rev. C}\ }\textbf {\bibinfo {volume} {101}},\ \bibinfo {pages}
  {025201} (\bibinfo {year} {2020})},\ \Eprint
  {http://arxiv.org/abs/1905.03636} {arXiv:1905.03636 [hep-ph]} \BibitemShut
  {NoStop}%
\bibitem [{\citenamefont {Liu}\ \emph {et~al.}(2018{\natexlab{b}})\citenamefont
  {Liu}, \citenamefont {Wu}, \citenamefont {Xie}, \citenamefont
  {Pavon~Valderrama},\ and\ \citenamefont {Geng}}]{Liu:2018bkx}%
  \BibitemOpen
  \bibfield  {author} {\bibinfo {author} {\bibfnamefont {M.-Z.}\ \bibnamefont
  {Liu}}, \bibinfo {author} {\bibfnamefont {T.-W.}\ \bibnamefont {Wu}},
  \bibinfo {author} {\bibfnamefont {J.-J.}\ \bibnamefont {Xie}}, \bibinfo
  {author} {\bibfnamefont {M.}~\bibnamefont {Pavon~Valderrama}}, \ and\
  \bibinfo {author} {\bibfnamefont {L.-S.}\ \bibnamefont {Geng}},\ }\href
  {\doibase 10.1103/PhysRevD.98.014014} {\bibfield  {journal} {\bibinfo
  {journal} {Phys. Rev.}\ }\textbf {\bibinfo {volume} {D98}},\ \bibinfo {pages}
  {014014} (\bibinfo {year} {2018}{\natexlab{b}})},\ \Eprint
  {http://arxiv.org/abs/1805.08384} {arXiv:1805.08384 [hep-ph]} \BibitemShut
  {NoStop}%
\bibitem [{\citenamefont {Liu}\ \emph {et~al.}(2019{\natexlab{b}})\citenamefont
  {Liu}, \citenamefont {Wu}, \citenamefont {Pavon~Valderrama}, \citenamefont
  {Xie},\ and\ \citenamefont {Geng}}]{Liu:2019stu}%
  \BibitemOpen
  \bibfield  {author} {\bibinfo {author} {\bibfnamefont {M.-Z.}\ \bibnamefont
  {Liu}}, \bibinfo {author} {\bibfnamefont {T.-W.}\ \bibnamefont {Wu}},
  \bibinfo {author} {\bibfnamefont {M.}~\bibnamefont {Pavon~Valderrama}},
  \bibinfo {author} {\bibfnamefont {J.-J.}\ \bibnamefont {Xie}}, \ and\
  \bibinfo {author} {\bibfnamefont {L.-S.}\ \bibnamefont {Geng}},\ }\href
  {\doibase 10.1103/PhysRevD.99.094018} {\bibfield  {journal} {\bibinfo
  {journal} {Phys. Rev.}\ }\textbf {\bibinfo {volume} {D99}},\ \bibinfo {pages}
  {094018} (\bibinfo {year} {2019}{\natexlab{b}})},\ \Eprint
  {http://arxiv.org/abs/1902.03044} {arXiv:1902.03044 [hep-ph]} \BibitemShut
  {NoStop}%
\bibitem [{\citenamefont {Calle~Cordon}\ and\ \citenamefont
  {Ruiz~Arriola}(2010)}]{Cordon:2009pj}%
  \BibitemOpen
  \bibfield  {author} {\bibinfo {author} {\bibfnamefont {A.}~\bibnamefont
  {Calle~Cordon}}\ and\ \bibinfo {author} {\bibfnamefont {E.}~\bibnamefont
  {Ruiz~Arriola}},\ }\href {\doibase 10.1103/PhysRevC.81.044002} {\bibfield
  {journal} {\bibinfo  {journal} {Phys. Rev.}\ }\textbf {\bibinfo {volume}
  {C81}},\ \bibinfo {pages} {044002} (\bibinfo {year} {2010})},\ \Eprint
  {http://arxiv.org/abs/0905.4933} {arXiv:0905.4933 [nucl-th]} \BibitemShut
  {NoStop}%
\bibitem [{\citenamefont {Meng}\ \emph {et~al.}(2017)\citenamefont {Meng},
  \citenamefont {Li},\ and\ \citenamefont {Zhu}}]{Meng:2017fwb}%
  \BibitemOpen
  \bibfield  {author} {\bibinfo {author} {\bibfnamefont {L.}~\bibnamefont
  {Meng}}, \bibinfo {author} {\bibfnamefont {N.}~\bibnamefont {Li}}, \ and\
  \bibinfo {author} {\bibfnamefont {S.-L.}\ \bibnamefont {Zhu}},\ }\href
  {\doibase 10.1103/PhysRevD.95.114019} {\bibfield  {journal} {\bibinfo
  {journal} {Phys. Rev.}\ }\textbf {\bibinfo {volume} {D95}},\ \bibinfo {pages}
  {114019} (\bibinfo {year} {2017})},\ \Eprint
  {http://arxiv.org/abs/1704.01009} {arXiv:1704.01009 [hep-ph]} \BibitemShut
  {NoStop}%
\bibitem [{\citenamefont {Pavon~Valderrama}(2020)}]{Valderrama:2019sid}%
  \BibitemOpen
  \bibfield  {author} {\bibinfo {author} {\bibfnamefont {M.}~\bibnamefont
  {Pavon~Valderrama}},\ }\href {\doibase 10.1140/epja/s10050-020-00099-8}
  {\bibfield  {journal} {\bibinfo  {journal} {Eur. Phys. J. A}\ }\textbf
  {\bibinfo {volume} {56}},\ \bibinfo {pages} {109} (\bibinfo {year} {2020})},\
  \Eprint {http://arxiv.org/abs/1906.06491} {arXiv:1906.06491 [hep-ph]}
  \BibitemShut {NoStop}%
\bibitem [{\citenamefont {Manohar}\ and\ \citenamefont
  {Wise}(1993)}]{Manohar:1992nd}%
  \BibitemOpen
  \bibfield  {author} {\bibinfo {author} {\bibfnamefont {A.~V.}\ \bibnamefont
  {Manohar}}\ and\ \bibinfo {author} {\bibfnamefont {M.~B.}\ \bibnamefont
  {Wise}},\ }\href {\doibase 10.1016/0550-3213(93)90614-U} {\bibfield
  {journal} {\bibinfo  {journal} {Nucl.Phys.}\ }\textbf {\bibinfo {volume}
  {B399}},\ \bibinfo {pages} {17} (\bibinfo {year} {1993})},\ \Eprint
  {http://arxiv.org/abs/hep-ph/9212236} {arXiv:hep-ph/9212236 [hep-ph]}
  \BibitemShut {NoStop}%
\bibitem [{\citenamefont {Falk}\ and\ \citenamefont
  {Luke}(1992)}]{Falk:1992cx}%
  \BibitemOpen
  \bibfield  {author} {\bibinfo {author} {\bibfnamefont {A.~F.}\ \bibnamefont
  {Falk}}\ and\ \bibinfo {author} {\bibfnamefont {M.~E.}\ \bibnamefont
  {Luke}},\ }\href {\doibase 10.1016/0370-2693(92)90618-E} {\bibfield
  {journal} {\bibinfo  {journal} {Phys. Lett.}\ }\textbf {\bibinfo {volume}
  {B292}},\ \bibinfo {pages} {119} (\bibinfo {year} {1992})},\ \Eprint
  {http://arxiv.org/abs/hep-ph/9206241} {arXiv:hep-ph/9206241 [hep-ph]}
  \BibitemShut {NoStop}%
\bibitem [{\citenamefont {Lu}\ \emph {et~al.}(2019)\citenamefont {Lu},
  \citenamefont {Geng},\ and\ \citenamefont {Valderrama}}]{Lu:2017dvm}%
  \BibitemOpen
  \bibfield  {author} {\bibinfo {author} {\bibfnamefont {J.-X.}\ \bibnamefont
  {Lu}}, \bibinfo {author} {\bibfnamefont {L.-S.}\ \bibnamefont {Geng}}, \ and\
  \bibinfo {author} {\bibfnamefont {M.~P.}\ \bibnamefont {Valderrama}},\ }\href
  {\doibase 10.1103/PhysRevD.99.074026} {\bibfield  {journal} {\bibinfo
  {journal} {Phys. Rev.}\ }\textbf {\bibinfo {volume} {D99}},\ \bibinfo {pages}
  {074026} (\bibinfo {year} {2019})},\ \Eprint
  {http://arxiv.org/abs/1706.02588} {arXiv:1706.02588 [hep-ph]} \BibitemShut
  {NoStop}%
\bibitem [{\citenamefont {Cho}(1993)}]{Cho:1992cf}%
  \BibitemOpen
  \bibfield  {author} {\bibinfo {author} {\bibfnamefont {P.~L.}\ \bibnamefont
  {Cho}},\ }\href {\doibase 10.1016/0550-3213(94)90522-3,
  10.1016/0550-3213(93)90263-O} {\bibfield  {journal} {\bibinfo  {journal}
  {Nucl. Phys.}\ }\textbf {\bibinfo {volume} {B396}},\ \bibinfo {pages} {183}
  (\bibinfo {year} {1993})},\ \bibinfo {note} {[Erratum: Nucl.
  Phys.B421,683(1994)]},\ \Eprint {http://arxiv.org/abs/hep-ph/9208244}
  {arXiv:hep-ph/9208244 [hep-ph]} \BibitemShut {NoStop}%
\bibitem [{\citenamefont {Durso}\ \emph {et~al.}(1984)\citenamefont {Durso},
  \citenamefont {Brown},\ and\ \citenamefont {Saarela}}]{Durso:1984um}%
  \BibitemOpen
  \bibfield  {author} {\bibinfo {author} {\bibfnamefont {J.}~\bibnamefont
  {Durso}}, \bibinfo {author} {\bibfnamefont {G.}~\bibnamefont {Brown}}, \ and\
  \bibinfo {author} {\bibfnamefont {M.}~\bibnamefont {Saarela}},\ }\href
  {\doibase 10.1016/0375-9474(84)90099-X} {\bibfield  {journal} {\bibinfo
  {journal} {Nucl.\ Phys.\ A}\ }\textbf {\bibinfo {volume} {430}},\ \bibinfo
  {pages} {653} (\bibinfo {year} {1984})}\BibitemShut {NoStop}%
\bibitem [{\citenamefont {Tanabashi}\ \emph {et~al.}(2018)\citenamefont
  {Tanabashi} \emph {et~al.}}]{Tanabashi:2018oca}%
  \BibitemOpen
  \bibfield  {author} {\bibinfo {author} {\bibfnamefont {M.}~\bibnamefont
  {Tanabashi}} \emph {et~al.} (\bibinfo {collaboration} {Particle Data
  Group}),\ }\href {\doibase 10.1103/PhysRevD.98.030001} {\bibfield  {journal}
  {\bibinfo  {journal} {Phys. Rev.}\ }\textbf {\bibinfo {volume} {D98}},\
  \bibinfo {pages} {030001} (\bibinfo {year} {2018})}\BibitemShut {NoStop}%
\bibitem [{\citenamefont {Ahmed}\ \emph {et~al.}(2001)\citenamefont {Ahmed}
  \emph {et~al.}}]{Ahmed:2001xc}%
  \BibitemOpen
  \bibfield  {author} {\bibinfo {author} {\bibfnamefont {S.}~\bibnamefont
  {Ahmed}} \emph {et~al.} (\bibinfo {collaboration} {CLEO Collaboration}),\
  }\href {\doibase 10.1103/PhysRevLett.87.251801} {\bibfield  {journal}
  {\bibinfo  {journal} {Phys.Rev.Lett.}\ }\textbf {\bibinfo {volume} {87}},\
  \bibinfo {pages} {251801} (\bibinfo {year} {2001})},\ \Eprint
  {http://arxiv.org/abs/hep-ex/0108013} {arXiv:hep-ex/0108013 [hep-ex]}
  \BibitemShut {NoStop}%
\bibitem [{\citenamefont {Anastassov}\ \emph {et~al.}(2002)\citenamefont
  {Anastassov} \emph {et~al.}}]{Anastassov:2001cw}%
  \BibitemOpen
  \bibfield  {author} {\bibinfo {author} {\bibfnamefont {A.}~\bibnamefont
  {Anastassov}} \emph {et~al.} (\bibinfo {collaboration} {CLEO
  Collaboration}),\ }\href {\doibase 10.1103/PhysRevD.65.032003} {\bibfield
  {journal} {\bibinfo  {journal} {Phys.Rev.}\ }\textbf {\bibinfo {volume}
  {D65}},\ \bibinfo {pages} {032003} (\bibinfo {year} {2002})},\ \Eprint
  {http://arxiv.org/abs/hep-ex/0108043} {arXiv:hep-ex/0108043 [hep-ex]}
  \BibitemShut {NoStop}%
\bibitem [{\citenamefont {Detmold}\ \emph {et~al.}(2012)\citenamefont
  {Detmold}, \citenamefont {Lin},\ and\ \citenamefont
  {Meinel}}]{Detmold:2012ge}%
  \BibitemOpen
  \bibfield  {author} {\bibinfo {author} {\bibfnamefont {W.}~\bibnamefont
  {Detmold}}, \bibinfo {author} {\bibfnamefont {C.~J.~D.}\ \bibnamefont {Lin}},
  \ and\ \bibinfo {author} {\bibfnamefont {S.}~\bibnamefont {Meinel}},\ }\href
  {\doibase 10.1103/PhysRevD.85.114508} {\bibfield  {journal} {\bibinfo
  {journal} {Phys. Rev.}\ }\textbf {\bibinfo {volume} {D85}},\ \bibinfo {pages}
  {114508} (\bibinfo {year} {2012})},\ \Eprint {http://arxiv.org/abs/1203.3378}
  {arXiv:1203.3378 [hep-lat]} \BibitemShut {NoStop}%
\bibitem [{\citenamefont {Yan}\ \emph {et~al.}(1992)\citenamefont {Yan},
  \citenamefont {Cheng}, \citenamefont {Cheung}, \citenamefont {Lin},
  \citenamefont {Lin},\ and\ \citenamefont {Yu}}]{Yan:1992gz}%
  \BibitemOpen
  \bibfield  {author} {\bibinfo {author} {\bibfnamefont {T.-M.}\ \bibnamefont
  {Yan}}, \bibinfo {author} {\bibfnamefont {H.-Y.}\ \bibnamefont {Cheng}},
  \bibinfo {author} {\bibfnamefont {C.-Y.}\ \bibnamefont {Cheung}}, \bibinfo
  {author} {\bibfnamefont {G.-L.}\ \bibnamefont {Lin}}, \bibinfo {author}
  {\bibfnamefont {Y.~C.}\ \bibnamefont {Lin}}, \ and\ \bibinfo {author}
  {\bibfnamefont {H.-L.}\ \bibnamefont {Yu}},\ }\href {\doibase
  10.1103/PhysRevD.46.1148, 10.1103/PhysRevD.55.5851} {\bibfield  {journal}
  {\bibinfo  {journal} {Phys. Rev.}\ }\textbf {\bibinfo {volume} {D46}},\
  \bibinfo {pages} {1148} (\bibinfo {year} {1992})},\ \bibinfo {note}
  {[Erratum: Phys. Rev.D55,5851(1997)]}\BibitemShut {NoStop}%
\bibitem [{\citenamefont {Gell-Mann}\ and\ \citenamefont
  {Levy}(1960)}]{GellMann:1960np}%
  \BibitemOpen
  \bibfield  {author} {\bibinfo {author} {\bibfnamefont {M.}~\bibnamefont
  {Gell-Mann}}\ and\ \bibinfo {author} {\bibfnamefont {M.}~\bibnamefont
  {Levy}},\ }\href {\doibase 10.1007/BF02859738} {\bibfield  {journal}
  {\bibinfo  {journal} {Nuovo Cim.}\ }\textbf {\bibinfo {volume} {16}},\
  \bibinfo {pages} {705} (\bibinfo {year} {1960})}\BibitemShut {NoStop}%
\bibitem [{\citenamefont {Sakurai}(1960)}]{Sakurai:1960ju}%
  \BibitemOpen
  \bibfield  {author} {\bibinfo {author} {\bibfnamefont {J.~J.}\ \bibnamefont
  {Sakurai}},\ }\href {\doibase 10.1016/0003-4916(60)90126-3} {\bibfield
  {journal} {\bibinfo  {journal} {Annals Phys.}\ }\textbf {\bibinfo {volume}
  {11}},\ \bibinfo {pages} {1} (\bibinfo {year} {1960})}\BibitemShut {NoStop}%
\bibitem [{\citenamefont {Casalbuoni}\ \emph {et~al.}(1993)\citenamefont
  {Casalbuoni}, \citenamefont {Deandrea}, \citenamefont {Di~Bartolomeo},
  \citenamefont {Gatto}, \citenamefont {Feruglio},\ and\ \citenamefont
  {Nardulli}}]{Casalbuoni:1992dx}%
  \BibitemOpen
  \bibfield  {author} {\bibinfo {author} {\bibfnamefont {R.}~\bibnamefont
  {Casalbuoni}}, \bibinfo {author} {\bibfnamefont {A.}~\bibnamefont
  {Deandrea}}, \bibinfo {author} {\bibfnamefont {N.}~\bibnamefont
  {Di~Bartolomeo}}, \bibinfo {author} {\bibfnamefont {R.}~\bibnamefont
  {Gatto}}, \bibinfo {author} {\bibfnamefont {F.}~\bibnamefont {Feruglio}}, \
  and\ \bibinfo {author} {\bibfnamefont {G.}~\bibnamefont {Nardulli}},\ }\href
  {\doibase 10.1016/0370-2693(93)90895-O} {\bibfield  {journal} {\bibinfo
  {journal} {Phys. Lett.}\ }\textbf {\bibinfo {volume} {B299}},\ \bibinfo
  {pages} {139} (\bibinfo {year} {1993})},\ \Eprint
  {http://arxiv.org/abs/hep-ph/9211248} {arXiv:hep-ph/9211248 [hep-ph]}
  \BibitemShut {NoStop}%
\bibitem [{\citenamefont {Can}\ \emph {et~al.}(2014)\citenamefont {Can},
  \citenamefont {Erkol}, \citenamefont {Isildak}, \citenamefont {Oka},\ and\
  \citenamefont {Takahashi}}]{Can:2013tna}%
  \BibitemOpen
  \bibfield  {author} {\bibinfo {author} {\bibfnamefont {K.~U.}\ \bibnamefont
  {Can}}, \bibinfo {author} {\bibfnamefont {G.}~\bibnamefont {Erkol}}, \bibinfo
  {author} {\bibfnamefont {B.}~\bibnamefont {Isildak}}, \bibinfo {author}
  {\bibfnamefont {M.}~\bibnamefont {Oka}}, \ and\ \bibinfo {author}
  {\bibfnamefont {T.~T.}\ \bibnamefont {Takahashi}},\ }\href {\doibase
  10.1007/JHEP05(2014)125} {\bibfield  {journal} {\bibinfo  {journal} {JHEP}\
  }\textbf {\bibinfo {volume} {05}},\ \bibinfo {pages} {125} (\bibinfo {year}
  {2014})},\ \Eprint {http://arxiv.org/abs/1310.5915} {arXiv:1310.5915
  [hep-lat]} \BibitemShut {NoStop}%
\bibitem [{\citenamefont {Detmold}\ \emph {et~al.}(2007)\citenamefont
  {Detmold}, \citenamefont {Orginos},\ and\ \citenamefont
  {Savage}}]{Detmold:2007wk}%
  \BibitemOpen
  \bibfield  {author} {\bibinfo {author} {\bibfnamefont {W.}~\bibnamefont
  {Detmold}}, \bibinfo {author} {\bibfnamefont {K.}~\bibnamefont {Orginos}}, \
  and\ \bibinfo {author} {\bibfnamefont {M.~J.}\ \bibnamefont {Savage}},\
  }\href {\doibase 10.1103/PhysRevD.76.114503} {\bibfield  {journal} {\bibinfo
  {journal} {Phys.Rev.}\ }\textbf {\bibinfo {volume} {D76}},\ \bibinfo {pages}
  {114503} (\bibinfo {year} {2007})},\ \Eprint
  {http://arxiv.org/abs/hep-lat/0703009} {arXiv:hep-lat/0703009 [HEP-LAT]}
  \BibitemShut {NoStop}%
\bibitem [{\citenamefont {Du}\ \emph {et~al.}(2020)\citenamefont {Du},
  \citenamefont {Baru}, \citenamefont {Guo}, \citenamefont {Hanhart},
  \citenamefont {Mei\ss{}ner}, \citenamefont {Oller},\ and\ \citenamefont
  {Wang}}]{Du:2019pij}%
  \BibitemOpen
  \bibfield  {author} {\bibinfo {author} {\bibfnamefont {M.-L.}\ \bibnamefont
  {Du}}, \bibinfo {author} {\bibfnamefont {V.}~\bibnamefont {Baru}}, \bibinfo
  {author} {\bibfnamefont {F.-K.}\ \bibnamefont {Guo}}, \bibinfo {author}
  {\bibfnamefont {C.}~\bibnamefont {Hanhart}}, \bibinfo {author} {\bibfnamefont
  {U.-G.}\ \bibnamefont {Mei\ss{}ner}}, \bibinfo {author} {\bibfnamefont
  {J.~A.}\ \bibnamefont {Oller}}, \ and\ \bibinfo {author} {\bibfnamefont
  {Q.}~\bibnamefont {Wang}},\ }\href {\doibase 10.1103/PhysRevLett.124.072001}
  {\bibfield  {journal} {\bibinfo  {journal} {Phys. Rev. Lett.}\ }\textbf
  {\bibinfo {volume} {124}},\ \bibinfo {pages} {072001} (\bibinfo {year}
  {2020})},\ \Eprint {http://arxiv.org/abs/1910.11846} {arXiv:1910.11846
  [hep-ph]} \BibitemShut {NoStop}%
\bibitem [{\citenamefont {Karliner}\ and\ \citenamefont
  {Rosner}(2016)}]{Karliner:2016ith}%
  \BibitemOpen
  \bibfield  {author} {\bibinfo {author} {\bibfnamefont {M.}~\bibnamefont
  {Karliner}}\ and\ \bibinfo {author} {\bibfnamefont {J.~L.}\ \bibnamefont
  {Rosner}},\ }\href {\doibase 10.1016/j.nuclphysa.2016.03.057} {\bibfield
  {journal} {\bibinfo  {journal} {Nucl. Phys. A}\ }\textbf {\bibinfo {volume}
  {954}},\ \bibinfo {pages} {365} (\bibinfo {year} {2016})},\ \Eprint
  {http://arxiv.org/abs/1601.00565} {arXiv:1601.00565 [hep-ph]} \BibitemShut
  {NoStop}%
\bibitem [{\citenamefont {Case}(1950)}]{Case:1950an}%
  \BibitemOpen
  \bibfield  {author} {\bibinfo {author} {\bibfnamefont {K.~M.}\ \bibnamefont
  {Case}},\ }\href {\doibase 10.1103/PhysRev.80.797} {\bibfield  {journal}
  {\bibinfo  {journal} {Phys. Rev.}\ }\textbf {\bibinfo {volume} {80}},\
  \bibinfo {pages} {797} (\bibinfo {year} {1950})}\BibitemShut {NoStop}%
\bibitem [{\citenamefont {Beane}\ \emph {et~al.}(2001)\citenamefont {Beane},
  \citenamefont {Bedaque}, \citenamefont {Childress}, \citenamefont
  {Kryjevski}, \citenamefont {McGuire},\ and\ \citenamefont {van
  Kolck}}]{Beane:2000wh}%
  \BibitemOpen
  \bibfield  {author} {\bibinfo {author} {\bibfnamefont {S.~R.}\ \bibnamefont
  {Beane}}, \bibinfo {author} {\bibfnamefont {P.~F.}\ \bibnamefont {Bedaque}},
  \bibinfo {author} {\bibfnamefont {L.}~\bibnamefont {Childress}}, \bibinfo
  {author} {\bibfnamefont {A.}~\bibnamefont {Kryjevski}}, \bibinfo {author}
  {\bibfnamefont {J.}~\bibnamefont {McGuire}}, \ and\ \bibinfo {author}
  {\bibfnamefont {U.}~\bibnamefont {van Kolck}},\ }\href {\doibase
  10.1103/PhysRevA.64.042103} {\bibfield  {journal} {\bibinfo  {journal} {Phys.
  Rev.}\ }\textbf {\bibinfo {volume} {A64}},\ \bibinfo {pages} {042103}
  (\bibinfo {year} {2001})},\ \Eprint {http://arxiv.org/abs/quant-ph/0010073}
  {arXiv:quant-ph/0010073 [quant-ph]} \BibitemShut {NoStop}%
\bibitem [{\citenamefont {Pavon~Valderrama}\ and\ \citenamefont
  {Ruiz~Arriola}(2005)}]{PavonValderrama:2005gu}%
  \BibitemOpen
  \bibfield  {author} {\bibinfo {author} {\bibfnamefont {M.}~\bibnamefont
  {Pavon~Valderrama}}\ and\ \bibinfo {author} {\bibfnamefont {E.}~\bibnamefont
  {Ruiz~Arriola}},\ }\href {\doibase 10.1103/PhysRevC.72.054002} {\bibfield
  {journal} {\bibinfo  {journal} {Phys. Rev.}\ }\textbf {\bibinfo {volume}
  {C72}},\ \bibinfo {pages} {054002} (\bibinfo {year} {2005})},\ \Eprint
  {http://arxiv.org/abs/nucl-th/0504067} {arXiv:nucl-th/0504067} \BibitemShut
  {NoStop}%
\bibitem [{\citenamefont {Pavon~Valderrama}\ and\ \citenamefont
  {Arriola}(2006)}]{PavonValderrama:2005wv}%
  \BibitemOpen
  \bibfield  {author} {\bibinfo {author} {\bibfnamefont {M.}~\bibnamefont
  {Pavon~Valderrama}}\ and\ \bibinfo {author} {\bibfnamefont {E.~R.}\
  \bibnamefont {Arriola}},\ }\href {\doibase 10.1103/PhysRevC.74.054001}
  {\bibfield  {journal} {\bibinfo  {journal} {Phys. Rev.}\ }\textbf {\bibinfo
  {volume} {C74}},\ \bibinfo {pages} {054001} (\bibinfo {year} {2006})},\
  \Eprint {http://arxiv.org/abs/nucl-th/0506047} {arXiv:nucl-th/0506047}
  \BibitemShut {NoStop}%
\bibitem [{\citenamefont {Pavon~Valderrama}\ and\ \citenamefont
  {Ruiz~Arriola}(2006)}]{PavonValderrama:2005uj}%
  \BibitemOpen
  \bibfield  {author} {\bibinfo {author} {\bibfnamefont {M.}~\bibnamefont
  {Pavon~Valderrama}}\ and\ \bibinfo {author} {\bibfnamefont {E.}~\bibnamefont
  {Ruiz~Arriola}},\ }\href {\doibase 10.1103/PhysRevC.74.064004} {\bibfield
  {journal} {\bibinfo  {journal} {Phys. Rev.}\ }\textbf {\bibinfo {volume}
  {C74}},\ \bibinfo {pages} {064004} (\bibinfo {year} {2006})},\ \Eprint
  {http://arxiv.org/abs/nucl-th/0507075} {arXiv:nucl-th/0507075} \BibitemShut
  {NoStop}%
\bibitem [{\citenamefont {Ecker}\ \emph {et~al.}(1989)\citenamefont {Ecker},
  \citenamefont {Gasser}, \citenamefont {Pich},\ and\ \citenamefont
  {de~Rafael}}]{Ecker:1988te}%
  \BibitemOpen
  \bibfield  {author} {\bibinfo {author} {\bibfnamefont {G.}~\bibnamefont
  {Ecker}}, \bibinfo {author} {\bibfnamefont {J.}~\bibnamefont {Gasser}},
  \bibinfo {author} {\bibfnamefont {A.}~\bibnamefont {Pich}}, \ and\ \bibinfo
  {author} {\bibfnamefont {E.}~\bibnamefont {de~Rafael}},\ }\href {\doibase
  10.1016/0550-3213(89)90346-5} {\bibfield  {journal} {\bibinfo  {journal}
  {Nucl. Phys. B}\ }\textbf {\bibinfo {volume} {321}},\ \bibinfo {pages} {311}
  (\bibinfo {year} {1989})}\BibitemShut {NoStop}%
\bibitem [{\citenamefont {Epelbaum}\ \emph {et~al.}(2002)\citenamefont
  {Epelbaum}, \citenamefont {Meissner}, \citenamefont {Gloeckle},\ and\
  \citenamefont {Elster}}]{Epelbaum:2001fm}%
  \BibitemOpen
  \bibfield  {author} {\bibinfo {author} {\bibfnamefont {E.}~\bibnamefont
  {Epelbaum}}, \bibinfo {author} {\bibfnamefont {U.~G.}\ \bibnamefont
  {Meissner}}, \bibinfo {author} {\bibfnamefont {W.}~\bibnamefont {Gloeckle}},
  \ and\ \bibinfo {author} {\bibfnamefont {C.}~\bibnamefont {Elster}},\ }\href
  {\doibase 10.1103/PhysRevC.65.044001} {\bibfield  {journal} {\bibinfo
  {journal} {Phys.Rev.}\ }\textbf {\bibinfo {volume} {C65}},\ \bibinfo {pages}
  {044001} (\bibinfo {year} {2002})},\ \Eprint
  {http://arxiv.org/abs/nucl-th/0106007} {arXiv:nucl-th/0106007 [nucl-th]}
  \BibitemShut {NoStop}%
\bibitem [{\citenamefont {Peng}\ \emph {et~al.}(2020)\citenamefont {Peng},
  \citenamefont {Liu}, \citenamefont {S\'anchez~S\'anchez},\ and\ \citenamefont
  {Pavon~Valderrama}}]{Peng:2020xrf}%
  \BibitemOpen
  \bibfield  {author} {\bibinfo {author} {\bibfnamefont {F.-Z.}\ \bibnamefont
  {Peng}}, \bibinfo {author} {\bibfnamefont {M.-Z.}\ \bibnamefont {Liu}},
  \bibinfo {author} {\bibfnamefont {M.}~\bibnamefont {S\'anchez~S\'anchez}}, \
  and\ \bibinfo {author} {\bibfnamefont {M.}~\bibnamefont {Pavon~Valderrama}},\
  }\href@noop {} {\  (\bibinfo {year} {2020})},\ \Eprint
  {http://arxiv.org/abs/2004.05658} {arXiv:2004.05658 [hep-ph]} \BibitemShut
  {NoStop}%
\bibitem [{\citenamefont {Uchino}\ \emph {et~al.}(2016)\citenamefont {Uchino},
  \citenamefont {Liang},\ and\ \citenamefont {Oset}}]{Uchino:2015uha}%
  \BibitemOpen
  \bibfield  {author} {\bibinfo {author} {\bibfnamefont {T.}~\bibnamefont
  {Uchino}}, \bibinfo {author} {\bibfnamefont {W.-H.}\ \bibnamefont {Liang}}, \
  and\ \bibinfo {author} {\bibfnamefont {E.}~\bibnamefont {Oset}},\ }\href
  {\doibase 10.1140/epja/i2016-16043-0} {\bibfield  {journal} {\bibinfo
  {journal} {Eur. Phys. J.}\ }\textbf {\bibinfo {volume} {A52}},\ \bibinfo
  {pages} {43} (\bibinfo {year} {2016})},\ \Eprint
  {http://arxiv.org/abs/1504.05726} {arXiv:1504.05726 [hep-ph]} \BibitemShut
  {NoStop}%
\bibitem [{\citenamefont {Yamaguchi}\ \emph {et~al.}(2020)\citenamefont
  {Yamaguchi}, \citenamefont {Garc\'\i{}a-Tecocoatzi}, \citenamefont
  {Giachino}, \citenamefont {Hosaka}, \citenamefont {Santopinto}, \citenamefont
  {Takeuchi},\ and\ \citenamefont {Takizawa}}]{Yamaguchi:2019seo}%
  \BibitemOpen
  \bibfield  {author} {\bibinfo {author} {\bibfnamefont {Y.}~\bibnamefont
  {Yamaguchi}}, \bibinfo {author} {\bibfnamefont {H.}~\bibnamefont
  {Garc\'\i{}a-Tecocoatzi}}, \bibinfo {author} {\bibfnamefont {A.}~\bibnamefont
  {Giachino}}, \bibinfo {author} {\bibfnamefont {A.}~\bibnamefont {Hosaka}},
  \bibinfo {author} {\bibfnamefont {E.}~\bibnamefont {Santopinto}}, \bibinfo
  {author} {\bibfnamefont {S.}~\bibnamefont {Takeuchi}}, \ and\ \bibinfo
  {author} {\bibfnamefont {M.}~\bibnamefont {Takizawa}},\ }\href {\doibase
  10.1103/PhysRevD.101.091502} {\bibfield  {journal} {\bibinfo  {journal}
  {Phys. Rev. D}\ }\textbf {\bibinfo {volume} {101}},\ \bibinfo {pages}
  {091502} (\bibinfo {year} {2020})},\ \Eprint
  {http://arxiv.org/abs/1907.04684} {arXiv:1907.04684 [hep-ph]} \BibitemShut
  {NoStop}%
\bibitem [{\citenamefont {Ali}\ \emph {et~al.}(2019)\citenamefont {Ali} \emph
  {et~al.}}]{Ali:2019lzf}%
  \BibitemOpen
  \bibfield  {author} {\bibinfo {author} {\bibfnamefont {A.}~\bibnamefont
  {Ali}} \emph {et~al.} (\bibinfo {collaboration} {GlueX}),\ }\href {\doibase
  10.1103/PhysRevLett.123.072001} {\bibfield  {journal} {\bibinfo  {journal}
  {Phys. Rev. Lett.}\ }\textbf {\bibinfo {volume} {123}},\ \bibinfo {pages}
  {072001} (\bibinfo {year} {2019})},\ \Eprint
  {http://arxiv.org/abs/1905.10811} {arXiv:1905.10811 [nucl-ex]} \BibitemShut
  {NoStop}%
\bibitem [{\citenamefont {Mutuk}(2019)}]{Mutuk:2019snd}%
  \BibitemOpen
  \bibfield  {author} {\bibinfo {author} {\bibfnamefont {H.}~\bibnamefont
  {Mutuk}},\ }\href {\doibase 10.1088/1674-1137/43/9/093103} {\bibfield
  {journal} {\bibinfo  {journal} {Chin. Phys. C}\ }\textbf {\bibinfo {volume}
  {43}},\ \bibinfo {pages} {093103} (\bibinfo {year} {2019})},\ \Eprint
  {http://arxiv.org/abs/1904.09756} {arXiv:1904.09756 [hep-ph]} \BibitemShut
  {NoStop}%
\bibitem [{\citenamefont {Guo}\ \emph {et~al.}(2014)\citenamefont {Guo},
  \citenamefont {Hidalgo-Duque}, \citenamefont {Nieves}, \citenamefont
  {Ozpineci},\ and\ \citenamefont {Valderrama}}]{Guo:2014hqa}%
  \BibitemOpen
  \bibfield  {author} {\bibinfo {author} {\bibfnamefont {F.~K.}\ \bibnamefont
  {Guo}}, \bibinfo {author} {\bibfnamefont {C.}~\bibnamefont {Hidalgo-Duque}},
  \bibinfo {author} {\bibfnamefont {J.}~\bibnamefont {Nieves}}, \bibinfo
  {author} {\bibfnamefont {A.}~\bibnamefont {Ozpineci}}, \ and\ \bibinfo
  {author} {\bibfnamefont {M.~P.}\ \bibnamefont {Valderrama}},\ }\href
  {\doibase 10.1140/epjc/s10052-014-2885-4} {\bibfield  {journal} {\bibinfo
  {journal} {Eur. Phys. J.}\ }\textbf {\bibinfo {volume} {C74}},\ \bibinfo
  {pages} {2885} (\bibinfo {year} {2014})},\ \Eprint
  {http://arxiv.org/abs/1404.1776} {arXiv:1404.1776 [hep-ph]} \BibitemShut
  {NoStop}%
\bibitem [{\citenamefont {Valderrama}(2012)}]{Valderrama:2012jv}%
  \BibitemOpen
  \bibfield  {author} {\bibinfo {author} {\bibfnamefont {M.~P.}\ \bibnamefont
  {Valderrama}},\ }\href {\doibase 10.1103/PhysRevD.85.114037} {\bibfield
  {journal} {\bibinfo  {journal} {Phys. Rev.}\ }\textbf {\bibinfo {volume}
  {D85}},\ \bibinfo {pages} {114037} (\bibinfo {year} {2012})},\ \Eprint
  {http://arxiv.org/abs/1204.2400} {arXiv:1204.2400 [hep-ph]} \BibitemShut
  {NoStop}%
\end{thebibliography}

%

\end{document}